\documentclass{style}
\usepackage{graphicx}
\usepackage{mathrsfs,amsmath,amssymb}

\newcommand{\lc}{\left<}
\newcommand{\rc}{\right>}
\newcommand{\lr}{\left|}
\newcommand{\rl}{\right|}
\newcommand{\lb}{\left(}
\newcommand{\rb}{\right)}
\newcommand{\ls}{\left[}
\newcommand{\rs}{\right]}
\newcommand{\Lb}{\left\{}
\newcommand{\Rb}{\right\}}

\begin{document}

\title{Pseudospin symmetry in nuclear structure and its supersymmetric representation}
\author{
  H Z Liang\email{haozhao.liang@riken.jp} \\
  \it RIKEN Nishina Center, Wako 351-0198, Japan
}



\date{}

\maketitle

\begin{abstract}
  The quasi-degeneracy between the single-particle states $(n,\,l,\,j=l+1/2)$ and $(n-1,\,l+2,\,j=l+3/2)$ indicates a special and hidden symmetry in atomic nuclei---the so-called pseudospin symmetry (PSS)---which is an important concept in both spherical and deformed nuclei.
  A number of phenomena in nuclear structure have been successfully interpreted directly or implicitly by this symmetry, including nuclear superdeformed configurations, identical bands, quantized alignment, pseudospin partner bands, and so on.
  Since the PSS was recognized as a relativistic symmetry in 1990s, there have been comprehensive efforts to understand its properties in various systems and potentials.
  In this Review, we mainly focus on the latest progress on the supersymmetric (SUSY) representation of PSS, and one of the key targets is to understand its symmetry-breaking mechanism in realistic nuclei in a quantitative and perturbative way.
  The SUSY quantum mechanics and its applications to the SU(2) and U(3) symmetries of the Dirac Hamiltonian are discussed in detail.
  It is shown that the origin of PSS and its symmetry-breaking mechanism, which are deeply hidden in the origin Hamiltonian, can be traced by its SUSY partner Hamiltonian.
  Essential open questions, such as the SUSY representation of PSS in the deformed system, are pointed out.
\end{abstract}


\section{Introduction}\label{Sect:1}

For celebrating the 40th anniversary of the Nobel Prize to nuclear structure studies in 1975, let us recall the pioneering works on the topic of pseudospin symmetry (PSS) in atomic nuclei by Bohr, Hamamoto, and Mottelson \cite{Bohr1982_PS26-267, Mottelson1990_NPA520-711c, Mottelson1991_NPA522-1c}, and highlight some relevant up-to-date progress, in particular, the supersymmetric (SUSY) representation of pseudospin symmetry.


The establishment of independent-particle shell model is one of the most important milestones in nuclear physics.
Similar to that of electrons orbiting in an atom, protons and neutrons in a nucleus generate shell structures, but different from the atomic systems, the corresponding nuclear magic numbers are found to be $2$, $8$, $20$, $28$, $50$, and $82$ for both protons and neutrons as well as $126$ for neutrons in stable nuclei.
In order to understand these magic numbers, simple models, such as the square-well or harmonic-oscillator (HO) potential, are not able to provide satisfactory answers.
Until 1949, independently, Haxel, Jensen, and Suess \cite{Haxel1949_PR075-1766} and Goeppert-Mayer \cite{Mayer1949_PR075-1969} introduced the strong spin-orbit (SO) interaction in the nuclear system, which largely splits the single-particle states $(n,\,l,\,j = l \pm 1/2)$ with high orbital angular momentum $l$ and excellently reproduces all traditional nuclear magic numbers.

Apart from the magic numbers, the nuclear shell model with strong spin-orbit interaction also provides wonderful descriptions for various kinds of nuclear ground-state properties and excited-state features.
By introducing the deformation-dependent oscillator length, Nilsson \textit{et al.} \cite{Nilsson1955_DMFM29-16, Nilsson1969_NPA131-1} extended this model to the deformed cases, and established the foundation for describing not only the deformed nuclei but also nuclear rotation phenomena.


\begin{figure}[tb]
\vspace{1.5em}
\centering\includegraphics[width=7cm]{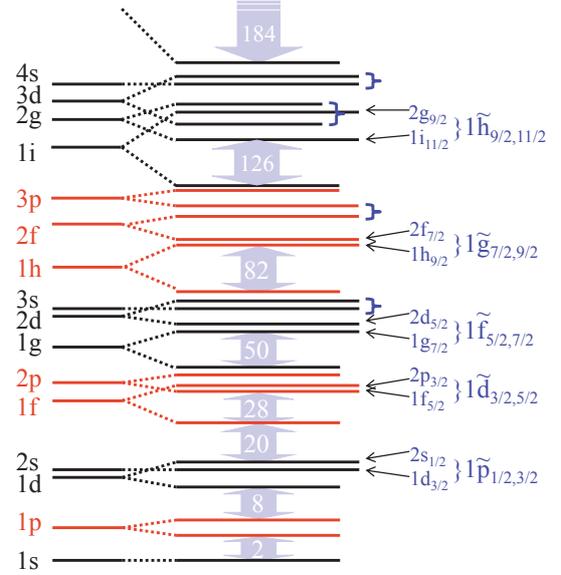}
\caption{(Color online) Schematic nuclear single-particle spectrum.
On one hand, the strong spin-orbit splitting between the spin doublets $(n,\,l,\,j = l \pm 1/2)$ shows the traditional magic numbers---$2$, $8$, $20$, $28$, $50$, $82$, and $126$.
On the other hand, the quasi-degeneracy is found between the pairs of single-particle states in braces, $(n,\,l,\,j = l + 1/2)$ and $(n-1,\,l + 2,\,j = l + 3/2)$.
The concept of pseudospin symmetry was introduced \cite{Hecht1969_NPA137-129, Arima1969_PLB30-517}, and the pseudospin doublets are denoted with quantum numbers $(\tilde{n}=n-1,\,\tilde{l}=l+1,\,j=\tilde{l}\pm1/2)$.
Taken from Ref.~\cite{Liang2015_PR570-1}.}
\label{Fig:1.PSS}
\end{figure}

In contrast to the large energy splitting between the spin doublets, by examining the single-particle spectra, in particular, those around the Fermi energy, Hecht and Adler \cite{Hecht1969_NPA137-129} and Arima, Harvey, and Shimizu \cite{Arima1969_PLB30-517} independently pointed out in 1969 the near degeneracy between pairs of single-particle states with quantum numbers $(n,\, l,\, j = l + 1/2)$ and $(n-1,\, l + 2,\, j = l + 3/2)$.
They introduced the concept of pseudospin symmetry and defined the pseudospin doublets as $(\tilde{n}=n-1,\, \tilde{l}=l+1,\, j=\tilde{l}\pm1/2)$ to illustrate such a near degeneracy.
A schematic nuclear single-particle spectrum with strong spin-orbit splitting and good pseudospin symmetry is illustrated in Fig.~\ref{Fig:1.PSS}.

The pseudospin symmetry remains an important concept in the axially deformed \cite{Bohr1982_PS26-267, Ratna-Raju1973_NPA202-433, Voigt1983_RMP55-949, Draayer1984_AP156-41} and even the triaxially deformed \cite{Blokhin1997_NPA612-163, Beuschel1997_NPA619-119} nuclei.
Based on this concept, a simple but useful pseudo-SU(3) model was proposed, and it was generalized to be the pseudo-symplectic model \cite{Rosensteel1976AP96-1, Rowe1985RPP48-1419, Troltenier1994_NPA576-351, Troltenier1995_NPA586-53}.
The concept of pseudospin symmetry has been also widely used in the odd-mass nuclei in the interacting Boson-Fermion model \cite{Iachello1981AP136-19}.

In the Symposium in honor of Akito Arima: Nuclear Physics in the 1990's, Mottelson \cite{Mottelson1991_NPA522-1c} preluded the link between the pseudospin symmetry and the experimental discoveries, when he introduced some themes in the study of very deformed rotating nuclei.
Almost from then on, a number of phenomena in nuclear structure have been successfully interpreted directly or implicitly by the pseudospin symmetry, including nuclear superdeformed configurations \cite{Dudek1987_PRL59-1405, Bahri1992_PRL68-2133, Dudek1992_PPNP28-131, Molique2000_PRC61-044304, Dudek2005_APPB36-975}, identical bands \cite{Byrski1990_PRL64-1650, Gelberg1990_JPG16-L143, Nazarewicz1990_PRL64-1654, Nazarewicz1990_NPA512-61, Zeng1991_PRC44-R1745}, quantized alignment \cite{Stephens1990_PRL65-301}, and pseudospin partner bands \cite{Xu2008_PRC78-064301, Hua2009_PRC80-034303}.
The pseudospin symmetry may also manifest itself in the magnetic moments and transitions \cite{Troltenier1994_NPA567-591, Ginocchio1999_PRC59-2487, Neumann-Cosel2000_PRC62-014308} and $\gamma$-vibrational states in nuclei \cite{Jolos2012_PRC86-044320}, as well as in nucleon-nucleus and nucleon-nucleon scatterings \cite{Ginocchio1999_PRL82-4599, Leeb2000_PRC62-024602, Ginocchio2002_PRC65-054002, Leeb2004_PRC69-054608}.
In addition, the role of pseudospin symmetry in the structure of halo nuclei \cite{Long2010_PRC81-031302R} and superheavy nuclei \cite{Jolos2007_PAN70-812, Li2014_PLB732-169} has been pointed out.

In the 21st century, it has been intensively discovered that the traditional magic numbers can change in nuclei far away from the stability line \cite{Sorlin2008_PPNP61-602, Wienholtz2013_Nature498-346, Steppenbeck2013_Nature502-207}.
This indicates the shell structure shown in Fig.~\ref{Fig:1.PSS} can evolve dramatically, where splitting of both spin and pseudospin doublets plays critical roles.
For example, the $N=28$ shell closure disappears due to the quenching of the spin-orbit splitting for the $\nu1f$ spin doublets \cite{Gaudefroy2006_PRL97-092501, Bastin2007_PRL99-022503, Tarpanov2008_PRC77-054316, Moreno-Torres2010_PRC81-064327}, whereas the $Z=64$ subshell closure is related to the conservation of pseudospin symmetry for the $\pi2\tilde p$ and $\pi1\tilde f$ pseudospin doublets \cite{Nagai1981_PRL47-1259, Long2007_PRC76-034314, Long2009_PLB680-428}.
The uncertainty with the proton magic number after $Z=82$ is related to the uncertainty with the strength of the spin-orbit interaction, and thus with the strength of the pseudospin-orbit (PSO) interaction in the superheavy nuclei.
Therefore, it is important to understand the nuclear shell evolution and the pseudospin symmetry on the same footing, in particular, near the limits of nucleus existence.


\begin{figure}[tb]
\centering\includegraphics[width=7cm]{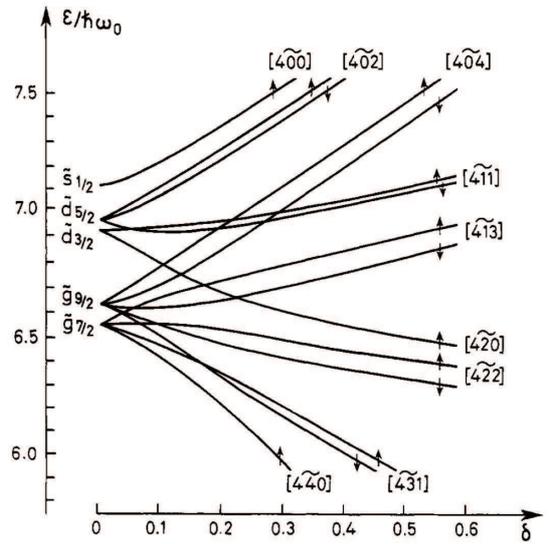}
\caption{Single-particle spectrum for the neutron shell $82<N<126$ with normal parity as a function of quadrupole deformation.
Pairs of states with asymptotic Nilsson quantum numbers $[N,\,n_3,\,\Lambda]\Omega=\Lambda+1/2$ and $[N,\,n_3,\,\Lambda+2]\Omega=\Lambda+3/2$ form the pseudospin partners as $[\tilde N=N-1,\,\tilde n_3,\,\tilde\Lambda=\Lambda+1]\tilde\Omega=\tilde\Lambda\pm1/2$ indicated by arrows $\uparrow$ and $\downarrow$.
Taken from Ref.~\cite{Bohr1982_PS26-267}.}
\label{Fig:1.BHM}
\end{figure}

Since the recognition of pseudospin symmetry in atomic nuclei, there have been comprehensive efforts to discover its origin.
One of the pioneering quantitative studies was carried out by Bohr, Hamamoto, and Mottelson \cite{Bohr1982_PS26-267} in 1982 in the scheme of rotating nuclear potentials.
For the large deformation, the asymptotic Nilsson quantum numbers $[N,\,n_3,\,\Lambda]\Omega$ are good quantum numbers.
Approximate degeneracy between the states $[N,\,n_3,\,\Lambda]\Omega=\Lambda+1/2$ and $[N,\,n_3,\,\Lambda+2]\Omega=\Lambda+3/2$ was discovered, and their corresponding pseudospin quantum numbers were denoted as $[\tilde N=N-1,\,\tilde n_3,\,\tilde\Lambda=\Lambda+1]\tilde\Omega=\tilde\Lambda\pm1/2$, as shown in Fig.~\ref{Fig:1.BHM}.
Based on this Nilsson Hamiltonian, Bohr, Hamamoto, and Mottelson tried to understand the origin of pseudospin symmetry in terms of the spin-orbit $v_{ls}$ and orbit-orbit $v_{ll}$ interactions.
It turned out that the origin of pseudospin symmetry was connected with a special ratio between the strengths of these two interactions, i.e., $v_{ls}/v_{ll}=4$ \cite{Bohr1982_PS26-267}.
They found that the pseudospin symmetry is helpful to qualitatively understand the properties of quasi-particle motions in the rotating potentials.
Furthermore, a weakness of the coupling of the odd nucleon pseudospin and the collective core rotational momenta in the odd-even deformed nuclei is manifested in the doublet structure of the ground-state rotational bands of these nuclei.
Several examples can be found in the review~\cite{Jolos2001_PPN32-113}.
This effect was discussed in Ref.~\cite{Bohr1982_PS26-267}, where it was clearly formulated that the pseudo-orbital momentum of the odd nucleon in the well deformed nuclei is strongly coupled to the core collective momentum, however, the pseudospin is decoupled.

This idea inspired the groups at Louisiana State University, University of California, and National Autonomous University of Mexico, and they proposed various explicit transformations from the normal scheme to the pseudospin scheme \cite{Bahri1992_PRL68-2133, Castanos1992_PLB277-238, Blokhin1995_PRL74-4149}.


The relation between the pseudospin symmetry and the relativistic mean-field (RMF) theory \cite{Ring1996_PPNP37-193, Vretenar2005_PR409-101, Meng2006_PPNP57-470, Niksic2011_PPNP66-519, Meng2011_PP31-199, Meng2013_FPC8-55, Meng2015_JPG42-093101, Meng2016} was first noted in Ref.~\cite{Bahri1992_PRL68-2133}, where the relativistic mean-field theory was used to explain such an approximate ratio $v_{ls}/v_{ll}\approx4$ between the strengths of the spin-orbit and orbit-orbit interactions.
In order to see the connection with the relativistic mean-field theory, it is illuminating to examine the Dirac equations as the equation of motion for nucleons in the relativistic framework.
The corresponding single-particle wave functions are expressed in the form of the Dirac spinors, which have both the upper and lower components.
For the spherical case, the upper and lower components have the same total angular momentum $j$ but their orbital angular momenta $l$ differ by one unit.

In 1997, Ginocchio \cite{Ginocchio1997_PRL78-436} revealed that the pseudospin symmetry is essentially a relativistic symmetry of the Dirac Hamiltonian, and the pseudo-orbital angular momentum $\tilde l$ is nothing but the orbital angular momentum of the lower component of Dirac spinor.
He also showed that the pseudospin symmetry in nuclei is exactly conserved when the scalar potential $S(\mathbf{r})$ and the vector potential $V(\mathbf{r})$ have the same size but opposite sign, i.e., $\Sigma(\mathbf{r}) \equiv S(\mathbf{r}) + V(\mathbf{r}) = 0$.

As a step further, one can reduce the Dirac equation into the Schr\"odinger-like second-order differential equation for either the upper or lower component.
There will be the corresponding spin-orbit and pseudospin-orbit potentials governing the relevant energy splitting for the spin and pseudospin doublets, respectively.
The pseudospin symmetry is exact if the derivative for the sum of the scalar and vector potentials vanishes, i.e., $d\Sigma(r)/dr=0$ \cite{Meng1998_PRC58-R628}.
Although this symmetry limit cannot be exactly fulfilled, because there are no longer bound states at such a limit, the condition $d\Sigma(r)/dr\approx0$ means that the pseudospin symmetry becomes better for exotic nuclei with highly diffused potentials \cite{Meng1999_PRC59-154}.

Following the discussions for spherical nuclei, the study of pseudospin symmetry within the relativistic framework was quickly extended to the deformed nuclei \cite{Lalazissis1998_PRC58-R45, Sugawara-Tanabe1998_PRC58-R3065, Sugawara-Tanabe2000_PRC62-054307, Sugawara-Tanabe2005_RMP55-277}.
As the pseudospin symmetry is a relativistic symmetry, the wave functions of the pseudospin partners satisfy certain relations \cite{Ginocchio1998_PRC57-1167, Ginocchio2002_PRC66-064312}.
These relations have been tested in both spherical and deformed nuclei, see, e.g., Refs.~\cite{Sugawara-Tanabe2002_PRC65-054313, Ginocchio2004_PRC69-034303}.


Following these works, extensive discussions about the pseudospin symmetry in the single-particle spectra have been made by exactly or approximately solving the Dirac equation with various potentials, for example, the spherical harmonic-oscillator \cite{Chen2003_CPL20-358},
Coulomb \cite{Lisboa2003_PRC67-054305},
Hulth\'en \cite{Guo2003_CPL20-602},
Morse \cite{Berkdemir2006_NPA770-32},
P\"oschl-Teller \cite{Jia2007_EPJA34-41},
and Woods-Saxon \cite{Guo2005_PLA338-90} potentials,
as well as the deformed harmonic-oscillator \cite{Ginocchio2004_PRC69-034318},
Manning-Rosen \cite{Asgarifar2013_PS87-025703},
and ring-shaped \cite{Zhang2009_CEJP7-768} potentials.
For details see, e.g., Section 2.2 in Ref.~\cite{Liang2015_PR570-1} and references therein.
Self-consistently, the pseudospin symmetry in spherical and deformed nuclei have been investigated within the relativistic mean-field and relativistic Hartree-Fock \cite{Bouyssy1987_PRC36-380, Long2006_PLB639-242, Long2006_PLB640-150, Liang2008_PRL101-122502, Long2010_PRC81-024308, Liang2012_PRC86-021302R, Niu2013_PLB723-172} theories.
One of interesting topics is the tensor effects on the pseudospin symmetry \cite{Lisboa2004_PRC69-024319, Alberto2005_PRC71-034313, deCastro2006_PRC73-054309, Long2007_PRC76-034314, Long2010_PRC81-031302R, Li2014_PLB732-169}.


For the Dirac equation, there exist not only the positive-energy states in the Fermi sea but also the negative-energy states in the Dirac sea, where the negative-energy states correspond to the anti-particle states.
When they developed the relativistic mean-field theory in the Dirac Woods-Saxon basis, Zhou, Meng, and Ring \cite{Zhou2003_PRC68-034323} examined carefully the negative-energy states in the Dirac sea and found that the pseudospin symmetry of those negative-energy states, or equivalently, the spin symmetry (SS) in the anti-nucleon spectra is well conserved \cite{Zhou2003_PRL91-262501}.
They further discovered that the spin symmetry in the anti-nucleon spectra is much better developed than the pseudospin symmetry in the usual nucleon spectra.
The spin symmetry in the anti-nucleon spectra was also tested by investigating relations between the Dirac wave functions \cite{He2006_EPJA28-265}.
Later, this symmetry was studied with the relativistic Hartree-Fock theory and the contribution from the Fock terms was analyzed \cite{Liang2010_EPJA44-119}.
It was discussed in Ref.~\cite{Zhou2003_PRL91-262501} that an open problem related to the experimental study of the spin symmetry in the anti-nucleon spectra is the polarization effect caused by the annihilation of anti-nucleons in a normal nucleus.
Some detailed calculations on the anti-baryon annihilation rates in nuclear environment showed that the in-medium annihilation rates may be strongly suppressed by a significant reduction of the reaction $Q$ values, leading to relatively long-lived anti-baryon-nucleus systems \cite{Mishustin2005_PRC71-035201}.
Alternatively, the spin symmetry in the anti-$\Lambda$ spectra of hypernuclei was studied \cite{Song2009_CPL26-122102, Song2010_ChinPhysC34-1425, Song2011_CPL28-092101}, which may be free from the problem of annihilation.
This kind of study would be of great interests for possible experimental tests.


In recent years, there has been an increasing interest in the exploration of continuum and resonant states, in particular, in the studies of exotic nuclei with extreme $N/Z$ ratios.
In exotic nuclei, the neutron or proton Fermi surface is close to the single-particle emission threshold, as a result the contribution of continuum and resonant states is important \cite{Meng1996_PRL77-3963, Meng1998_PRL80-460, Meng1998_NPA635-3, Zhou2010_PRC82-011301R, Chen2012_PRC85-067301}.
Many methods have been developed for the studies of resonances \cite{Kukulin1989}, for example, the analytical continuation in coupling constant method \cite{Yang2001_CPL18-196, Zhang2004_PRC70-034308}, the real stabilization method \cite{Zhang2008_PRC77-014312, Zhou2009_JPB42-245001}, the complex scaling method \cite{Guo2010_PRC82-034318, Liu2012_PRC86-054312}, the coupled channels method \cite{Hagino2004_NPA735-55, Li2010_PRC81-034311}, and so on.
Therefore, the exploration of symmetries in resonant states is certainly interesting \cite{Guo2005_PRC72-054319, Guo2006_PRC74-024320, Liu2013_PRA87-052122}.

Recently, Lu, Zhao, and Zhou \cite{Lu2012_PRL109-072501} gave a rigorous verification of the pseudospin symmetry in the single-particle resonant states.
They discovered that the pseudospin symmetry in the single-particle resonant states is exactly conserved under the same condition discussed for the bound states, i.e., $\Sigma(r) = 0$ or $d\Sigma(r)/dr = 0$.
By examining the zeros of Jost functions corresponding to the lower component of Dirac spinor, general properties of pseudospin-symmetry breaking in energy and width were examined, and the pseudospin-symmetry-breaking part can be separated from other parts in the Jost functions \cite{Lu2013_PRC88-024323}.


Works are also in progress for understanding the origin of pseudospin symmetry and its symmetry-breaking mechanism in a perturbative and quantitative way.
The perturbation theory was used in Refs.~\cite{Liang2011_PRC83-041301R, Li2011_ChinPhysC35-825} to investigate the symmetries of the Dirac Hamiltonian and their symmetry breaking in realistic nuclei.
An illuminating example is that the energy splitting of the pseudospin doublets can be regarded as a result of perturbation from the Dirac Hamiltonian with a relativistic harmonic-oscillator (RHO) potential, where the pseudospin doublets are exactly degenerate \cite{Liang2011_PRC83-041301R}.


Alternatively, the supersymmetric quantum mechanics \cite{Cooper1995_PR251-267, Cooper2001} was used to investigate the symmetries of the Dirac Hamiltonian \cite{Leviatan2004_PRL92-202501, Typel2008_NPA806-156, Leviatan2009_PRL103-042502}.
In particular, by employing both the exact and broken patterns in supersymmetry, the special feature---all states with $\tilde l > 0$ have their own pseudospin partners except for the so-called intruder states---can be interpreted within a unified scheme.
In Ref.~\cite{Leviatan2004_PRL92-202501}, Leviatan showed three kinds of symmetries of the Dirac Hamiltonian by using the supersymmetric scheme, i.e, the Coulomb, spin, and pseudospin symmetries.
In Ref.~\cite{Typel2008_NPA806-156}, Typel derived a regular pseudospin-symmetry-breaking potential with the supersymmetric technique, in contrast singularities appear when the Dirac equation is reduced to a Schr\"odinger-like equation for the lower component of Dirac spinor.
However, by reducing the Dirac equation to a Schr\"odinger-like equation \cite{Typel2008_NPA806-156}, the corresponding effective Hamiltonian thus obtained is not Hermitian.
Such a fact prevents us from being able to carry out the perturbation calculations in a quantitative way.


Recent works by Guo and coauthors \cite{Guo2012_PRC85-021302R, Li2013_PRC87-044311, Guo2014_PRL112-062502} bridged the gap between the perturbation calculations and the supersymmetric description of pseudospin symmetry by using the similarity renormalization group (SRG) \cite{Wegner1994_AP506-77, Bylev1998_PLB428-329, Wegner2001_PR348-77} for transforming the Dirac Hamiltonian into a diagonal form.
The effective Hamiltonian expanded in a series of $1/M$ is Hermitian, which makes the perturbation calculations feasible.
Therefore, it is promising to understand the origin of pseudospin symmetry and its symmetry breaking in the realistic nuclear systems by combining the supersymmetric quantum mechanics, the perturbation theory, and the similarity renormalization group technique, as carried out in Refs.~\cite{Liang2013_PRC87-014334, Shen2013_PRC88-024311}.


In this Review, we will mainly focus on the latest progress in the studies of pseudospin symmetry, in particular, its supersymmetric representation.
We will outline the general formalism in Section~\ref{Sect:2}, and discuss some essential progress in detail in Section~\ref{Sect:3}.
A summary, but more importantly, the relevant open questions will be emphasized in Section~\ref{Sect:4}.
Note that some other topics covered in the previous reviews \cite{Ginocchio2005_PR414-165, Liang2015_PR570-1, Guo2016} will not be repeated here.


\section{General Formalism}\label{Sect:2}

In this Section, we will outline the essential formalism for the following discussions, in particular, the single-particle Dirac equation and its equivalent Schr\"odinger-like equations, as well as the basic idea of the supersymmetric quantum mechanics.
The key symbols and notations used in this paper follow those recommended in Ref.~\cite{Liang2015_PR570-1}.

\subsection{Dirac and Schr\"odinger-like equations}\label{Sect:2.1}

\subsubsection{Dirac equations}

In the relativistic or the so-called covariant framework, the motion of nucleons is described by the Dirac equation.
Originating from the minimal coupling of the scalar and vector mesons to the nucleons in the covariant density functional theory \cite{Meng2016}, the single-particle Dirac equation reads
\begin{equation}\label{Eq:2.1.HDirac}
  \{\boldsymbol{\alpha}\cdot\mathbf{p} + \beta[M+S(\mathbf{r})]+V(\mathbf{r})\}\psi(\mathbf{r})
  =\epsilon\psi(\mathbf{r})\,,
\end{equation}
where $\epsilon=E+M$ is the single-particle energy including the rest mass of nucleon $M$, and $\hbar=c=1$ are set in this paper.
Here $\boldsymbol{\alpha}$ and $\beta$ are the Dirac matrices, while $S(\mathbf{r})$ and $V(\mathbf{r})$ are the scalar and vector potentials, respectively.

When the spherical symmetry is adopted, the single-particle eigenstates are specified by a set of quantum numbers $\alpha=(a, m_a)=(n_a, l_a, j_a, m_a)$, and the single-particle wave functions can be factorized as
\begin{equation}\label{Eq:2.1.spwfR1}
    \psi_\alpha(\mathbf{r}) =
    \frac{1}{r}
    \lb \begin{array}{c}
        iG_a(r) \\ F_a(r)\hat{\boldsymbol{\sigma}}\cdot\hat{\mathbf{r}}
    \end{array} \rb \mathscr Y^{l_a}_{j_am_a}(\hat{\mathbf{r}})\,,
\end{equation}
with the spherical harmonics spinor $\mathscr Y^{l}_{jm}(\hat{\mathbf{r}})$ for the angular and spin parts \cite{Varshalovich1988}.
The corresponding normalization condition reads
\begin{equation}\label{Eq:2.1.Normalization}
  \int \psi^\dag_\alpha(\mathbf{r})\psi_\alpha(\mathbf{r})d^3\mathbf{r}
  =\int \ls G_a^2(r)+F_a^2(r)\rs dr=1\,.
\end{equation}

It is important that, for the lower component of Dirac spinor (\ref{Eq:2.1.spwfR1}), one holds $\hat{\boldsymbol{\sigma}}\cdot\hat{\mathbf{r}}\mathscr Y^{l_a}_{j_am_a}(\hat{\mathbf{r}}) = -\mathscr Y^{\tilde l_a}_{j_am_a}(\hat{\mathbf{r}})$ with $\tilde l = l\pm 1$ for the $j = l\pm1/2$ states.
Thus, the single-particle wave functions can be rewritten as
\begin{equation}\label{Eq:2.1.spwfR}
    \psi_\alpha(\mathbf{r}) =
    \frac{1}{r}
    \lb \begin{array}{c}
        iG_a(r) \mathscr Y^{l_a}_{j_am_a}(\hat{\mathbf{r}})
        \\ -F_a(r)\mathscr Y^{\tilde l_a}_{j_am_a}(\hat{\mathbf{r}})
    \end{array} \rb\,.
\end{equation}
In such a way, the pseudo-orbital angular momentum $\tilde{l}$ is found to be the orbital angular momentum of the lower component of Dirac spinor \cite{Ginocchio1997_PRL78-436}.

The corresponding radial Dirac equation reads
\begin{equation}\label{Eq:2.1.DiraceqR}
    \lb\begin{array}{cc}
    M+\Sigma(r) & -\frac{d}{dr}+\frac{\kappa}{r} \\
        \frac{d}{dr}+\frac{\kappa}{r} & -M+\Delta(r)
    \end{array}\rb
    \lb\begin{array}{c}
        G(r) \\ F(r)
    \end{array}\rb
    =\epsilon
    \lb\begin{array}{c}
        G(r) \\ F(r)
    \end{array}\rb\,,
\end{equation}
where $\Sigma(r)\equiv S(r)+V(r)$ and $\Delta(r)\equiv V(r)-S(r)$ denote the combinations of the scalar and vector potentials, and $\kappa$ is another relativistic good quantum number defined as $\kappa=\mp(j+1/2)$ for the $j=l\pm1/2$ states.
For brevity, we omit the subscripts if there is no confusion.

The symmetries of the Dirac Hamiltonian in Eq.~(\ref{Eq:2.1.HDirac}) or (\ref{Eq:2.1.DiraceqR}) can be studied by the Bell-Reugg condition \cite{Bell1975_NPB98-151} or the SUSY scheme \cite{Leviatan2004_PRL92-202501, Leviatan2009_PRL103-042502}.
Alternatively, they can be investigated by reducing the Dirac equation to the corresponding Schr\"odinger-like second-order differential equations as follows.

\subsubsection{Schr\"odinger-like equations}

Focusing on the spherical case, one can derive the Schr\"odinger-like equation for the upper component $G(r)$ of Dirac spinor by substituting
\begin{equation}\label{Eq:2.1.FG}
  F(r) = \frac{1}{M-\Delta(r)+\epsilon}\lb\frac{d}{dr}+\frac{\kappa}{r}\rb G(r)
\end{equation}
in Eq.~(\ref{Eq:2.1.DiraceqR}), and obtain
\begin{align}\label{Eq:2.1.SchrG}
    &\Lb -\frac{1}{M_+}\frac{d^2}{dr^2}
  +\frac{1}{M_+^2}\frac{dM_+}{dr}\frac{d}{dr}
  +\ls (M+\Sigma)+\frac{1}{M_+}\frac{\kappa(\kappa+1)}{r^2}\right.\right.\nonumber\\
  &\left.\left.+\frac{1}{M_+^2}\frac{dM_+}{dr}\frac{\kappa}{r}
  \rs \Rb G = \epsilon G\,,
\end{align}
with the energy-dependent effective mass $M_+(r)=M-\Delta(r)+\epsilon$.
In analogy with the usual Schr\"{o}dinger equations, $\Sigma(r)$ is the central potential in which particles move, the term proportional to $l(l+1)=\kappa(\kappa+1)$ corresponds to the centrifugal barrier (CB), and the last term corresponds to the spin-orbit potential, which leads to the substantial spin-orbit splitting in the nuclear single-particle spectra, i.e.,
\begin{subequations}\label{Eq:2.1.VCBandVSO}
\begin{align}
  V_{\rm CB}(r) &=\frac{1}{M_+(r)}\frac{\kappa(\kappa+1)}{r^2}\,,\\
  V_{\rm SO}(r) &=\frac{1}{M_+^2(r)}\frac{dM_+(r)}{dr}\frac{\kappa}{r}\,.
\end{align}
\end{subequations}
Vanishing $V_{\rm SO}$ leads to zero spin-orbit splitting, that is,
\begin{equation}\label{Eq:2.1.SSlimit}
  -\frac{dM_+(r)}{dr}=\frac{d\Delta(r)}{dr}=0\,,
\end{equation}
which is the spin-symmetry limit.

Similarly, one can derive the Schr\"odinger-like equation for the lower component $F(r)$ by substituting
\begin{equation}\label{Eq:2.1.GF}
  G(r) = \frac{1}{-M-\Sigma(r)+\epsilon}\lb-\frac{d}{dr}+\frac{\kappa}{r}\rb F(r)\,,
\end{equation}
and obtain
\begin{align}\label{Eq:2.1.SchrF}
    &\Lb -\frac{1}{M_-}\frac{d^2}{dr^2}
    +\frac{1}{M_-^2}\frac{dM_-}{dr}\frac{d}{dr}
    +\ls (-M+\Delta)+\frac{1}{M_-}\frac{\kappa(\kappa-1)}{r^2}\right.\right.\nonumber\\
    &\left.\left.
    -\frac{1}{M_-^2}\frac{dM_-}{dr}\frac{\kappa}{r}
    \rs \Rb F= \epsilon F\,,
\end{align}
with the energy-dependent effective mass $M_-(r)=-M-\Sigma(r)+\epsilon$.
It has been shown that either Eq.~(\ref{Eq:2.1.SchrG}) or (\ref{Eq:2.1.SchrF}), together with its charge conjugated equation, is equivalent to the original Dirac equation (\ref{Eq:2.1.DiraceqR}) \cite{Zhang2010_IJMPE19-55, Zhang2009_ChinPC33S1-113, Zhang2009_CPL26-092401, Li2011_SciChinaPMA54-231, Tanimura2015_PTEP2015-073D01}.

When one focuses on the Schr\"odinger-like equation (\ref{Eq:2.1.SchrF}) for the lower component instead of the upper one, although $\Delta(r)$ does not stand for the potential in which particles move, all terms except one, $-(1/M_-^2)(dM_-/dr)(\kappa/r)$, are identical for the pseudospin doublets $a$ and $b$ because of $\kappa_a(\kappa_a-1)=\kappa_b(\kappa_b-1)$.
As pointed out in Ref.~\cite{Meng1998_PRC58-R628}, if such a term vanishes, i.e.,
\begin{equation}\label{Eq:2.1.PSSlimit}
  -\frac{dM_-(r)}{dr}=\frac{d\Sigma(r)}{dr}=0\,,
\end{equation}
each pair of pseudospin doublets should be degenerate and the PSS should be exactly conserved.
This is the PSS limit, which is more general and includes the symmetry limit $\Sigma(r)=0$ discussed in Ref.~\cite{Ginocchio1997_PRL78-436}.
From the physical point of view, $\Sigma(r)=0$ is never fulfilled in realistic nuclei, since in which there exist no bound states for nucleons \cite{Leviatan2001_PLB518-214}, but $d\Sigma(r)/dr\sim0$ can be approximately satisfied in exotic nuclei with highly diffuse potentials \cite{Meng1999_PRC59-154}.
Analogically, such a term is regarded as the pseudospin-orbit potential, while the term proportional to $\tilde l(\tilde l+1)=\kappa(\kappa-1)$ is regarded as the pseudo-centrifugal barrier (PCB), i.e.,
\begin{subequations}\label{Eq:2.1.VPSOandVPCB}
\begin{align}
  V_{\rm PCB}(r) &= \frac{1}{M_-(r)}\frac{\kappa(\kappa-1)}{r^2}\,,\\
  V_{\rm PSO}(r) &= -\frac{1}{M_-^2(r)}\frac{dM_-(r)}{dr}\frac{\kappa}{r}\,.
\end{align}
\end{subequations}

The PSS limit shown in Eq.~(\ref{Eq:2.1.PSSlimit}) and the special features of the PSO potential in Eq.~(\ref{Eq:2.1.VPSOandVPCB}) have been intensively discussed in various systems and potentials during the past two decades, such as from stable to exotic nuclei, from the non-confining to confining potentials, from the local to non-local potentials, from the central to tensor potentials, from the bound to resonant states, from the nucleon to anti-nucleon spectra, from the nucleon to hyperon spectra, and from spherical to deformed nuclei.
Readers are referred to, e.g., Section 3 in Ref.~\cite{Liang2015_PR570-1} for some interesting discussions.

\subsection{Supersymmetric quantum mechanics}\label{Sect:2.2}

For the supersymmetric representation of PSS, let us also recall some basic formalism of SUSY quantum mechanics \cite{Cooper1995_PR251-267, Cooper2001}.

It has been shown that every second-order differential Hamiltonian can be factorized in a product of two Hermitian conjugate first-order differential operators \cite{Infeld1951_RMP23-21}, i.e.,
\begin{equation}\label{Eq:4.3.SUSYH1}
    H_1=B^+B^-\,,
\end{equation}
with $B^-=[B^+]^\dag$.
Its SUSY partner Hamiltonian can thus be constructed as \cite{Cooper2001}
\begin{equation}\label{Eq:4.3.SUSYH2}
    H_2=B^-B^+\,.
\end{equation}
The Hermitian operators
\begin{equation}\label{Eq:4.3.SUSYQ}
  Q_1=
  \lb\begin{array}{cc}
    0 & B^+ \\ B^- & 0
  \end{array}\rb\,,\quad
  Q_2=
  \lb\begin{array}{cc}
    0 & -iB^+ \\ iB^- & 0
  \end{array}\rb
\end{equation}
are the so-called supercharges with the involution
\begin{equation}\label{Eq:4.3.SUSYtau}
  \tau=\tau^\dag=
  \lb\begin{array}{cc}
    1 & 0 \\ 0 & -1
  \end{array}\rb\,,
\end{equation}
satisfying $\{Q_1,\tau\}=\{Q_2,\tau\}=0$.
The extended SUSY Hamiltonian $H_S$ is the square of these Hermitian supercharges,
\begin{equation}\label{Eq:4.3.SUSYHS}
    H_S=Q^2_1=Q^2_2=
    \lb\begin{array}{cc}
        H_1 & 0 \\ 0 & H_2
    \end{array}\rb\,.
\end{equation}
The supercharges $Q_1,\,Q_2$ and the extended Hamiltonian $H_S$, together with the commutators $[H_S,Q_1]=[H_S,Q_2]=0$ and anti-commutator $\{Q_1,Q_2\}=0$, form one of the simplest examples of supersymmetric algebra.

Since the extended SUSY Hamiltonian $H_S$ is the square of the supercharges, all of its eigenvalues $E_S(n)$ in eigenequation
\begin{equation}\label{Eq:4.3.SUSYeq}
    H_S\Psi_S(n)=E_S(n)\Psi_S(n)
\end{equation}
are non-negative.
The two-component wave function reads
\begin{equation}\label{Eq:4.3.SUSYPsiS}
    \Psi_S(n)=
    \lb\begin{array}{c}
        \psi_1(n) \\ \psi_2(n)
    \end{array}\rb\,,
\end{equation}
and $\psi_1(n)$ and $\psi_2(n)$ are the eigenfunctions of $H_1$ and $H_2$, respectively.

For each eigenstate with a positive eigenvalue $E_S(n)>0$, it is the eigenstate for both $H_1$ and $H_2$, and the corresponding eigenfunctions satisfy
\begin{subequations}\label{Eq:4.3.SUSYWFtran}
\begin{align}
    \psi_2(n)&=\frac{B^-}{\sqrt{E_S(n)}}\psi_1(n)\,,\\
    \psi_1(n)&=\frac{B^+}{\sqrt{E_S(n)}}\psi_2(n)\,,
\end{align}
\end{subequations}
with the normalization factor $1/\sqrt{E_S(n)}$.

\begin{figure}[tb]
\begin{center}
\includegraphics[width=8cm]{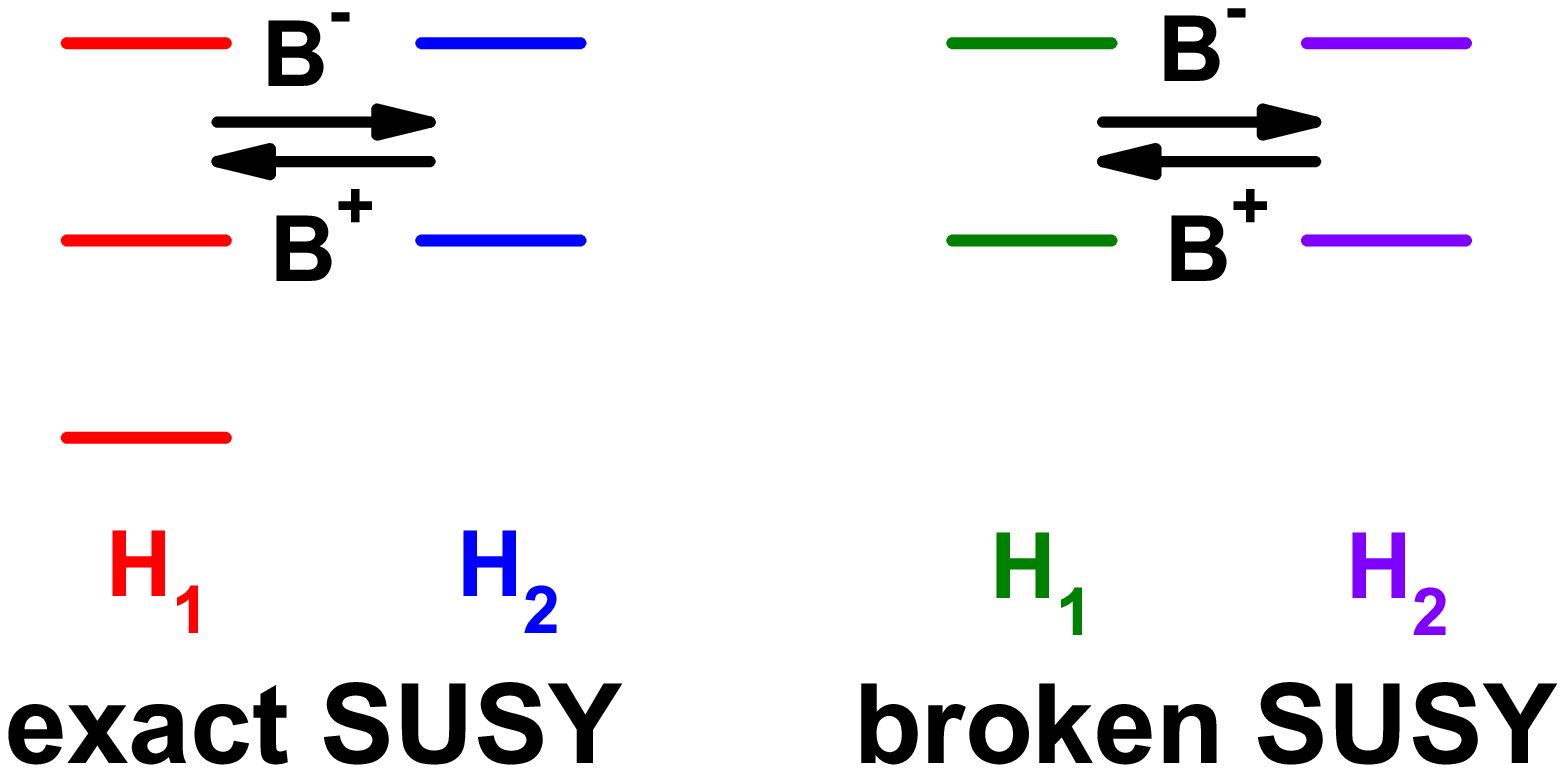}
\end{center}
\caption{(Color online) Schematic patterns of the exact and broken supersymmetries.
Taken from Ref.~\cite{Liang2013_PRC87-014334}.}
\label{Fig:2.2.SUSY}
\end{figure}

The SUSY can be either exact (also called unbroken) or broken \cite{Cooper2001}.
The SUSY is exact when the eigenvalue equation~(\ref{Eq:4.3.SUSYeq}) has a zero energy eigenstate $E_S(0)=0$.
In this case, as a usual convention, the Hamiltonian $H_1$ has an additional eigenstate at zero energy that does not appear in its partner Hamiltonian $H_2$, because $B^-\psi_1(0)=0$ means $\psi_2(0)=0$, i.e., the trivial eigenfunction of $H_2$ identically equals zero.
The SUSY is broken when the eigenvalue equation~(\ref{Eq:4.3.SUSYeq}) does not have any zero energy eigenstate.
In this case, the SUSY partner Hamiltonians $H_1$ and $H_2$ have the identical spectra.
The schematic patterns of the exact and broken SUSY are illustrated in Fig.~\ref{Fig:2.2.SUSY}.

In short, the eigenstates of Hamiltonians $H_1$ and $H_2$ are exactly one-to-one identical except for the so-called intruder states.
In such a way, the origin of symmetries deeply hidden in $H_1$ can be traced in its SUSY partner Hamiltonian $H_2$.

\section{Supersymmetric Representation of Pseudospin Symmetry and its Perturbative Nature}\label{Sect:3}

In this Section, we will discuss in detail the up-to-date progress in the supersymmetric representation of pseudospin symmetry.
Within this scheme, we will also focus on one of the longstanding issues---\textit{Whether or not the nature of pseudospin symmetry is perturbative?}

\subsection{Supersymmetry for Dirac equations and SU(2) symmetries}\label{Sect:3.1}

In this Subsection, we will show the supersymmetric quantum mechanics for the Dirac equations.
In this scheme, three kinds of symmetries of the Dirac Hamiltonian, i.e, the Coulomb, spin, and pseudospin symmetries, were discovered in Ref.~\cite{Leviatan2004_PRL92-202501}, where the spin and pseudospin symmetries correspond to the SU(2) symmetries of the Dirac Hamiltonian \cite{Ginocchio1998_PLB425-1}.
It is known that in realistic nuclei these SU(2) symmetries are broken, nevertheless, from the perturbation point of view, the SU(2) spin-symmetry breaking is perturbative whereas the SU(2) pseudospin-symmetry breaking is not \cite{Liang2011_PRC83-041301R}.

\subsubsection{SUSY for Dirac equations}

One of the first discussions on the PSS in the framework of SUSY quantum mechanics was presented by Leviatan in 2004 \cite{Leviatan2004_PRL92-202501}.
Instead of using the above mentioned scheme for the second-order differential or the so-called factorizable Hamiltonian, he employed a SUSY scheme directly for the first-order differential Dirac Hamiltonian by using the intertwining relation.

In Section~\ref{Sect:2.2}, one starts from a factorizable Hamiltonian $H_1$, then identifies the pair of Hermitian conjugate operators $B^+$ and $B^-$ in Eq.~(\ref{Eq:4.3.SUSYH1}), and eventually generates its SUSY partner Hamiltonian $H_2$ in Eq.~(\ref{Eq:4.3.SUSYH2}).
Alternatively, this procedure can be carried out in a different way.
Assuming one holds the so-called intertwining relation between $H_1$ and $H_2$ \cite{Nieto2003_AP305-151},
\begin{equation}\label{Eq:4.3.SUSYDitw}
  B^-H_1 = H_2 B^-\,,
\end{equation}
this intertwining relation ensures that, if $\psi_1(n)$ is an eigenstate of $H_1$ in Eq.~(\ref{Eq:4.3.SUSYPsiS}), $\psi_2(n)\propto B^-\psi_1(n)$ shown in Eq.~(\ref{Eq:4.3.SUSYWFtran}) is also an eigenstate of $H_2$ with the same energy $E_S(n)$, unless $B^-\psi_1(n)$ vanishes or produces an unphysical state, e.g., non-normalizable.
In other words, the SUSY patterns shown in Fig.~\ref{Fig:2.2.SUSY} can be set up as long as the intertwining relation is satisfied, but Hamiltonians $H_1$ and $H_2$ are not necessarily factorizable.

As a result, one can insist that both SUSY partner Hamiltonians $H_1\equiv H(\kappa_a)$ and $H_2\equiv H(\kappa_b)$ be the Dirac Hamiltonian of the form prescribed in Eq.~(\ref{Eq:2.1.DiraceqR}), and search for possible solutions of $B^-_\kappa$ that satisfy
\begin{align}\label{Eq:4.3.SUSYDHB}
  &B^-_\kappa
  \lb\begin{array}{cc}
    M+\Sigma(r) & - \frac{d}{dr}+\frac{\kappa_a}{r} \\
    \frac{d}{dr}+\frac{\kappa_a}{r} & -M+\Delta(r)
  \end{array}\rb\nonumber\\
  =&
  \lb\begin{array}{cc}
    M+\Sigma(r) & - \frac{d}{dr}+ \frac{\kappa_b}{r} \\
     \frac{d}{dr}+ \frac{\kappa_b}{r} & -M+\Delta(r)
  \end{array}\rb
  B^-_\kappa\,.
\end{align}

\begin{figure}[tb]
\begin{center}
  \includegraphics[width=8cm]{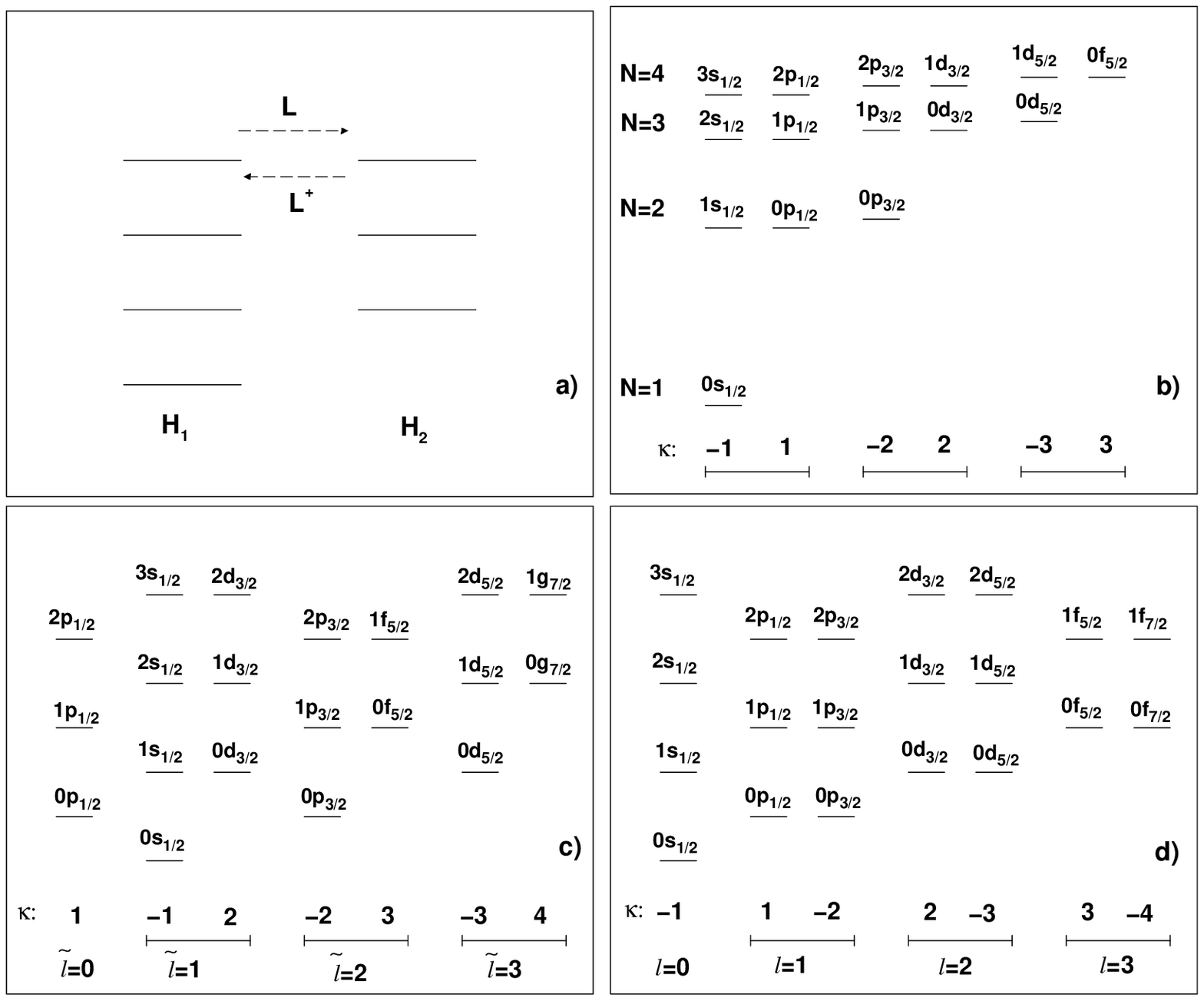}
\end{center}
\caption{Schematic patterns in (a) SUSY quantum mechanics and those with the (b) Coulomb symmetry, (c) pseudospin symmetry, and (d) spin symmetry of the Dirac Hamiltonian.
Taken from Ref.~\cite{Leviatan2004_PRL92-202501}.}
\label{Fig:3.1.SUSYD}
\end{figure}

By considering a matrical Darboux transformation operator,
\begin{equation}\label{Eq:4.3.SUSYDB}
  B^-_\kappa = P_\kappa(r)\frac{d}{dr}+Q_\kappa(r)\,,
\end{equation}
where $P_\kappa$ and $Q_\kappa$ are $2\times2$ matrices, and assuming certain forms in the functions $[P_\kappa(r)]_{ij}$ and $[Q_\kappa(r)]_{ij}$, three different kinds of solutions were found in Ref.~\cite{Leviatan2004_PRL92-202501}.
They correspond to three different kinds of symmetry limits: (i) Coulomb symmetry, (ii) spin symmetry, and (iii) pseudospin symmetry.
The schematic patterns at these symmetry limits are illustrated in Fig.~\ref{Fig:3.1.SUSYD}.

In the Coulomb-symmetry limit, the partner states $a$ and $b$ that form the degenerate doublets are $\kappa_a+\kappa_b=0$, for example, $(2s_{1/2},\,1p_{1/2})$, $(2p_{3/2},\,1d_{3/2})$.
The corresponding scalar and vector potentials are in the forms of $S(r) = \alpha_S/r$ and $V(r) = \alpha_V/r$, respectively.
The transformation operator reads \cite{Leviatan2004_PRL92-202501}
\begin{equation}\label{Eq:4.3.SUSYDCB}
  B^-_\kappa =
  \lb\begin{array}{cc}
     \frac{d}{dr}+ \frac{\varepsilon_+}{r}+ \frac{M\alpha_+}{\kappa_a} & \quad- \frac{\alpha_S}{\kappa_a} \frac{d}{dr}+ \frac{\alpha_V}{r} \\
     \frac{\alpha_S}{\kappa_a} \frac{d}{dr}- \frac{\alpha_V}{r} & \quad \frac{d}{dr}- \frac{\varepsilon_-}{r}- \frac{M\alpha_-}{\kappa_a}
  \end{array}\rb\,,
\end{equation}
where $\varepsilon_\pm=\kappa_a+\alpha_S\alpha_\pm/\kappa_a$ and $\alpha_\pm=\alpha_S\pm\alpha_V$.

The solution corresponding to the spin-symmetry limit is exactly that shown in Eq.~(\ref{Eq:2.1.SSlimit}), i.e., $V(r)-S(r)=\Delta_0$.
The transformation operator reads \cite{Leviatan2004_PRL92-202501}
\begin{equation}\label{Eq:4.3.SUSYDSSB}
  B^-_\kappa =
  \lb\begin{array}{cc}
    2M+\Sigma(r)-\Delta_0 & \quad- \frac{d}{dr}+ \frac{\kappa_a}{r} \\
     \frac{d}{dr}+ \frac{\kappa_b}{r} & \quad 0
  \end{array}\rb\,,
\end{equation}
with $\kappa_a+\kappa_b=-1$ for the spin doublets, e.g., $(1p_{1/2},\,1p_{3/2})$, $(1d_{3/2},\,1d_{5/2})$.

The solution corresponding to the PSS limit is that shown in Eq.~(\ref{Eq:2.1.PSSlimit}), i.e., $V(r)+S(r)=\Sigma_0$, and the transformation operator reads \cite{Leviatan2004_PRL92-202501}
\begin{equation}\label{Eq:4.3.SUSYDPSSB}
  B^-_\kappa =
  \lb\begin{array}{cc}
    0 & \quad- \frac{d}{dr}+ \frac{\kappa_b}{r} \\
     \frac{d}{dr}+ \frac{\kappa_a}{r} & \quad-2M+\Delta(r)-\Sigma_0
  \end{array}\rb\,,
\end{equation}
with $\kappa_a+\kappa_b=1$ for the pseudospin doublets, e.g., $(2s_{1/2},\,1d_{3/2})$, $(2p_{3/2},\,1f_{5/2})$.

\subsubsection{SU(2) symmetries}

The spin- and pseudospin-symmetry limits discussed above are found to be the SU(2)-symmetry limits of the Dirac Hamiltonian \cite{Ginocchio1998_PLB425-1}.

In Ref.~\cite{Bell1975_NPB98-151}, Bell and Ruegg discussed the symmetries in a general single-particle Dirac Hamiltonian, which are called the Bell-Ruegg symmetries in recent literatures.
As a special case, the Dirac Hamiltonian shown in Eq.~(\ref{Eq:2.1.HDirac}) holds an SU(2) symmetry if $S^2(\mathbf{r})=V^2(\mathbf{r})$ \cite{Bell1975_NPB98-151, Ginocchio2005_PR414-165}.

Note that the conclusions concerning the properties of symmetries discussed hereafter remain valid even if either the scalar or vector potential is modified by an arbitrary constant, i.e.,
\begin{equation}\label{Eq:2.1.modifySV}
  S(\mathbf{r}) \rightarrow S(\mathbf{r})+c_S\,,\quad V(\mathbf{r}) \rightarrow V(\mathbf{r})+c_V\,,
\end{equation}
because one can simply adjust the mass or energy by the same constant so that the Dirac equation remains unchanged,
\begin{equation}\label{Eq:2.1.modifyME}
  M \rightarrow M - c_S\,,\quad \epsilon \rightarrow \epsilon + c_V\,.
\end{equation}

Therefore, it is shown that at the exactly spin-symmetry limit, $V(r)-S(r)=\Delta_0$, the SU(2) generators read \cite{Ginocchio1998_PLB425-1}
\begin{equation}\label{Eq:2.3.SSgenerator}
  \mathbf{S}
  = \lb\begin{array}{cc}
    \mathbf{s} & 0 \\ 0 &\tilde{\mathbf{s}}
    \end{array}\rb\,,
\end{equation}
with $\mathbf{s} = \boldsymbol{\sigma}/2$ and $\tilde{\mathbf{s}} = (\boldsymbol{\sigma}\cdot\hat{\mathbf{p}}) \mathbf{s} (\boldsymbol{\sigma}\cdot\hat{\mathbf{p}})$.
These generators satisfy the SU(2) algebra and commute with the Dirac Hamiltonian,
\begin{equation}
  [S_i, S_j] = i\epsilon_{ijk}S_k\,,
  \quad
  [\mathbf{S}, H] = 0\,.
\end{equation}
As a step further, the single-particle wave functions in Eq.~(\ref{Eq:2.1.spwfR}) of spin partners satisfy the conditions \cite{Ginocchio2005_PR414-165, He2006_EPJA28-265}
\begin{equation}\label{Eq:2.3.GatSS}
  G_a(r) = G_b(r)
\end{equation}
and
\begin{equation}\label{Eq:2.3.FatSS}
    \lb-\frac{d}{dr}+\frac{\kappa_a}{r}\rb F_a(r) = \lb-\frac{d}{dr}+\frac{\kappa_b}{r}\rb F_b(r)\,,
\end{equation}
for the upper and lower components of Dirac spinor.

Similarly, at the exact PSS limit, $V(r)+S(r)=\Sigma_0$, the SU(2) generators read \cite{Ginocchio1998_PLB425-1}
\begin{equation}\label{Eq:2.3.PSSgenerator}
  \tilde{\mathbf{S}}
  = \lb\begin{array}{cc}
    \tilde{\mathbf{s}} & 0 \\ 0 & \mathbf{s}
    \end{array}\rb\,,
\end{equation}
and
\begin{equation}
  [\tilde S_i, \tilde S_j] = i\epsilon_{ijk}\tilde S_k\,,
  \quad
  [\tilde{\mathbf{S}}, H] = 0\,.
\end{equation}
The single-particle wave functions in Eq.~(\ref{Eq:2.1.spwfR}) of pseudospin partners satisfy the conditions \cite{Ginocchio2002_PRC66-064312}
\begin{equation}\label{Eq:2.3.FatPSS}
  F_a(r) = F_b(r)
\end{equation}
and
\begin{equation}\label{Eq:2.3.GatPSS}
    \lb\frac{d}{dr}+\frac{\kappa_a}{r}\rb G_a(r) = \lb\frac{d}{dr}+\frac{\kappa_b}{r}\rb G_b(r)\,,
\end{equation}
for the lower and upper components of Dirac spinor.
These relations have been tested in realistic nuclei \cite{Ginocchio2002_PRC66-064312, He2006_EPJA28-265, Liang2015_PR570-1}.

\subsubsection{Perturbative and non-perturbative behaviors}

Since the PSS was recognized as a relativistic symmetry of the Dirac Hamiltonian \cite{Ginocchio1997_PRL78-436}, the perturbative nature of this symmetry has become a hot topic.
The main concern is that there are no bound states at the exact SU(2) PSS limit (\ref{Eq:2.1.PSSlimit}) discussed above, and thus the PSS is always broken in realistic nuclei.
The non-perturbative behaviors of PSS have been considered since the study in Ref.~\cite{Marcos2001_PLB513-30}.
Following Arima's definition of dynamical symmetry \cite{Arima1999_RIKEN-AF-NP-276}, such non-perturbative behavior is related to the dynamical nature of the PSS \cite{Alberto2001_PRL86-5015, Alberto2002_PRC65-034307}.
However, there was no quantitative investigations based on the exact perturbation theory until that in Ref.~\cite{Liang2011_PRC83-041301R}.

In Ref.~\cite{Liang2011_PRC83-041301R}, the perturbation theory was used for the first time to investigate the spin and pseudospin symmetries of the Dirac Hamiltonian and their symmetry-breaking in realistic nuclei.
The perturbation corrections to the single-particle energies and wave functions were calculated order by order.
In such a way, the link between the single-particle states in the realistic nuclear systems and their counterparts at the symmetry limits can be constructed explicitly.

Following the idea of Rayleigh-Schr\"odinger perturbation theory, the Dirac Hamiltonian $H$ in Eq.~(\ref{Eq:2.1.HDirac}) or (\ref{Eq:2.1.DiraceqR}) is divided as
\begin{equation}\label{Eq:4.1.HH0W}
    H = H_0 + W\,,
\end{equation}
or equivalently
\begin{equation}\label{Eq:4.1.H0HW}
    H_0 = H - W\,,
\end{equation}
where $H_0$ leads to the exact spin or pseudospin symmetry, and $W$ is identified as the corresponding symmetry-breaking potential.
The conditions,
\begin{equation}\label{Eq:4.1.PTcondition}
    \left|\frac{W_{mk}}{E_k-E_m}\right|\ll 1
    \quad\mbox{for}\quad m\neq k
\end{equation}
with $W_{mk}=\left< \psi_m | W | \psi_k \right>$ that govern the convergence of the perturbation series, determine whether or not $W$ can be treated as a small perturbation.

For the SU(2) spin- and pseudospin-symmetry limits shown in Eqs.~(\ref{Eq:2.1.SSlimit}) and (\ref{Eq:2.1.PSSlimit}), the Dirac Hamiltonians with the exact symmetries read
\begin{subequations}\label{Eq:4.1.H0SU2}
\begin{align}
    H_0^{\rm SS}&=
    \lb\begin{array}{cc}
        M+\Sigma & - \frac{d}{dr}+ \frac{\kappa}{r} \\
         \frac{d}{dr}+ \frac{\kappa}{r} & -M+\Delta_0
    \end{array}\rb\,,\\
    H_0^{\rm PSS}&=
    \lb\begin{array}{cc}
        M+\Sigma_0 & - \frac{d}{dr}+ \frac{\kappa}{r} \\
         \frac{d}{dr}+ \frac{\kappa}{r} & -M+\Delta
    \end{array}\rb\,,
\end{align}
\end{subequations}
respectively, whose eigenenergies are denoted as $E_0$ and the corresponding symmetry-breaking potentials are
\begin{subequations}\label{Eq:4.1.WSU2}
\begin{align}
    W^{\rm SS}&=
    \lb\begin{array}{cc}
        0 & 0 \\
        0 & \Delta-\Delta_0
    \end{array}\rb\,,\\
    W^{\rm PSS}&=
    \lb\begin{array}{cc}
        \Sigma-\Sigma_0 & 0 \\
        0 & 0
    \end{array}\rb\,.
\end{align}
\end{subequations}

In contrast to using the Schr\"{o}dinger-like equations in the previous studies \cite{Marcos2001_PLB513-30, Alberto2001_PRL86-5015, Alberto2002_PRC65-034307}, it is remarkable that all operators involved here, $H$, $H_0$, and $W$, are Hermitian, and they do not contain any singularity.
This allows us to perform the order-by-order perturbation calculations.
It is also crucial that within the present decomposition the $W$ term is the only symmetry-breaking potential, thus the ambiguity caused by the strong cancellations among the different terms in the Schr\"{o}dinger-like equations can also be avoided.

Therefore, this method is able to provide an explicit and quantitative way for investigating the perturbative nature of spin and pseudospin symmetries.
For the symmetry of perturbative nature, the link between the single-particle states in realistic nuclei and their counterparts at the symmetry limits can be constructed quantitatively.
For the symmetry of non-perturbative nature, the divergence of the perturbation series will be found explicitly.

\begin{figure}[tb]
\begin{center}
  \includegraphics[width=7cm]{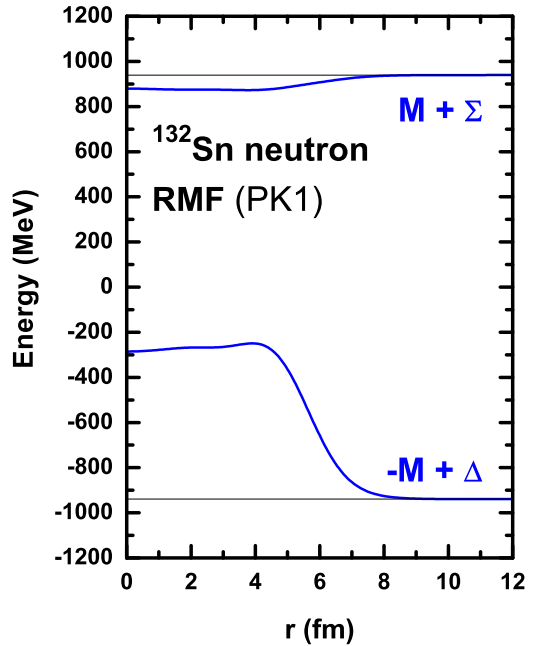}
\end{center}
\caption{(Color online) Single-particle mean-field potentials for neutrons in $^{132}$Sn calculated by the RMF theory with the effective interaction PK1 \cite{Long2004_PRC69-034319}.
Taken from Ref.~\cite{Liang2015_PR570-1}.
\label{Fig:2.3.132potl}}
\end{figure}

\begin{figure}[tb]
\begin{center}
  \includegraphics[width=8cm]{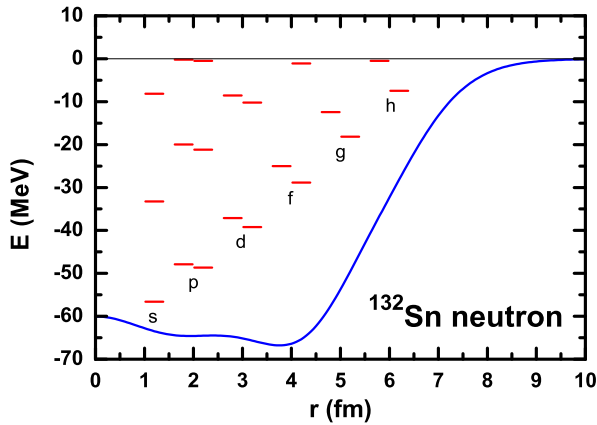}
\end{center}
\caption{(Color online) Single-particle spectrum of neutrons in $^{132}$Sn calculated by the RMF theory with PK1.
For each pair of spin doublets, the left state is that with $j_<=l-1/2$ and the right one with $j_>=l+1/2$.
Potential $\Sigma(r)$ is shown with the solid line.
Taken from Ref.~\cite{Liang2015_PR570-1}.
\label{Fig:2.3.132spectra}}
\end{figure}

In Ref.~\cite{Liang2011_PRC83-041301R}, the neutrons in $^{132}$Sn were taken as examples.
The corresponding mean-field potentials and single-particle energies, $E=\epsilon-M$ excluding the rest mass of nucleon, calculated by the self-consistent RMF theory with the effective interaction PK1 \cite{Long2004_PRC69-034319} are shown in Figs.~\ref{Fig:2.3.132potl} and \ref{Fig:2.3.132spectra}, respectively.
The depths of potentials are of $\Sigma(r)\sim70$~MeV and $\Delta(r)\sim700$~MeV, respectively.

\begin{figure}[tb]
\begin{center}
\includegraphics[width=8cm]{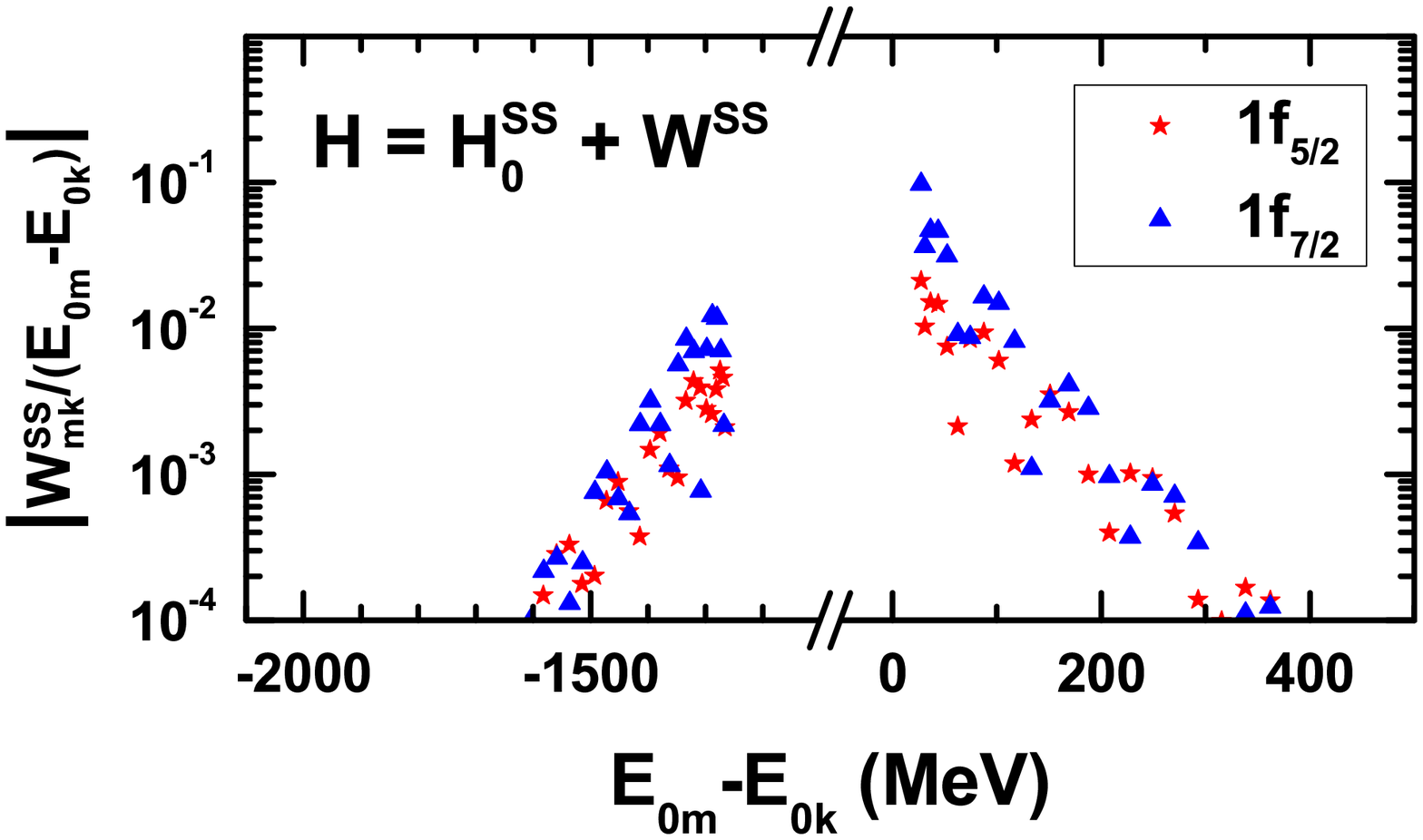}
\includegraphics[width=8cm]{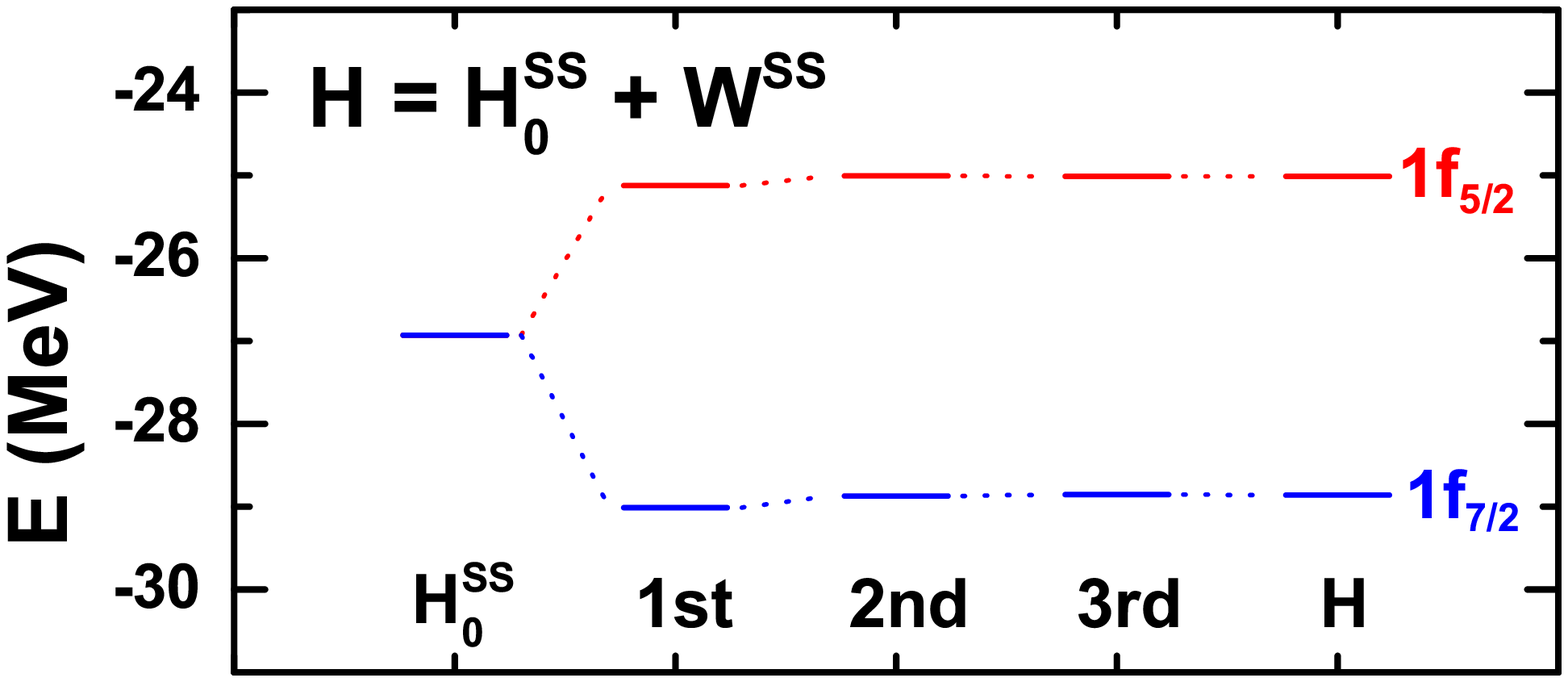}
\end{center}
\caption{(Color online) Upper panel: Values of $\left|W_{mk}/(E_m-E_k)\right|$ versus the energy differences $E_m-E_k$ for the $k=1f$ spin doublets.
The unperturbed eigenstates are chosen as those of $H_0^{\rm SS}$, and the single-particle states $m$ include the states in both the Fermi and Dirac sea.
Lower panel: Single-particle energies of the $1f$ spin doublets obtained at the exact SU(2) spin-symmetry limit, and by the first-, second-, and third-order perturbation calculations, as well as those by the RMF theory.
Taken from Ref.~\cite{Liang2011_PRC83-041301R}.
\label{Fig:4.1.SU2}}
\end{figure}

For the case of spin symmetry, taking the spin doublets $(1f_{5/2},\,1f_{7/2})$ as an example, the values of $\left|W_{mk}/(E_m-E_k)\right|$ are plotted as a function of the energy difference $E_m-E_k$ in the upper panel of Fig.~\ref{Fig:4.1.SU2}.
In this case, the unperturbed eigenstates are chosen as those of $H_0^{\rm SS}$ in Eq.~(\ref{Eq:4.1.H0SU2}), and the constant potential is chosen as $-M+\Delta_0=-350$~MeV.
It is verified that the convergence of the perturbation series is not sensitive to the value of $\Delta_0$.
For the completeness of the basis, the single-particle states $m$ must include not only the states in the Fermi sea but also those in the Dirac sea.

It is shown that the values of $\left|W_{mk}/(E_m-E_k)\right|$ decrease as a general tendency when the energy difference $|E_m-E_k|$ increases.
This feature provides natural cut-offs of the single-particle states in the perturbation calculations.
It is crucial to find that the largest value of $\left|W_{mk}/(E_m-E_k)\right|$ is around $0.1$, which indicates the criterion in Eq.~(\ref{Eq:4.1.PTcondition}) can be nicely fulfilled.

The perturbation corrections to the single-particle energies of the $1f$ spin doublets are then examined.
In the lower panel of Fig.~\ref{Fig:4.1.SU2}, by choosing the unperturbed eigenstates as those of $H_0^{\rm SS}$, the single-particle energies obtained at the exact SU(2) spin-symmetry limit, and their counterparts obtained by the first-, second-, and third-order perturbation calculations, as well as those obtained by the self-consistent RMF theory, are shown from left to right.
It is shown that the spin-orbit splitting is well reproduced by the second-order perturbation calculations.
Equivalently, the perturbation corrections can be performed with $H_0=H-W$, i.e., by choosing the unperturbed eigenstates as those of $H$.
The perturbation corrections to the single-particle wave functions can be examined in the same way, and the same conclusions hold \cite{Liang2011_PRC83-041301R}.

In other words, the nature of spin-symmetry breaking from the SU(2) limit is perturbative, even though the spin-orbit splitting in realistic nuclei is substantial, in particular, for the states with high orbital angular momentum \cite{Liang2011_PRC83-041301R}.

\begin{figure}[tb]
\begin{center}
\includegraphics[width=8cm]{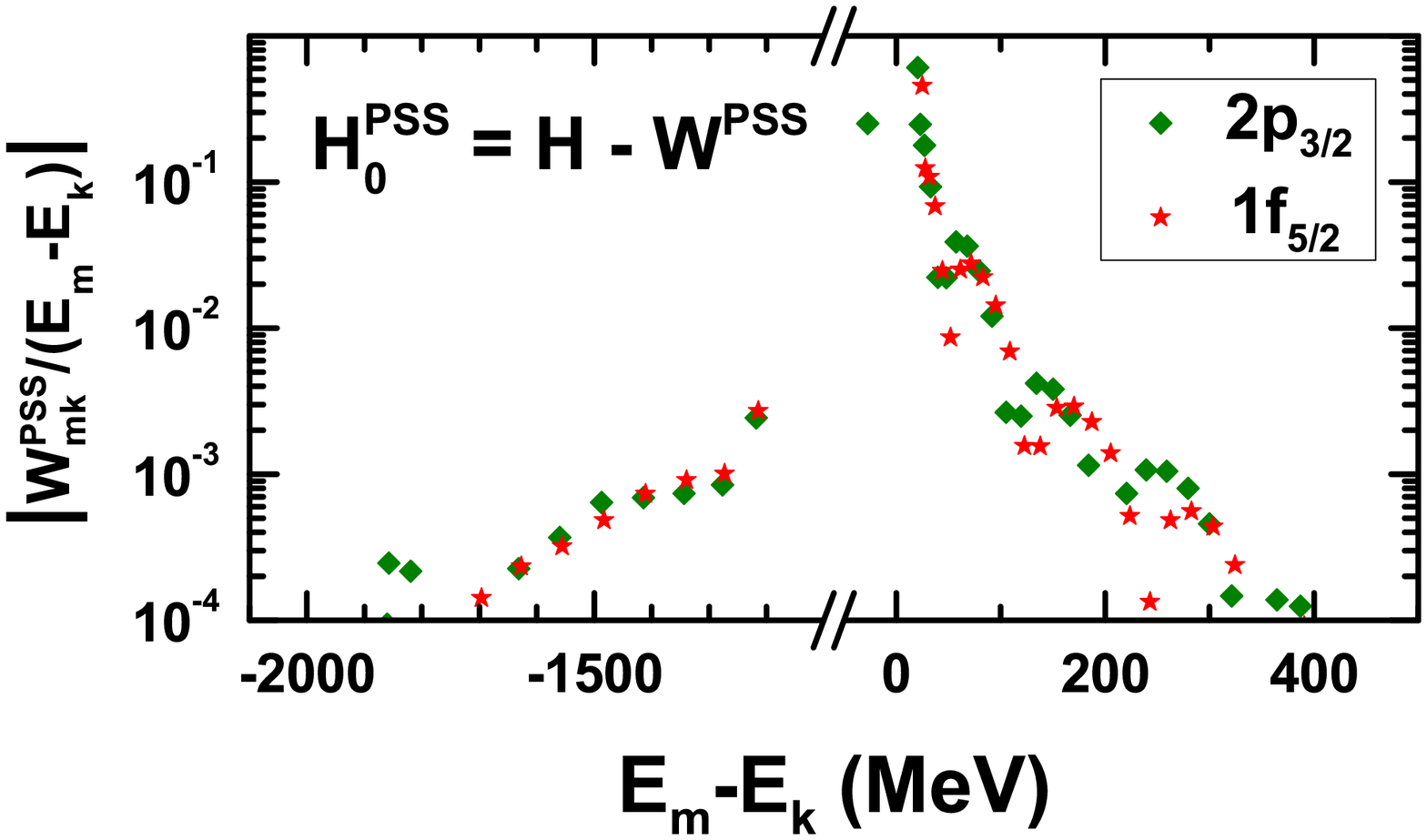}
\includegraphics[width=8cm]{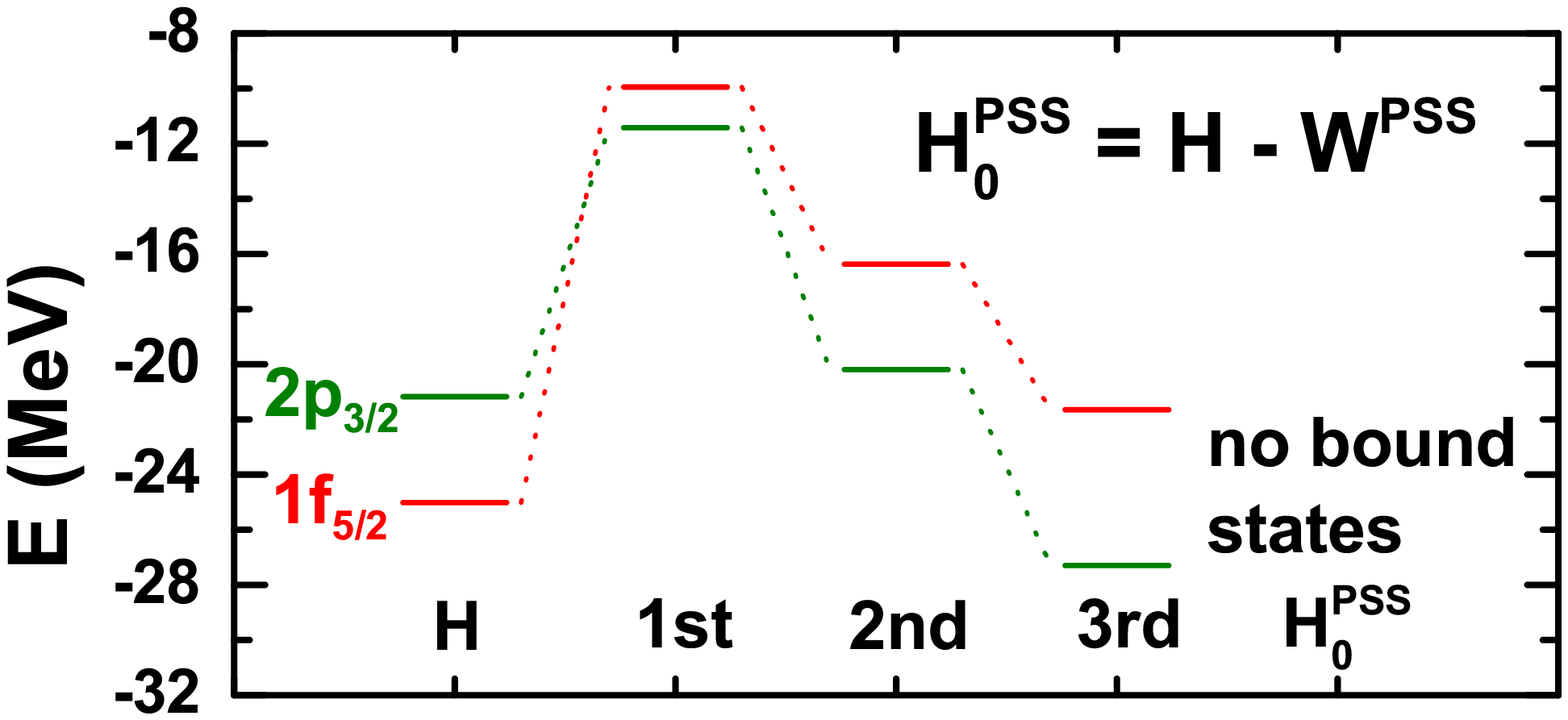}
\end{center}
\caption{(Color online) Upper panel: Values of $\left|W_{mk}/(E_m-E_k)\right|$ versus the energy differences $E_m-E_k$ for the $k=1\tilde d$ pseudospin doublets.
The unperturbed eigenstates are chosen as those of $H$.
Lower panel: Single-particle energies of the $1\tilde d$ pseudospin doublets obtained by the RMF theory, and by the first-, second-, and third-order perturbation calculations.
There exist no bound states at the exact SU(2) PSS limit.
Taken from Ref.~\cite{Liang2011_PRC83-041301R}.
\label{Fig:3.1.SU2p}}
\end{figure}

For the PSS case, the pseudospin doublets $(2p_{3/2},\,1f_{5/2})$ are taking as an example.
Since there are no bound states at the exact SU(2) PSS limit $H_0^{\rm PSS}$, the perturbation calculations are only performed from $H$ to $H_0^{\rm PSS}$, i.e., the unperturbed eigenstates are chosen as those of $H$ and the perturbation is taken as $-W^{\rm PSS}$ in Eq.~(\ref{Eq:4.1.WSU2}).
The values of $\left|W_{mk}/(E_m-E_k)\right|$ are plotted as a function of the energy difference $E_m-E_k$ in the upper panel of Fig.~\ref{Fig:3.1.SU2p}, where $M+\Sigma_0=900$~MeV.

It is critical to find that the largest value of $\left|W_{mk}/(E_m-E_k)\right|$ is about $0.6$ for the PSS case, compared to $0.1$ for the SS case.
It is not completely surprising if one keeps in mind that different components of Dirac spinor are involved,
\begin{subequations}
\begin{align}
    W^{\rm SS}_{mk} &= \lc F_m\rl (\Delta-\Delta_0)\lr F_k\rc\,,\\
    W^{\rm PSS}_{mk} &= \lc G_m\rl (\Sigma-\Sigma_0) \lr G_k\rc\,.
\end{align}
\end{subequations}
Although the potentials obviously satisfy $\left|\Delta-\Delta_0\right|\gg \left|\Sigma-\Sigma_0\right|$, the upper component is of $G(r)\sim O(1)$ and the lower component $F(r)\sim O(1/10)$ for the states of nucleons in the Fermi sea.

Formally, the perturbation corrections to the single-particle energies of the $1\tilde d$ pseudospin doublets can be performed.
In the lower panel of Fig.~\ref{Fig:3.1.SU2p}, by choosing the unperturbed eigenstates as those of $H$, the single-particle energies obtained by the self-consistent RMF theory, and their counterparts obtained by the first-, second-, and third-order perturbation calculations are shown.
It is seen that the energy corrections do not converge, meanwhile there exist no bound states at the exact SU(2) PSS limit.

Therefore, it is confirmed in an explicit way that the bridge connecting the Dirac Hamiltonian in realistic nuclei and that with the exact SU(2) spin symmetry can be constructed, but the behavior of PSS is non-perturbative if the PSS SU(2) solution shown in Eq.~(\ref{Eq:2.1.PSSlimit}) or (\ref{Eq:4.3.SUSYDPSSB}) is regarded as its symmetry limit \cite{Liang2011_PRC83-041301R}.

However, Ginocchio \cite{Ginocchio2005_PRL95-252501} presented another kind of symmetry of the Dirac Hamiltonian---the U(3) symmetry---in which the energies of pseudospin doublets are strictly degenerate.
Based on this symmetry limit, the nature of PSS is indeed perturbative, as discussed in the next Subsection.

\subsection{U(3) symmetry and supersymmetry for Schr\"odinger-like equations}\label{Sect:3.2}

In this Subsection, we will first discuss the U(3) symmetry of the Dirac Hamiltonian \cite{Ginocchio2005_PRL95-252501}.
Based on this symmetry limit, the perturbative nature of pseudospin symmetry will be shown explicitly \cite{Liang2011_PRC83-041301R}.
One of the open questions is concerning about the supersymmetric representation of the Dirac Hamiltonian with such U(3) symmetry.
A possible but yet incomplete answer is the supersymmetric quantum mechanics for the Schr\"odinger-like equations \cite{Typel2008_NPA806-156}.
In such a way, the U(3) pseudospin-symmetry limit can be derived but the symmetry-breaking term, if it presents, is not Hermitian.

\subsubsection{U(3) symmetry}

It is well known that the non-relativistic harmonic oscillator in spherical systems has degeneracies in addition to those due to the rotational invariance.
The energy spectrum depends only on the total harmonic-oscillator quantum number $N=2n+l$, thus the single-particle states in a whole major shell have the same energy.
These degeneracies are produced by the U(3) symmetry \cite{Elliott1958PRSA245-128}.
In addition, the energy does not depend on the orientation of spin and hence the non-relativistic harmonic oscillator holds the spin symmetry as well.

The relativistic harmonic oscillator does have the same kind of U(3) symmetry, i.e., the energy spectrum depends only on the total harmonic-oscillator quantum number $N$, although the energy spectrum for the RHO in general does not have a linear dependence on $N$ as does in the non-relativistic case.
The Dirac Hamiltonian in Eq.~(\ref{Eq:2.1.HDirac}) for a spherical RHO potential with the spin symmetry reads
\begin{equation}\label{Eq:3.2.HRHO}
  H = \boldsymbol{\alpha}\cdot\mathbf{p} + \beta[M+S(\mathbf{r})]+V(\mathbf{r})
\end{equation}
with
\begin{equation}
  S(\mathbf{r}) = V(\mathbf{r}) = \frac{M\omega^2}{2}r^2\,.
\end{equation}
The corresponding U(3) generators were derived by Ginocchio in Ref.~\cite{Ginocchio2005_PRL95-252501}.

For that let us first recall the U(3) symmetry in the non-relativistic case.
The non-relativistic U(3) generators are the orbital angular momenta $\mathbf{l}=\mathbf{r}\times\mathbf{p}$, the quadrupole operators $q_m = \frac{1}{M\omega} \sqrt{\frac{3}{2}} \lb M^2\omega^2[rr]^{(2)}_m + [pp]^{(2)}_m\rb$, and the monopole generator $\mathcal{N}_{\rm NR} = \frac{1}{2\sqrt{2}M\omega} (M^2\omega^2r^2+p^2)-\frac{3}{2}$, where $[\cdot\cdot]^{(2)}_m$ means coupled to angular momentum rank $2$ and projection $m$.
They form the closed U(3) algebra as
\begin{subequations}
\begin{align}
  [\mathcal{N}_{\rm NR}, \mathbf{l}] &= [\mathcal{N}_{\rm NR}, q_m] = 0\,,\\
  [\mathbf{l},\mathbf{l}]^{(t)} &= -\sqrt{2} \mathbf{l} \delta_{t1}\,,\\
  [\mathbf{l},q]^{(t)} &= -\sqrt{6} q \delta_{t2}\,,\\
  [q,q]^{(t)} &= 3\sqrt{10}\mathbf{l}\delta_{t1}\,,
\end{align}
\end{subequations}
with $\mathcal{N}_{\rm NR}$ generating a U(1) algebra, whose eigenvalues are the total number of quanta $N$, and $(\mathbf{l},\,q)$ generating an SU(3) algebra.

Analogously, the main task for the relativistic case is to identify the corresponding generators $\mathbf{L}$, $Q_m$, and $\mathcal{N}$.
Similar to the spin operator $\mathbf{S}$ shown in Eq.~(\ref{Eq:2.3.SSgenerator}), the orbital angular momentum $\mathbf{L}$ in the relativistic case reads
\begin{equation}\label{Eq:3.2.Lgenerator}
  \mathbf{L}
  = \lb\begin{array}{cc}
    \mathbf{l} & 0 \\ 0 & (\boldsymbol{\sigma}\cdot\hat{\mathbf{p}})\mathbf{l}(\boldsymbol{\sigma}\cdot\hat{\mathbf{p}})
    \end{array}\rb\,.
\end{equation}
For $Q_m$ and $\mathcal{N}$, they are assumed as the forms of \cite{Ginocchio2005_PRL95-252501}
\begin{equation}
  Q_m
  = \lb\begin{array}{cc}
    (Q_m)_{11} &
    (Q_m)_{12}(\boldsymbol{\sigma}\cdot \mathbf{p}) \\
    (\boldsymbol{\sigma}\cdot \mathbf{p})(Q_m)_{21} &
    (\boldsymbol{\sigma}\cdot \mathbf{p})(Q_m)_{22}(\boldsymbol{\sigma}\cdot \mathbf{p})
    \end{array}\rb
\end{equation}
and
\begin{equation}
  \mathcal{N}
  = \lb\begin{array}{cc}
    (\mathcal{N})_{11} &
    (\mathcal{N})_{12}(\boldsymbol{\sigma}\cdot \mathbf{p}) \\
    (\boldsymbol{\sigma}\cdot \mathbf{p})(\mathcal{N})_{21} &
    (\boldsymbol{\sigma}\cdot \mathbf{p})(\mathcal{N})_{22}(\boldsymbol{\sigma}\cdot \mathbf{p})
    \end{array}\rb
    -\mathcal{N}_0\,,
\end{equation}
and one of the solutions found in Ref.~\cite{Ginocchio2005_PRL95-252501} is
\begin{align}
  &Q_m = \sqrt{\frac{3}{M\omega^2(H+M)}}\times\nonumber\\
  &\lb\begin{array}{cc}
    \frac{M\omega^2}{2}(\frac{M\omega^2}{2}r^2+2M) [rr]^{(2)}_m + [pp]^{(2)}_m &
    \frac{M\omega^2}{2}[rr]^{(2)}_m(\boldsymbol{\sigma}\cdot \mathbf{p}) \\
    (\boldsymbol{\sigma}\cdot \mathbf{p})\frac{M\omega^2}{2}[rr]^{(2)}_m &
    [pp]^{(2)}_m
    \end{array}\rb
\end{align}
and
\begin{align}
  &\mathcal{N} = -\frac{3}{2} + \frac{1}{2\sqrt{M\omega^2(H+M)}}\nonumber\\
  &\times\lb\begin{array}{cc}
    \frac{M\omega^2}{2}(\frac{M\omega^2}{2}r^2+2M) r^2 + p^2 &
    \frac{M\omega^2}{2}r^2(\boldsymbol{\sigma}\cdot \mathbf{p}) \\
    (\boldsymbol{\sigma}\cdot \mathbf{p})\frac{M\omega^2}{2}r^2 &
    p^2
    \end{array}\rb\,.
\end{align}
The commutation relations are then those of the U(3) algebra,
\begin{subequations}
\begin{align}
  [\mathcal{N}, \mathbf{L}] &= [\mathcal{N}, Q_m] = 0\,,\\
  [\mathbf{L},\mathbf{L}]^{(t)} &= -\sqrt{2} \mathbf{L} \delta_{t1}\,,\\
  [\mathbf{L},Q]^{(t)} &= -\sqrt{6} Q \delta_{t2}\,,\\
  [Q,Q]^{(t)} &= 3\sqrt{10}\mathbf{L}\delta_{t1}\,.
\end{align}
\end{subequations}

The spin generators $\mathbf{S}$ in Eq.~(\ref{Eq:2.3.SSgenerator}) commute with the U(3) generators as well as the Dirac Hamiltonian $H$ in Eq.~(\ref{Eq:3.2.HRHO}), so the invariance group is $U(3)\times SU(2)$.

\subsubsection{Perturbative nature of PSS}

As the U(3) symmetry of the Dirac Hamiltonian conserves strictly the degeneracy in energy of pseudospin doublets, it is important to investigate in a quantitative way the perturbative nature of PSS based on such a symmetry limit.

For that the Dirac Hamiltonian $H$ in Eq.~(\ref{Eq:2.1.DiraceqR}) is split as
\begin{equation}\label{Eq:4.1.HU3}
    H = H_0^{\rm RHO} + W^{\rm RHO}\,,
\end{equation}
with the symmetry-conserving Hamiltonian
\begin{equation}\label{Eq:4.1.H0U3}
H_0^{\rm RHO}=
    \lb\begin{array}{cc}
        M+\Sigma_{\rm HO} & - \frac{d}{dr}+ \frac{\kappa}{r} \\
         \frac{d}{dr}+ \frac{\kappa}{r} & -M+\Delta_0
    \end{array}\rb
\end{equation}
and the symmetry-breaking potential
\begin{equation}\label{Eq:4.1.WU3}
    W^{\rm RHO} =
    \lb\begin{array}{cc}
        \Sigma-\Sigma_{\rm HO} & 0 \\
        0 & \Delta-\Delta_0
    \end{array}\rb\,,
\end{equation}
where $\Sigma_{\rm HO}(r) = c_0 + c_2 r^2$ with a harmonic-oscillator form.
The constants $-M+\Delta_0=-350$~MeV and $M+c_0=865$~MeV are chosen in $H_0^{\rm RHO}$, which holds to the energy degeneracy of the whole major shell.
As discussed before, the perturbative properties are not sensitive to the choice of these two constants.
Meanwhile, the coefficient $c_2$ is chosen as $1.00$~MeV/fm$^2$ to minimize the perturbations to the $pf$ major shell.

\begin{figure}[tb]
\begin{center}
\includegraphics[width=8cm]{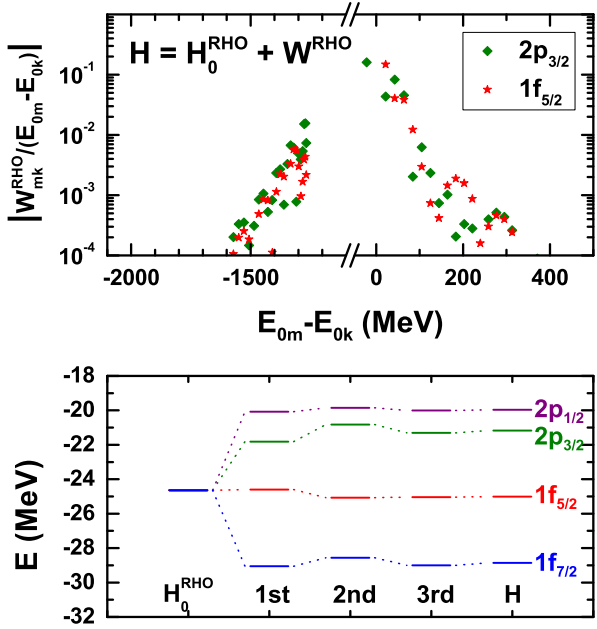}
\end{center}
\caption{(Color online) Same as Fig.~\ref{Fig:4.1.SU2}, but for the case of the relativistic harmonic-oscillator potential, and the whole $pf$ major shell.
The unperturbed eigenstates are chosen as those of $H_0^{\rm RHO}$.
Taken from Ref.~\cite{Liang2011_PRC83-041301R}.
\label{Fig:4.1.U3}}
\end{figure}

In the upper panel of Fig.~\ref{Fig:4.1.U3}, the values of $\left|W^{\rm RHO}_{mk}/(E_m-E_k)\right|$ for the $k=1\tilde{d}$ pseudospin doublets are shown as a function of the energy difference $E_m-E_k$.
It is found that its general patterns are the same as those shown in Fig.~\ref{Fig:4.1.SU2}, and the largest perturbation correction is around $0.16$.
This indicates that the criterion in Eq.~(\ref{Eq:4.1.PTcondition}) is fulfilled.

In the lower panel of Fig.~\ref{Fig:4.1.U3}, the perturbation corrections to the single-particle energies of the states in the whole $pf$ major shell are shown.
It is shown that both the spin-orbit and pseudospin-orbit splitting are well reproduced by the third-order perturbation calculations.
Furthermore, the single-particle wave functions of $H$ can also be reproduced by the second-order perturbation calculations starting from $H_0^{\rm RHO}$ \cite{Liang2011_PRC83-041301R}.
Thus, the link between the single-particle states in realistic nuclei and their counterparts at the U(3)-symmetry limit can be established explicitly.

In short, the quantitative connection between the Dirac Hamiltonian in realistic nuclei and that with the RHO potential has been constructed by using the perturbation theory.
The energy splitting of the pseudospin doublets can be regarded as a result of small perturbation around the Dirac Hamiltonian with the U(3) symmetry, where the exact energy degeneracy of the pseudospin doublets holds.
This indicates the nature of PSS is indeed perturbative \cite{Liang2011_PRC83-041301R}.

\subsubsection{SUSY for Schr\"odinger-like equations}\label{Sect:3.2.3}

Note that the U(3) symmetry was not included in the supersymmetries of the Dirac Hamiltonian discussed in Ref.~\cite{Leviatan2004_PRL92-202501}.
Thus, the next but yet unsolved question will be \textit{Whether or not the supersymmetric representation of PSS based on the U(3)-symmetry limit can be found?}

One of possible but yet incomplete solutions is the SUSY quantum mechanics for the Schr\"odinger-like equations, which will be discussed here.
Another possible but also incomplete solution is the SUSY quantum mechanics for the Dirac equations with SRG, which will be discussed in Section~\ref{Sect:3.3}.

By using the SUSY quantum mechanics for the Schr\"odinger-like equation~(\ref{Eq:2.1.SchrG}) for the upper component of Dirac spinor, Typel \cite{Typel2008_NPA806-156} investigated the properties of PSS and concluded with a regular symmetry-breaking potential.

In Ref.~\cite{Typel2008_NPA806-156}, the effects of the tensor interaction were also taken into account, but we do not repeat this tensor part here for simplicity.
In the Schr\"odinger-like equation (\ref{Eq:2.1.SchrG}) for the upper component of Dirac spinor,
\begin{equation}\label{Eq:4.3.SchrG}
  H_G(\kappa) G_{n\kappa} = \epsilon_{n\kappa} G_{n\kappa}\,,
\end{equation}
the effective Hamiltonian reads
\begin{align}\label{Eq:4.3.SchrHG}
  &H_G(\kappa)\nonumber\\
  =&\frac{1}{M_+} \ls -\frac{d^2}{dr^2} +\frac{\kappa(\kappa+1)}{r^2}+\frac{M'_+}{M_+} \lb\frac{d}{dr} +\frac{\kappa}{r}\rb\rs + (M+\Sigma)\,,
\end{align}
with $M_+(r)=\epsilon_{n\kappa}+ M-V(r)+ S(r)$.
The main task is to construct the operators $B^+_\kappa$ and $B^-_\kappa$.
The particular ansatz for the Hamiltonian in Eq.~(\ref{Eq:4.3.SchrHG}) reads
\begin{subequations}\label{Eq:4.3.SUSY1B+B-}
\begin{align}
    B^+_\kappa &= \ls Q_\kappa(r)-\frac{d}{dr}\rs\frac{1}{\sqrt{M_+(r)}}\,,\\
    B^-_\kappa &= \frac{1}{\sqrt{M_+(r)}}\ls Q_\kappa(r)+\frac{d}{dr}\rs\,,
\end{align}
\end{subequations}
where the superpotentials $Q_\kappa(r)$ are the functions of $r$ to be determined.
Then, the SUSY partner Hamiltonians read
\begin{subequations}\label{Eq:4.3.SUSY1HQ}
\begin{align}
  &H_1(\kappa) = B^+_\kappa B^-_\kappa\nonumber\\
    =& \frac{1}{M_+}\ls Q^2_\kappa - Q'_\kappa -\frac{d^2}{dr^2} +\frac{M'_+}{M_+}\lb Q_\kappa +\frac{d}{dr}\rb\rs
\end{align}
and
\begin{align}
  &H_2(\kappa) = B^-_\kappa B^+_\kappa\nonumber\\
    =& \frac{1}{M_+}\ls Q^2_\kappa + Q'_\kappa -\frac{d^2}{dr^2} +\frac{M'_+}{M_+}\frac{d}{dr} +\frac{M''_+}{2M_+} -\frac{3(M'_+)^2}{4M_+^2}\rs\,.
\end{align}
\end{subequations}
In order to identify the structure $\kappa(\kappa+1)$ in Eq.~(\ref{Eq:4.3.SchrHG}), the reduced superpotentials $q_\kappa(r)$ are introduced as \cite{Typel2008_NPA806-156}
\begin{equation}\label{Eq:4.3.SUSY1qQ}
    q_\kappa(r) = Q_\kappa(r) - \frac{\kappa}{r}\,.
\end{equation}
The Hamiltonians $H_1$ and $H_2$ are further rewritten as
\begin{subequations}\label{Eq:4.3.SUSY1Hq}
\begin{align}
  H_1(\kappa) =
    \frac{1}{M_+}\bigg[& -\frac{d^2}{dr^2} +\frac{\kappa(\kappa+1)}{r^2} +q^2_\kappa +2q_\kappa\frac{\kappa}{r} -q'_\kappa \nonumber\\
    & +\frac{M'_+}{M_+}\lb q_\kappa +\frac{d}{dr} +\frac{\kappa}{r}\rb\bigg]\,,\\
  H_2(\kappa)
    = \frac{1}{M_+}\bigg[& -\frac{d^2}{dr^2} +\frac{\kappa(\kappa-1)}{r^2} +q^2_\kappa +2q_\kappa\frac{\kappa}{r} +q'_\kappa \nonumber\\
    &+\frac{M'_+}{M_+}\frac{d}{dr} +\frac{M''_+}{2M_+} -\frac{3(M'_+)^2}{4M_+^2}\bigg]\,.
\end{align}
\end{subequations}

In general, the effective Hamiltonian $H_G$ in Eq.~(\ref{Eq:4.3.SchrHG}) differs from the SUSY Hamiltonian $H_1$ in Eq.~(\ref{Eq:4.3.SUSY1Hq}) by a constant, i.e.,
\begin{equation}\label{Eq:4.3.Eshift}
    H_G(\kappa) = H_1(\kappa)+e(\kappa)\,,
\end{equation}
where $e(\kappa)$ is the so-called energy shift \cite{Cooper2001}.
The reduced superpotentials $q_\kappa(r)$ then satisfy the first-order differential equation,
\begin{equation}\label{Eq:4.3.SUSY1q}
  q^2_\kappa +\lb 2\frac{\kappa}{r} +\frac{M'_+(\kappa)}{M_+(\kappa)}\rb q_\kappa -q'_\kappa = -M_+(\kappa) N(\kappa)\,.
\end{equation}
Note that $N(\kappa)=e(\kappa)-M-\Sigma(r)$ depends on the energy shift, whereas $M_+(\kappa)=\epsilon_{n\kappa}+M-\Delta(r)$ depends on the single-particle energy.
For the regular nuclear potentials, a boundary condition for the reduced superpotentials reads
\begin{equation}\label{Eq:4.3.SUSY1q0}
  q_\kappa(0) = 0\,.
\end{equation}
At small radius, $q_\kappa(r)$ behaves asymptotically as a linear function of $r$,
\begin{equation}\label{Eq:4.3.SUSY1qsmall}
  \lim_{r\rightarrow0} q_\kappa(r)= \frac{M_+(\kappa) N(\kappa)}{1-2\kappa} r\,,
\end{equation}
and at large radius, $q_\kappa(r)$ becomes a constant,
\begin{equation}\label{Eq:4.3.SUSY1qlarge}
  \lim_{r\rightarrow\infty} q_\kappa(r)= \sqrt{(M+\epsilon_{n\kappa})(M-e(\kappa))}\,,
\end{equation}
if the nuclear potentials vanish there.

It is important to examine the asymptotic behaviors of the full superpotentials $Q_\kappa(r)$, because they determine the type of SUSY \cite{Cooper2001}.
If $Q_\kappa(r)$ changes its sign from $r\rightarrow0$ to $r\rightarrow\infty$, it corresponds to the exact SUSY, and thus there exists a single non-degenerate state at zero energy.
In contrast, if $Q_\kappa(r)$ keeps its sign from $r\rightarrow0$ to $r\rightarrow\infty$, it corresponds to the broken SUSY, and thus all eigenstates are doubly degenerate with positive energies.

In the present case, $Q_\kappa(r)$ are always positive at $r\rightarrow\infty$, while $Q_\kappa(r)$ at $r\rightarrow0$ is determined by the angular-momentum term $\kappa/r$, i.e., the sign of $\kappa$.
In other words, the SUSY is exact for all the cases of $\kappa<0$, whereas SUSY is broken for all the cases of $\kappa>0$.
This is crucial to understand the intruder states in the PSS \cite{Typel2008_NPA806-156, Liang2013_PRC87-014334}.

The $\kappa$-dependent energy shifts $e(\kappa)$ can be determined as follows:
(i) For the case of $\kappa_a<0$, the SUSY is exact, and it requires
\begin{equation}\label{Eq:4.3.SUSY1e1}
    e(\kappa_a) = \epsilon_{1\kappa_a}\,.
\end{equation}
(ii) For the case of $\kappa_b>0$, the SUSY is broken, and thus the corresponding energy shift can be, in principle, any number which makes the whole set of $H_1$ eigenstates positive.
In practice, the energy shifts are determined by assuming that the PSO potentials vanish as $r\rightarrow0$.
This behavior is similar to that of the usual surface-peaked spin-orbit potentials.
Considering $M_+(\kappa_a)$ and $M_+(\kappa_b)$ are almost identical as $\epsilon_a\approx \epsilon_b$, the energy shifts read
\begin{equation}\label{Eq:4.3.SUSY1e2}
    e(\kappa_b) = 2\left.(M+\Sigma)\right|_{r=0} - e(\kappa_a)\,.
\end{equation}

Finally, one can derive the corresponding PSS-breaking potential.
The Hamiltonians $H_2(\kappa_a) + e(\kappa_a)$ and $H_2(\kappa_b) + e(\kappa_b)$ for the pseudospin doublets are almost identical, and their difference is given by the potential \cite{Typel2008_NPA806-156}
\begin{align}\label{Eq:4.3.SUSY1PSO}
  \tilde W^{\rm PSS} &= [H_2(\kappa_a) + e(\kappa_a)] - [H_2(\kappa_b) + e(\kappa_b)]\nonumber\\
  &= \frac{2}{\sqrt{M_+}}\frac{d}{dr}\frac{q_{\kappa_a}-q_{\kappa_b}}{\sqrt{M_+}}\,,
\end{align}
where the difference between $M_+(\kappa_a)$ and $M_+(\kappa_b)$ is neglected.
It is crucial that this symmetry-breaking potential is a regular function of $r$ without singularity, in contrast to that shown in Eq.~(\ref{Eq:2.1.VPSOandVPCB}).

One of the simplest cases that such a symmetry-breaking potential vanishes is nothing but the relativistic harmonic-oscillator potential without the tensor term, i.e.,
\begin{equation}\label{Eq:4.3.SUSY1RHO}
  S(r)=V(r)=\frac{M}{2}\omega^2 r^2\,.
\end{equation}
This is exactly the Dirac Hamiltonian with the U(3) symmetry shown in Eq.~(\ref{Eq:3.2.HRHO}).

In other words, the U(3)-symmetry limit of the Dirac Hamiltonian can be derived by using the SUSY quantum mechanics for the Schr\"odinger-like equation.
However, neither the effective Hamiltonian $H_G$ in Eq.~(\ref{Eq:4.3.SchrHG}) nor its SUSY partner in Eq.~(\ref{Eq:4.3.SUSY1Hq}) is Hermitian, since the upper component wave functions alone, as the solutions of the Schr\"{o}dinger-like equation, are not orthogonal to each other.
This prevents us from being able to perform the quantitative perturbation calculations, when the PSS-breaking potential is non-zero.

\subsection{Supersymmetric representation of PSS with SRG}\label{Sect:3.3}

Another possible but also yet incomplete solution for the supersymmetric representation of pseudospin symmetry based on the U(3)-symmetry limit is that with the similarity renormalization group \cite{Liang2013_PRC87-014334, Shen2013_PRC88-024311}.
In this Subsection, we will first introduce the basic idea of similarity renormalization group for the Dirac Hamiltonian \cite{Guo2012_PRC85-021302R, Li2013_PRC87-044311, Guo2014_PRL112-062502}, then present the perturbative nature of pseudospin symmetry by combining the supersymmetric quantum mechanics, the similarity renormalization group, and the perturbation calculations \cite{Liang2013_PRC87-014334, Shen2013_PRC88-024311}.

\subsubsection{Similarity renormalization group}

Recent works in Refs.~\cite{Guo2012_PRC85-021302R, Li2013_PRC87-044311, Guo2014_PRL112-062502} bridged the gap between the perturbation calculations and the SUSY description of PSS by using the SRG technique.

The idea of SRG \cite{Wegner1994_AP506-77, Bylev1998_PLB428-329, Wegner2001_PR348-77} is to drive the Hamiltonian toward a band-diagonal form via the so-called flow equation and unitary transformations that suppress off-diagonal matrix elements.
In recent years, the SRG has been also widely used in the nuclear effective field theory and \textit{ab initio} calculations.
Recent reviews on the relevant topics can be found in, e.g., Refs.~\cite{Bogner2010_PPNP65-94, Hammer2013_RMP85-197}.

For the Dirac Hamiltonian shown in Eq.~(\ref{Eq:2.1.HDirac}), it can be transformed with the SRG into a diagonal form and expanded in a series of $1/M$.
It is very important that the effective Hamiltonian in the Schr\"odinger-like equation thus obtained is Hermitian, which makes the perturbation calculations feasible.

For that, the Dirac Hamiltonian in Eq.~(\ref{Eq:2.1.HDirac}) is separated into the diagonal $\varepsilon$ and off-diagonal $o$ parts, $H=\varepsilon+o$, with $[\varepsilon,\beta] = 0$ and $\{o,\beta\} = 0$.
In order to obtain the equivalent Schr\"odinger-like equation for nucleons, the main task is to decouple the eigenvalue equations for the upper and lower components of Dirac spinor.
One of the possible ways is to make the off-diagonal part of the Dirac Hamiltonian vanish with a proper unitary transformation.

According to the SRG \cite{Wegner2001_PR348-77}, the Hamiltonian $H$ is transformed by a unitary operator $U(l)$ with a flow parameter $l$ as
\begin{equation}\label{Eq:4.3.SRGHl}
    H(l) = U(l)HU^\dag(l)\,,
\end{equation}
where $H(l)=\varepsilon(l)+o(l)$ with the initial condition $H(0) = H$.
By taking the differential of the above equation, the so-called flow equation for the Hamiltonian reads
\begin{equation}\label{Eq:4.3.SRGdHl}
    \frac{d}{dl}H(l) = [\eta(l),H(l)]\,,
\end{equation}
with an anti-Hermitian generator
\begin{equation}\label{Eq:4.3.SRGeta}
  \eta(l) = \frac{dU(l)}{dl}U^\dag(l)\,.
\end{equation}
As discussed in Ref.~\cite{Bylev1998_PLB428-329}, one of the proper choices of $\eta(l)$ for letting the off-diagonal part $o(l)\rightarrow 0$ when $l\rightarrow\infty$ reads
\begin{equation}\label{Eq:4.3.SRGetaD}
  \eta(l) = [\beta M,H(l)]\,.
\end{equation}
Finally, the diagonal part of the Dirac Hamiltonian $\varepsilon(l)$ at the $l\rightarrow\infty$ limit can be derived analytically in a series of $1/M$ \cite{Guo2012_PRC85-021302R},
\begin{align}\label{Eq:4.3.SRGSchr}
    &~\varepsilon(\infty) \nonumber\\
    =&~ M\varepsilon_0(\infty) +\varepsilon_1(\infty) +\frac{\varepsilon_2(\infty)}{M} +\frac{\varepsilon_3(\infty)}{M^2} +\frac{\varepsilon_4(\infty)}{M^3} +\cdots\nonumber\\
    =&~
    \beta M+(\beta S+V)+\frac{1}{2M}\beta (\boldsymbol{\alpha}\cdot\mathbf{p})^2\nonumber\\
    &~+\frac{1}{8M^2} \ls\ls \boldsymbol{\alpha}\cdot\mathbf{p},\,(\beta S+V)\rs,\, \boldsymbol{\alpha}\cdot\mathbf{p}\rs \nonumber\\
    &~+\frac{1}{32M^3}\beta \bigg( -4(\boldsymbol{\alpha}\cdot\mathbf{p})^4 -2\ls \boldsymbol{\alpha}\cdot\mathbf{p},\, (\beta S+V)\rs^2\nonumber\\
    &~\left.\qquad\qquad +\Lb\boldsymbol{\alpha}\cdot\mathbf{p},\, \ls\ls \boldsymbol{\alpha}\cdot\mathbf{p},\, (\beta S+V)\rs,\, (\beta S+V)\rs\Rb\right)
    \nonumber \\
    &~+\cdots
\end{align}
In such a way, the eigenvalue equations for the upper and lower components of Dirac spinor are decoupled.
The equivalent Schr\"odinger-like equation for nucleons with Hermitian effective Hamiltonian can be obtained.
For details see Refs.~\cite{Guo2012_PRC85-021302R, Li2013_PRC87-044311} for the spherical case and Ref.~\cite{Guo2014_PRL112-062502} for the axially deformed case.

For the spherical case, the effective Hamiltonian for the nucleons in the Fermi sea expanded up to the $(1/M^3)$-th order reads \cite{Guo2012_PRC85-021302R}
\begin{align}\label{Eq:4.3.SRGHFermi}
  H_F =&~ M +\Sigma +\frac{p_F^2}{2M} -\frac{1}{2M^2}\lb Sp_F^2 -S'\frac{d}{dr}\rb -\frac{\kappa}{r}\frac{\Delta'}{4M^2} \nonumber\\
  &~+\frac{\Sigma''}{8M^2}
  +\frac{S}{2M^3}\lb Sp_F^2 -2S'\frac{d}{dr}\rb +\frac{\kappa}{r}\frac{S\Delta'}{2M^3} \nonumber\\
  &~ -\frac{(\Sigma')^2 -2\Sigma'\Delta' +4S\Sigma''}{16M^3} -\frac{p_F^4}{8M^3}\,,
\end{align}
with the operator $p_F^2 = -d^2/(dr^2) + \kappa(\kappa+1)/r^2$.
This Hamiltonian is decomposed into five Hermitian components: the non-relativistic term, the spin-orbit term, the dynamical term, the relativistic modification of kinetic energy, and the Darwin term.
Since all these terms are Hermitian, one can calculate the contribution of each term to the single-particle energies, which is very useful to disclose the origin of the relativistic symmetries.

\subsubsection{SUSY with SRG}

Gathering all the pieces presented above, it is promising to understand the PSS and its symmetry-breaking mechanism in a quantitative way by combining the SRG, SUSY quantum mechanics, and perturbation calculations \cite{Liang2013_PRC87-014334, Shen2013_PRC88-024311}.

Up to the $1/M$-th order, Eq.~(\ref{Eq:4.3.SRGHFermi}) corresponds to a usual Schr\"odinger equation.
Within the spherical symmetry, the radial Schr\"odinger equation is written in the form of
\begin{equation}\label{Eq:4.3.SUSY2Schr}
    H(\kappa) R(r) = E R(r)\,,
\end{equation}
with the single-particle Hamiltonian
\begin{equation}\label{Eq:4.3.SUSY2HSchr}
    H(\kappa) = -\frac{d^2}{2Mdr^2} + \frac{\kappa(\kappa+1)}{2Mr^2} + V(r)
\end{equation}
and the single-particle wave functions
\begin{equation}\label{Eq:4.3.SUSY2psi}
    \psi(\mathbf{r}) = \frac{R(r)}{r}\mathscr Y_{j m}^{l}(\hat{\mathbf{r}})\,.
\end{equation}
In this Subsection, $V(r)$ is the non-relativistic central potential standing for the sum of the scalar and vector potentials $\Sigma(r)$ in Eq.~(\ref{Eq:4.3.SRGHFermi}).

It is clear that $H$ conserves the spin symmetry.
In order to investigate the origin of PSS and its symmetry breaking, it is crucial to identify the pseudo-centrifugal barrier that is proportional to $\tilde{l}(\tilde{l}+1)=\kappa(\kappa-1)$.
The SUSY quantum mechanics is one of promising approaches for identifying such structure.

Following the similar procedures shown in Section~\ref{Sect:3.2.3}, one starts with a couple of Hermitian conjugate first-order operators
\begin{subequations}\label{Eq:4.3.SUSY2B+B-}
\begin{align}
    B^+_\kappa &= \ls Q_\kappa(r)-\frac{d}{dr}\rs\frac{1}{\sqrt{2M}}\,,\\
    B^-_\kappa &= \frac{1}{\sqrt{2M}}\ls Q_\kappa(r)+\frac{d}{dr}\rs\,,
\end{align}
\end{subequations}
and the reduced superpotentials
\begin{equation}\label{Eq:4.3.SUSY2q}
    q_\kappa(r) = Q_\kappa(r) - \frac{\kappa}{r}\,,
\end{equation}
and ends up with the SUSY partner Hamiltonians
\begin{align}\label{Eq:4.3.SUSY2Hq}
    H_1(\kappa) &= B^+_\kappa B^-_\kappa \nonumber\\
    &= \frac{1}{2M} \ls-\frac{d^2}{dr^2} +\frac{\kappa(\kappa+1)}{r^2} +q_\kappa^2 +\frac{2\kappa}{r}q_\kappa -q'_\kappa\rs\,,\\
    H_2(\kappa) &= B^-_\kappa B^+_\kappa \nonumber\\
    &= \frac{1}{2M} \ls-\frac{d^2}{dr^2} +\frac{\kappa(\kappa-1)}{r^2} +q_\kappa^2 +\frac{2\kappa}{r}q_\kappa +q'_\kappa\rs\,.
\end{align}
It is important to note that these Hamiltonians are Hermitian, but not those in Eqs.~(\ref{Eq:4.3.SUSY1Hq}).

The reduced superpotentials $q_\kappa(r)$ satisfy the first-order differential equation \cite{Liang2013_PRC87-014334},
\begin{equation}\label{Eq:4.3.SUSY2qeq}
    \frac{1}{2M}\ls q_\kappa^2(r)+\frac{2\kappa}{r}q_\kappa(r)-q'_\kappa(r)\rs + e(\kappa) = V(r)\,,
\end{equation}
with the asymptotic behaviors
\begin{equation}\label{Eq:4.3.SUSY2qasy}
    \lim_{r\rightarrow\infty}q_\kappa(r) = \sqrt{-2Me(\kappa)}
\end{equation}
and
\begin{equation}\label{Eq:4.3.SUSY2qasy0}
    \lim_{r\rightarrow0}q_{\kappa}(r)=\frac{2M(e(\kappa)-V)}{(1-2\kappa)}r\,.
\end{equation}
The energy shifts are determined in the same way as that shown in Eqs.~(\ref{Eq:4.3.SUSY1e1}) and (\ref{Eq:4.3.SUSY1e2}), i.e,
\begin{equation}\label{Eq:4.3.SUSY2e}
    e(\kappa_a) = E_{1\kappa_a}
    \quad\mbox{and}\quad
    e(\kappa_b) = 2\left.V\right|_{r=0} - e(\kappa_a)\,,
\end{equation}
for the states with $\kappa_a<0$ and $\kappa_b>0$, respectively.

Before we show some numerical results, it is worthwhile to search analytically a possible exact PSS limit within the present scheme.
The SUSY partner Hamiltonian reads
\begin{equation}\label{Eq:4.3.SUSY2Htil}
    \tilde{H}(\kappa)=H_2(\kappa) + e(\kappa)
    = -\frac{d^2}{2Mdr^2} + \frac{\kappa(\kappa-1)}{2Mr^2} + \tilde{V}_\kappa(r)\,,
\end{equation}
with
\begin{equation}\label{Eq:4.3.SUSY2Vtil}
    \tilde{V}_\kappa(r)=V(r)+q'_\kappa(r)/M\,.
\end{equation}
By definition the exact PSS limit holds $E_{n\kappa_a} = E_{(n-1) \kappa_b}$ with $\kappa_a<0$ and $\kappa_a+\kappa_b=1$,
which indicates $H_2(\kappa_a) + e(\kappa_a) = H_2(\kappa_b) + e(\kappa_b)$.
By combining Eqs.~(\ref{Eq:4.3.SUSY2Hq}) and (\ref{Eq:4.3.SUSY2qeq}), as well as the boundary condition $q_\kappa(0)=0$, one is ready to have
\begin{equation}\label{Eq:4.3.SUSY2PSSq}
    q_{\kappa_a}(r) = q_{\kappa_b}(r) = M\omega_{\kappa} r\,,
\end{equation}
with a known constant $\omega_{\kappa} \equiv (e(\kappa_a)-e(\kappa_b))/(\kappa_b-\kappa_a)$.
As the reduced superpotentials $q_\kappa(r)$ are simply linear functions of $r$, the central potential $V(r)$ reads
\begin{equation}\label{Eq:4.3.SUSY2VHO}
    V_{\rm HO}(r) = \frac{M}{2}\omega_{\kappa}^2 r^2+V(0)\,.
\end{equation}
Such a PSS limit is nothing but the U(3) symmetry of the Schr\"odinger Hamiltonian, which leads to the energy degeneracy of the whole major shell as discussed in the previous Subsection.

\subsubsection{Perturbative nature of PSS}

In this Subsection, we will employ some numerical results to show explicitly the perturbative nature of PSS breaking in realistic nuclei.
For that, the mass of nucleon takes $M=939.0$~MeV, and the central potential $V(r)$ adopts the Woods-Saxon form
\begin{equation}\label{Eq:4.3.SUSY2WS}
    V(r) = \frac{V_0}{1+e^{(r-R)/a}}\,,
\end{equation}
with the parameters $V_0=-63.297$~MeV, $R=6.278$~fm, and $a=0.615$~fm, which correspond to the neutron mean-field potential provided in Ref.~\cite{Koepf1991_ZPA339-81} by taking $N=82$ and $Z=50$.
This potential is illustrated as the solid line in Fig.~\ref{Fig:4.3.SUSY2V} shown below.
Note that in this Subsection we use a tilde to denote the operators, potentials, and wave functions belonging to $\tilde{H}$.

\begin{figure}[tb]
\begin{center}
  \includegraphics[width=8cm]{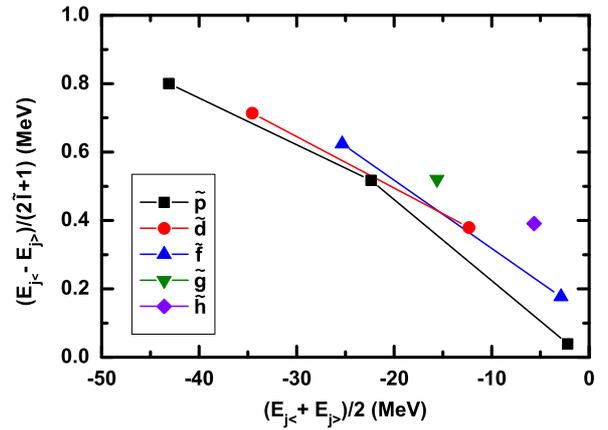}
\end{center}
\caption{(Color online) Reduced pseudospin-orbit splitting $(E_{j_<}-E_{j_>})/(2\tilde{l}+1)$ versus their average values of single-particle energy $(E_{j_<}+E_{j_>})/2$.
Taken from Ref.~\cite{Liang2013_PRC87-014334}.}
\label{Fig:4.3.SUSY2dE}
\end{figure}

In Fig.~\ref{Fig:4.3.SUSY2dE}, we plot the reduced PSO splitting $(E_{j_<}-E_{j_>})/(2\tilde{l}+1)$ versus their average  values of single-particle energy $E_{\rm av}=(E_{j_<}+E_{j_>})/2$, where $j_<$ and $j_>$ denote the states with $j=\tilde{l}-1/2$ and $j=\tilde{l}+1/2$, respectively.
It is found that the amplitudes of the reduced PSO splitting are smaller than $1$~MeV.
More importantly, in general the splitting becomes smaller with the increasing single-particle energies.
To investigate the physical mechanism for such energy-dependent behavior is helpful to figure out whether the PSS is an accidental symmetry.

\begin{figure}[tb]
\begin{center}
  \includegraphics[width=8cm]{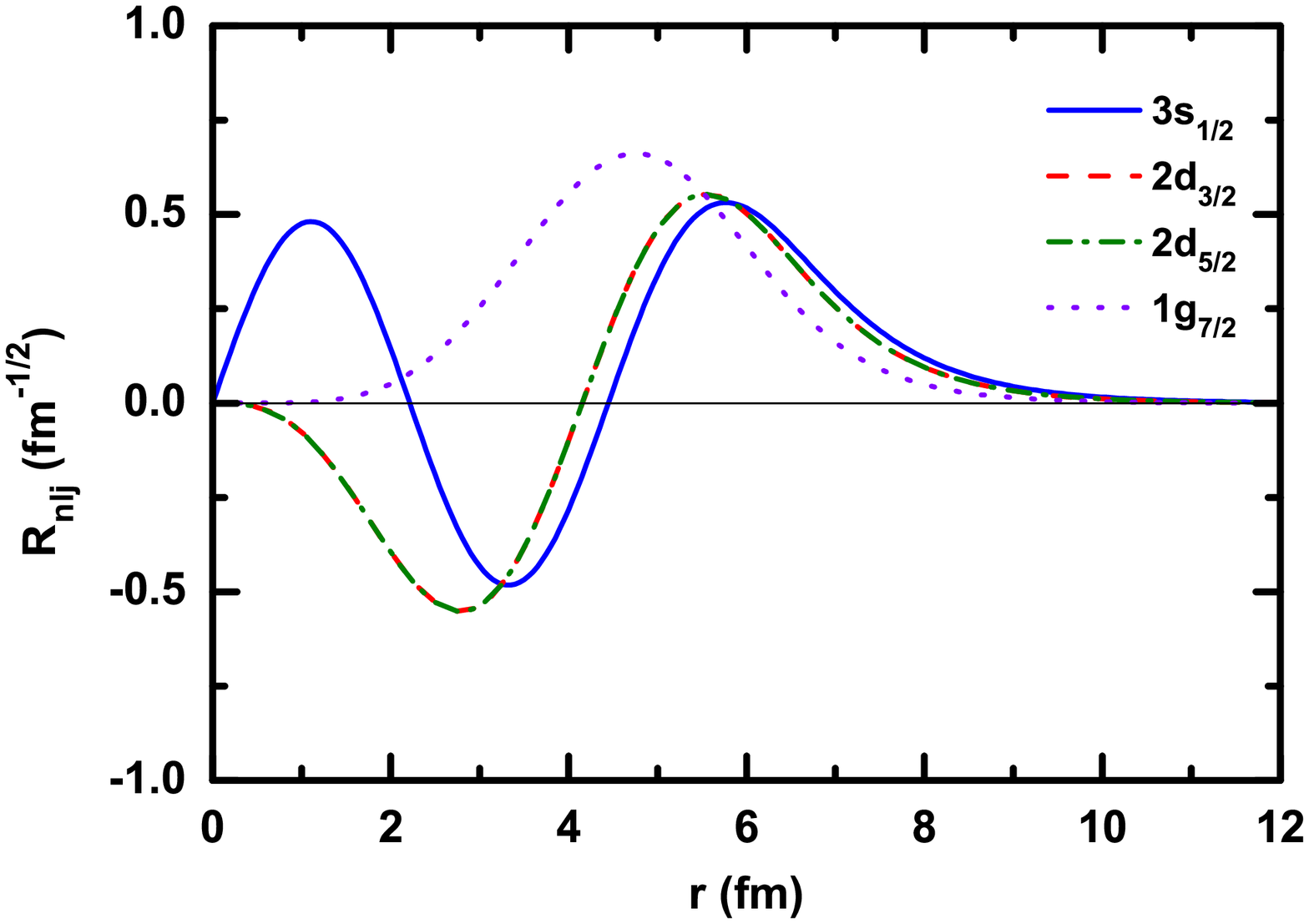}\\
  \includegraphics[width=8cm]{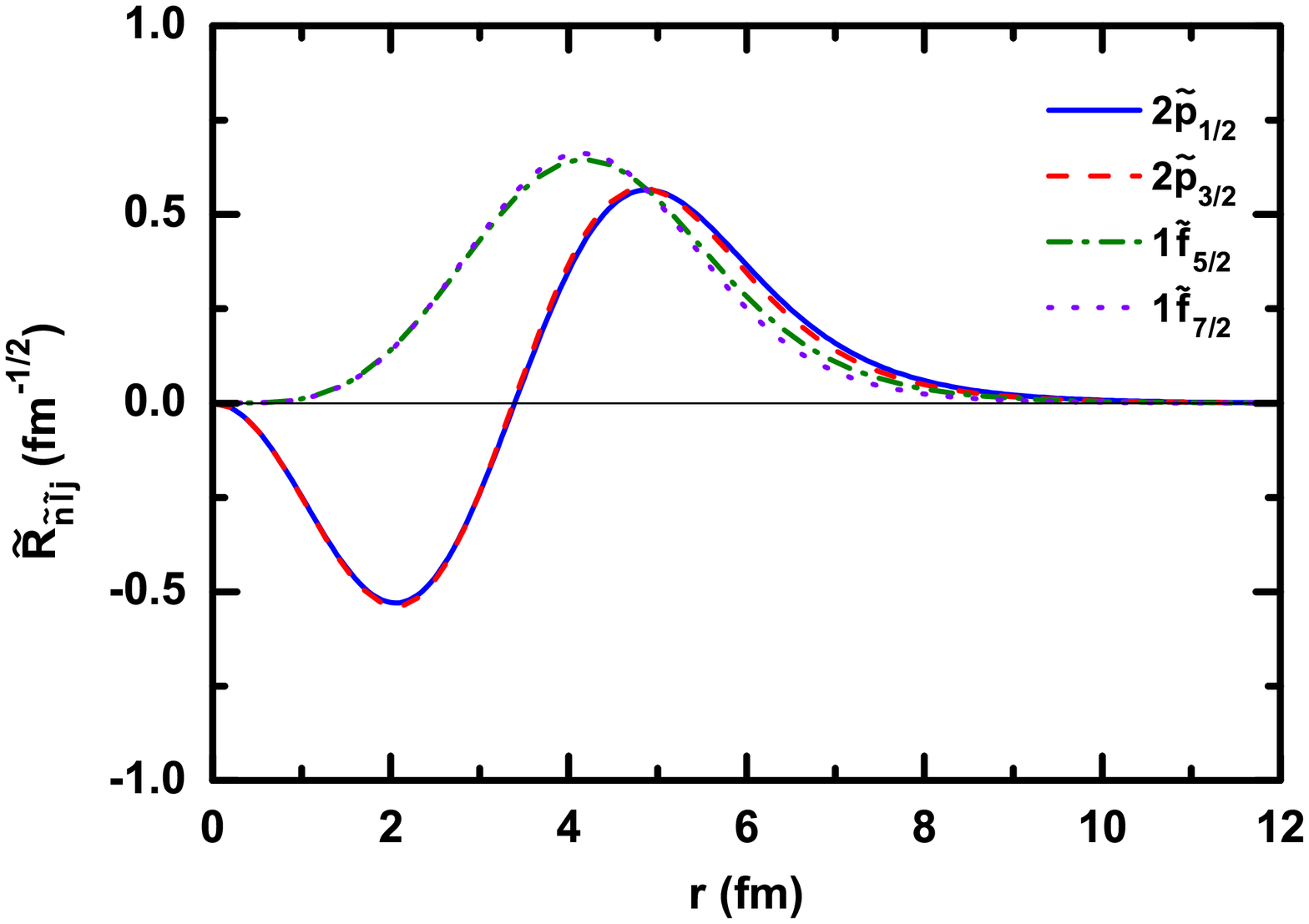}
\end{center}
\caption{(Color online) Single-particle radial wave functions $R_{nlj}(r)$ of $H$ (the upper panel) and $\tilde{R}_{\tilde{n}\tilde{l}j}(r)$ of $\tilde{H}$ (the lower panel) for the $3s_{1/2}$, $2d_{3/2}$, $2d_{5/2}$, and $1g_{7/2}$ states.
Taken from Ref.~\cite{Liang2013_PRC87-014334}.}
\label{Fig:4.3.SUSY2WF}
\end{figure}

In the upper panel of Fig.~\ref{Fig:4.3.SUSY2WF}, the single-particle radial wave functions $R_{nlj}(r)$ of $H$ are shown by taking the $2\tilde{p}$ and $1\tilde{f}$ pseudospin doublets as examples.
Since there is no spin-orbit term in $H$, the wave functions of the spin doublets are identical.
In contrast, the wave functions of the pseudospin doublets are very different from each other, which makes it difficult to trace the origin of PSS and analyze its symmetry-breaking mechanism.

\begin{table}[tb]
\begin{center}
\caption{Contributions from the kinetic term (kin.), centrifugal barrier (CB), and central potential (cen.) to the single-particle energies $E$ and the corresponding PSO splitting $\Delta E_{\rm PSO}$ for the $2\tilde{p}$ and $1\tilde{f}$ pseudospin doublets.
All units are in MeV.
Data are taken from Ref.~\cite{Liang2013_PRC87-014334}.
\label{Tab:4.3.SUSY2dE1}}
\begin{tabular}{@{}lrrrr@{}} \hline\hline
State & \multicolumn{1}{c}{$E_{\rm kin.}$} & \multicolumn{1}{c}{$E_{\rm CB}$} & \multicolumn{1}{c}{$E_{\rm cen.}$} & \multicolumn{1}{c}{$E$} \\ \hline
            $3s_{1/2}$ & $28.953$ &   $0.000$ & $-50.545$ & $-21.591$ \\
            $2d_{3/2}$ & $16.845$ &  $11.758$ & $-51.746$ & $-23.143$ \\
  $\Delta E_{\rm PSO}$ & $12.109$ & $-11.758$ &   $1.201$ &   $1.552$ \\ \hline
            $2d_{5/2}$ & $16.845$ &  $11.758$ & $-51.746$ & $-23.143$ \\
            $1g_{7/2}$ &  $6.197$ &  $20.483$ & $-54.188$ & $-27.508$ \\
  $\Delta E_{\rm PSO}$ & $10.648$ &  $-8.725$ &   $2.442$ &   $4.365$ \\ \hline\hline
\end{tabular}
\end{center}
\end{table}

Before the quantitative analysis done in Ref.~\cite{Liang2011_PRC83-041301R} by using the perturbation theory, investigations of the pseudospin-orbit splitting were usually done by decomposing the contributions term by term.
Contribution from each term is calculated as
\begin{equation}\label{Eq:4.3.SUSY2Ei}
    E_i = \int R^*(r)\hat O_i R(r) dr\,.
\end{equation}
with the corresponding operator $\hat O_i$.
In the representation of $H$ in Eq.~(\ref{Eq:4.3.SUSY2HSchr}), the operators of the kinetic term, centrifugal barrier, and central potential read $-d^2/(2Mdr^2)$, $\kappa(\kappa+1)/(2Mr^2)$, and $V(r)$, respectively.
Their contributions to the single-particle energies $E$ as well as the corresponding PSO splitting $\Delta E_{\rm PSO}$ are shown in Table~\ref{Tab:4.3.SUSY2dE1} for the $2\tilde{p}$ and $1\tilde{f}$ pseudospin doublets.
It is not surprising that, in this representation, the contributions to $\Delta E_{\rm PSO}$ come from all terms and they substantially cancel to each other in a sophisticated way.

The phenomenon of such strong cancellations among different terms was usually associated with the dynamical nature \cite{Alberto2001_PRL86-5015, Alberto2002_PRC65-034307} and even the non-perturbative nature \cite{Marcos2001_PLB513-30, Lisboa2010_PRC81-064324, Ginocchio2011_JPCS267-012037} of PSS.
However, such connection is sometimes misleading.
Indeed, as shown in Section~\ref{Sect:3.2}, the nature of PSS is perturbative from the U(3)-symmetry limit \cite{Liang2011_PRC83-041301R}.

By using the SUSY quantum mechanics, what is much more important here is that the origin of PSS and its symmetry-breaking mechanism can be studied explicitly in the representation of the SUSY partner Hamiltonian $\tilde{H}$ \cite{Liang2013_PRC87-014334}.

\begin{figure}[tb]
\begin{center}
  \includegraphics[width=8cm]{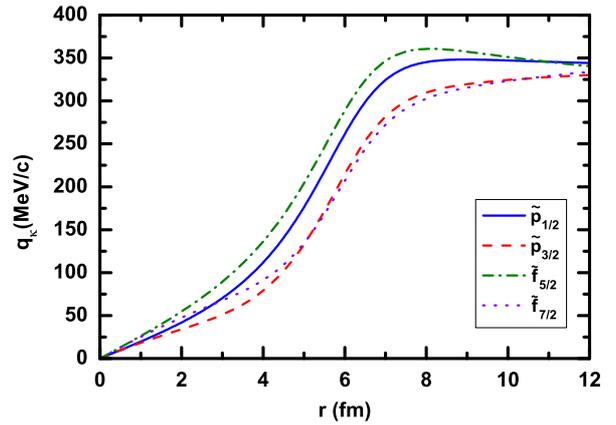}
\end{center}
\caption{(Color online) Reduced superpotentials $q_\kappa(r)$ for the $\tilde{p}$ and $\tilde{f}$ states.
Taken from Ref.~\cite{Liang2013_PRC87-014334}.}
\label{Fig:4.3.SUSY2q}
\end{figure}

Let us start with the reduced superpotentials $q_\kappa(r)$ by solving the first-order differential equation~(\ref{Eq:4.3.SUSY2qeq}) with the boundary condition $q_\kappa(0)=0$.
The $\kappa$-dependent $q_\kappa(r)$ are shown in Fig.~\ref{Fig:4.3.SUSY2q} in the unit of MeV/$c$.
Note that, although they depend on $\kappa$, the reduced superpotentials $q_\kappa(r)$ do not depend on the main quantum number $n$ for a given $\kappa$.
That is essential for understanding the general pattern of $\Delta E_{\rm PSO}$ versus $E_{\rm av}$ as shown in Fig.~\ref{Fig:4.3.SUSY2dE}.

\begin{figure}[tb]
\begin{center}
  \includegraphics[width=8cm]{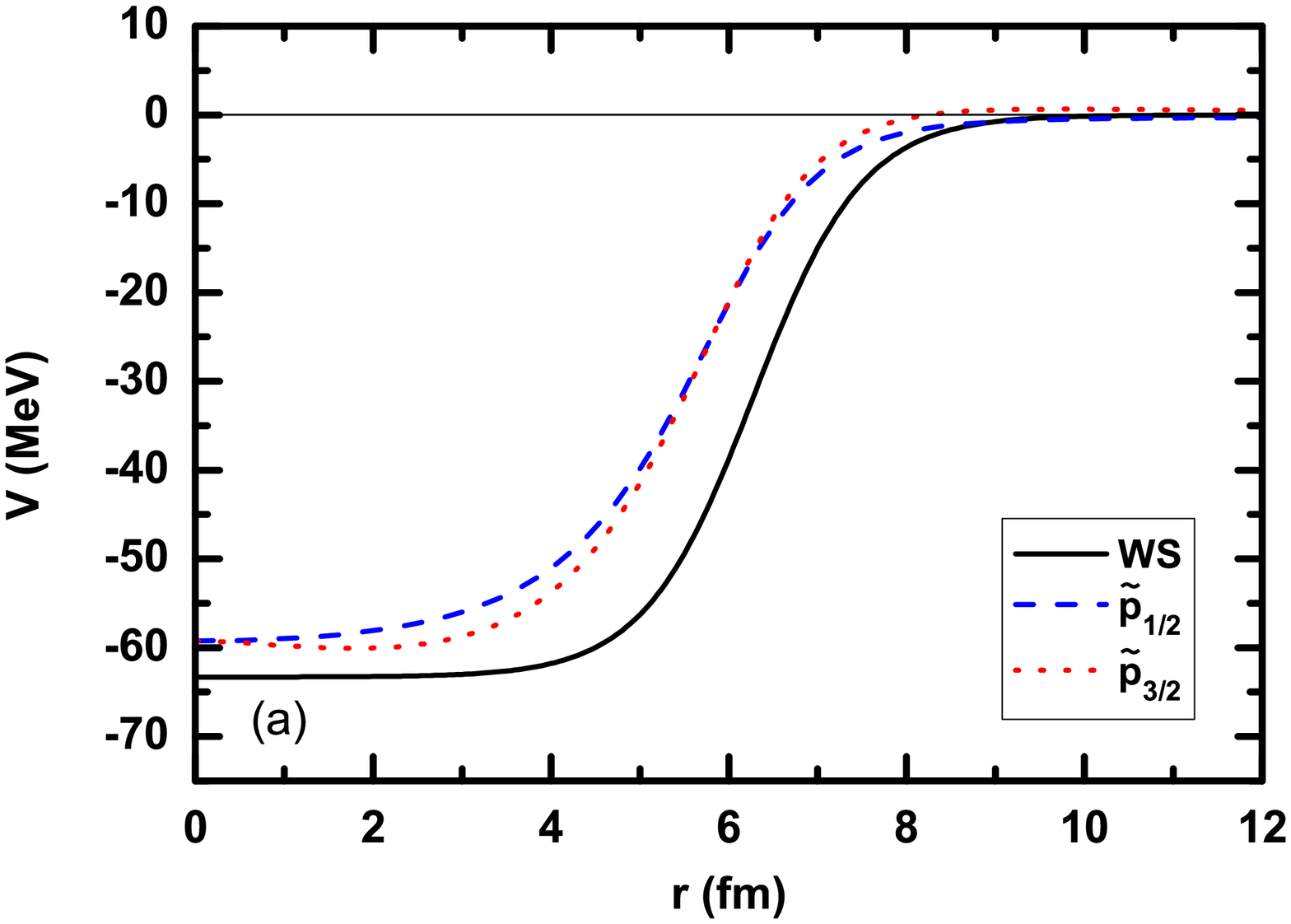}\\
  \includegraphics[width=8cm]{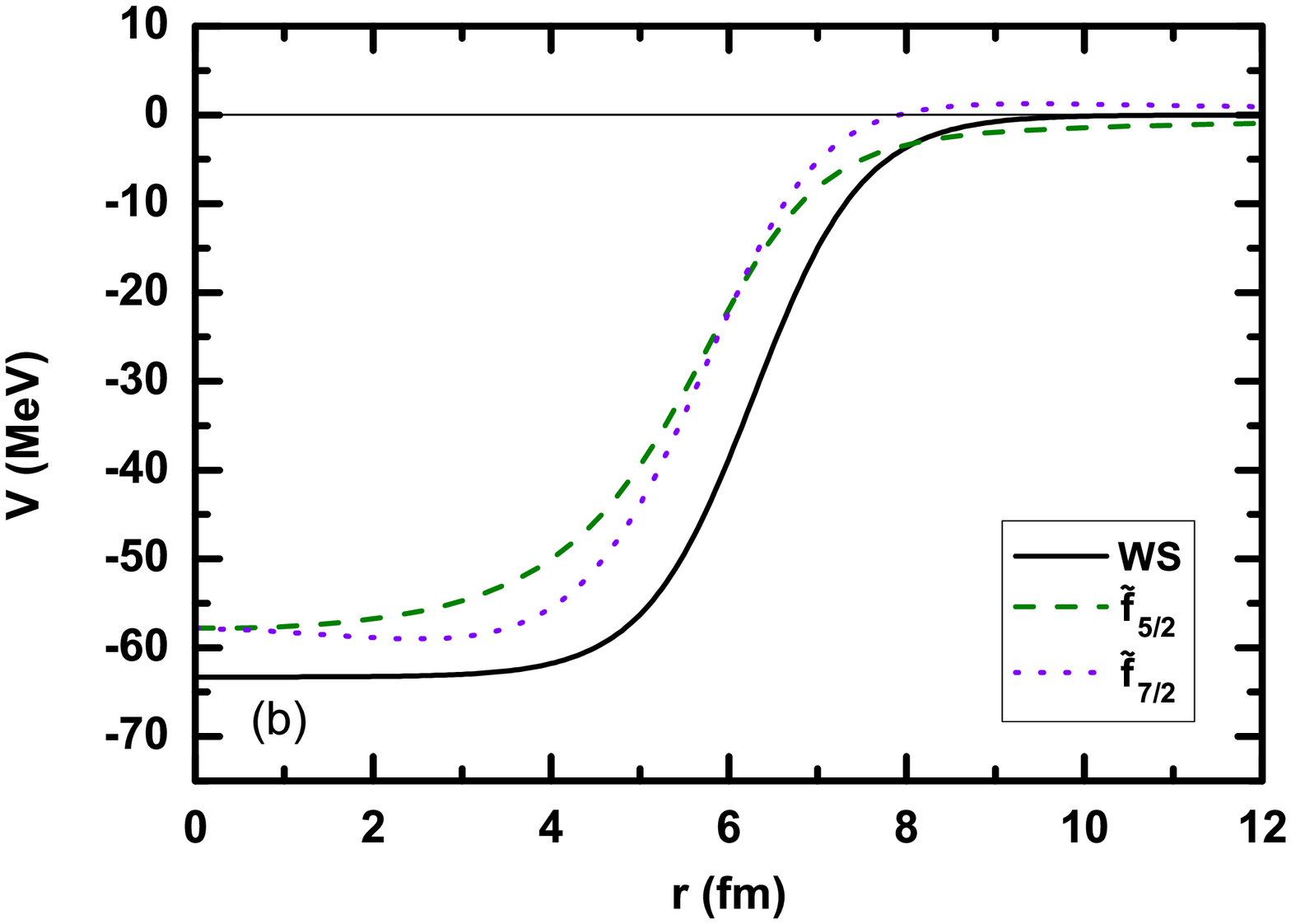}
\end{center}
\caption{(Color online) $\kappa$-dependent central potentials $\tilde{V}_\kappa(r)$ in $\tilde{H}$ for the $\tilde{p}$ and $\tilde{f}$ states.
The Woods-Saxon potential in $H$ is shown for comparison.
Taken from Ref.~\cite{Liang2013_PRC87-014334}.}
\label{Fig:4.3.SUSY2V}
\end{figure}

The $\kappa$-dependent central potentials $\tilde V_\kappa(r)$ in $\tilde{H}$ can be then obtained, and their asymptotic behaviors satisfy
\begin{equation}\label{Eq:4.3.SUSY2Vasym0}
    \lim_{r\rightarrow\infty}\tilde{V}_{\kappa}(r) = 0
\end{equation}
and
\begin{equation}\label{Eq:4.3.SUSY2Vasym}
    \lim_{r\rightarrow0}\tilde{V}_{\kappa}(r)=V+\frac{2(e(\kappa)-V)}{(1-2\kappa)}\,.
\end{equation}
It is important that these potentials are regular and converge at both $r\rightarrow0$ and $r\rightarrow\infty$.
In Fig.~\ref{Fig:4.3.SUSY2V}, these central potentials $\tilde V_\kappa(r)$ are shown for the $\tilde{p}$ and $\tilde{f}$ states, while the Woods-Saxon potential $V(r)$ in $H$ is shown for comparison.
For all $\kappa$, the potentials $\tilde V_\kappa(r)$ approximately remain a Woods-Saxon shape, and they are shallower than the original potential $V(r)$.
By comparing the two panels, it is found that the amplitude of the difference between $\tilde V_\kappa(r)$ for a pair of pseudospin partners increases with the difference of their quantum numbers $|\kappa_a-\kappa_b|$.

\begin{figure}[tb]
\begin{center}
  \includegraphics[width=7cm]{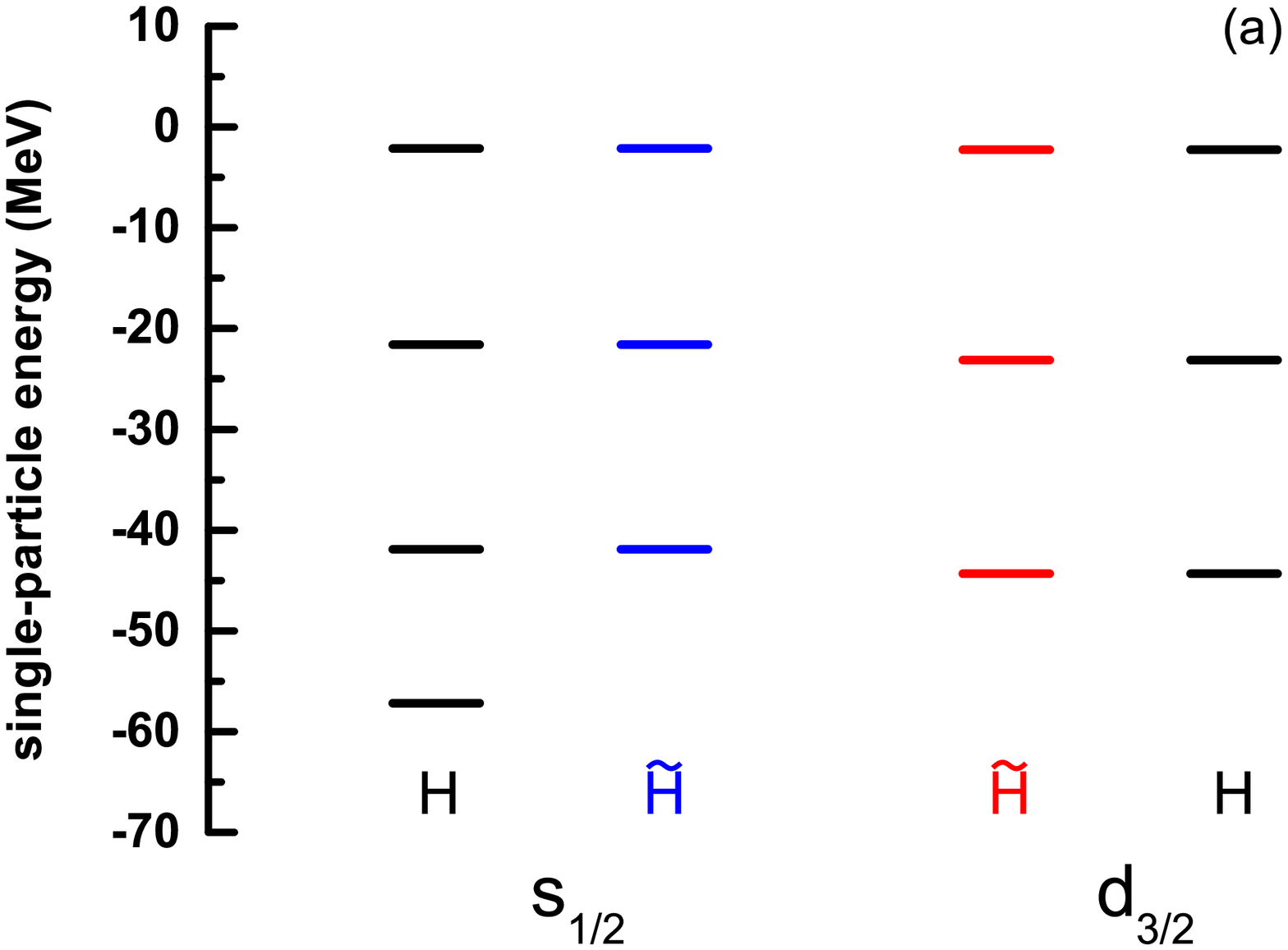}\\
  \includegraphics[width=7cm]{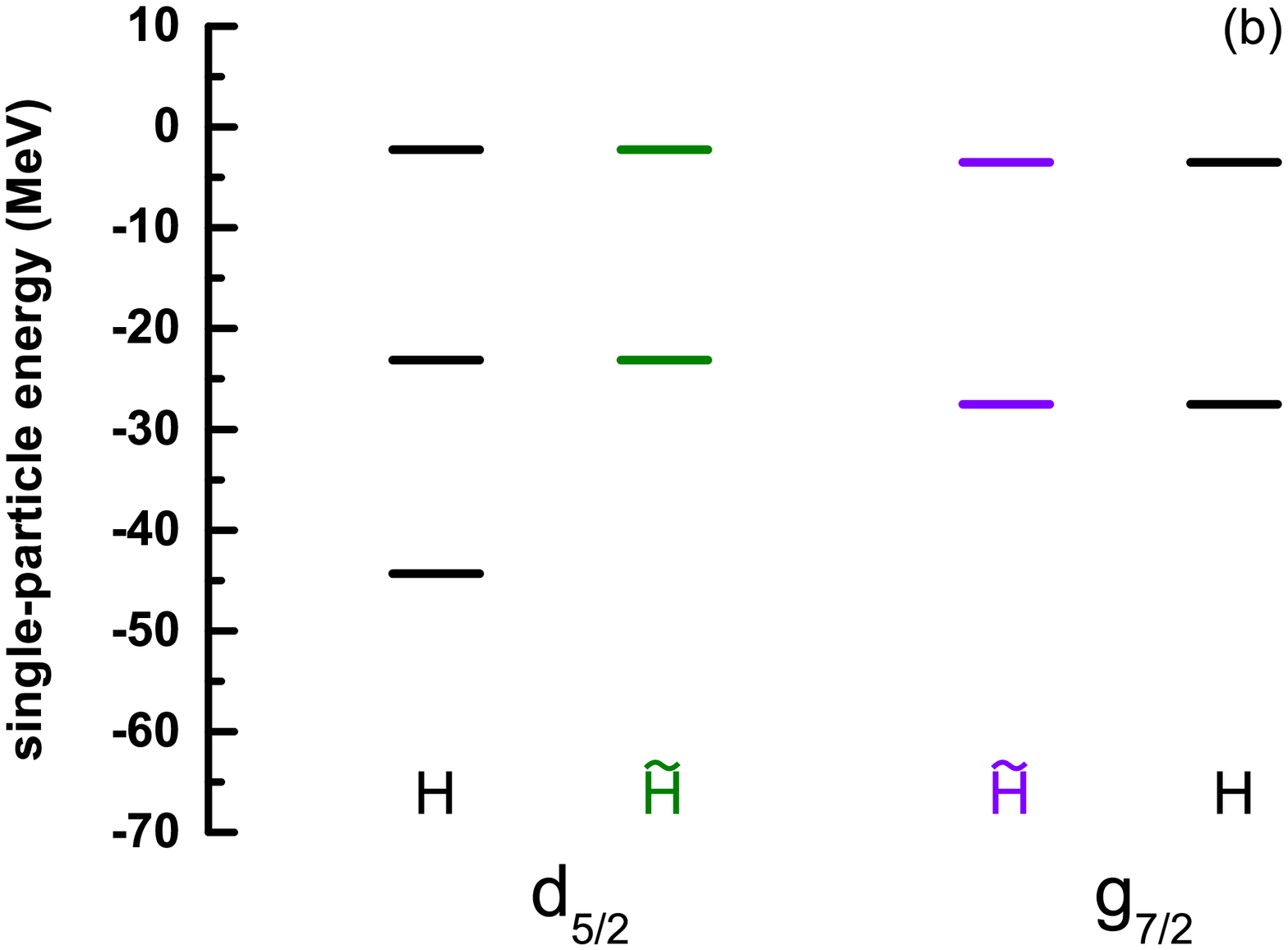}
\end{center}
\caption{(Color online) Single-particle energies obtained with $H$ and $\tilde{H}$ for the $\tilde{p}$ and $\tilde{f}$ states.
Taken from Ref.~\cite{Liang2013_PRC87-014334}.}
\label{Fig:4.3.SUSY2E}
\end{figure}

With the central potentials $\tilde V_\kappa(r)$, it is straightforward to calculate the single-particle energies and wave functions of the SUSY partner Hamiltonians $\tilde{H}(\kappa)$.
In Fig.~\ref{Fig:4.3.SUSY2E}, the energies of bound states obtained with $\tilde{H}$ are compared with those obtained with the original $H$.
It is seen explicitly that the eigenstates of $H$ and $\tilde{H}$ are identical, except for the lowest eigenstates with $\kappa<0$ in $H$, which are the so-called intruder states.
In other words, the fact that the intruder states have no pseudospin partners can be interpreted as a natural result of the exact SUSY for $\kappa<0$ and broken SUSY for $\kappa>0$ \cite{Typel2008_NPA806-156, Liang2013_PRC87-014334}.
By holding this one-to-one relation in the two sets of spectra, the origin of PSS, which is deeply hidden in $H$, can be now traced by using its SUSY partner Hamiltonian $\tilde{H}$.

The single-particle radial wave functions $\tilde{R}_{\tilde{n}\tilde{l}j}(r)$ of $\tilde{H}$ for the $2\tilde{p}$ and $1\tilde{f}$ pseudospin doublets are shown in the lower panel of Fig.~\ref{Fig:4.3.SUSY2WF}.
In contrast to that shown in the upper panel, it is found that, in the SUSY representation, the radial wave functions of pseudospin doublets are almost identical to each other.
Therefore, the quasi-degeneracy of pseudospin doublets is closely related to the similarity of their wave functions, and vice versa \cite{Liang2013_PRC87-014334}.

\begin{table}[tb]
\begin{center}
\caption{Contributions from the kinetic term (kin.), pseudo-centrifugal barrier (PCB), and central potential (cen.) to the single-particle energies $E$ and the corresponding PSO splitting $\Delta E_{\rm PSO}$ for the $2\tilde{p}$ and $1\tilde{f}$ pseudospin doublets.
All units are in MeV.
Data are taken from Ref.~\cite{Liang2013_PRC87-014334}.
\label{Tab:4.3.SUSY2dE2}}
\begin{tabular}{@{}lrrrr@{}} \hline\hline
State & \multicolumn{1}{c}{$E_{\rm kin.}$} & \multicolumn{1}{c}{$E_{\rm PCB}$} & \multicolumn{1}{c}{$E_{\rm cen.}$} & \multicolumn{1}{c}{$E$} \\ \hline
    $2\tilde{p}_{1/2}$ & $16.602$ &  $6.723$ & $-44.916$ & $-21.591$ \\
    $2\tilde{p}_{3/2}$ & $17.331$ &  $6.857$ & $-47.332$ & $-23.143$ \\
  $\Delta E_{\rm PSO}$ & $-0.729$ & $-0.134$ &   $2.415$ &   $1.552$ \\ \hline
    $1\tilde{f}_{5/2}$ &  $5.710$ & $16.286$ & $-45.139$ & $-23.143$ \\
    $1\tilde{f}_{7/2}$ &  $6.293$ & $16.591$ & $-50.392$ & $-27.508$ \\
  $\Delta E_{\rm PSO}$ & $-0.584$ & $-0.305$ &   $5.253$ &   $4.365$ \\ \hline\hline
\end{tabular}
\end{center}
\end{table}

The same strategy as done in Table~\ref{Tab:4.3.SUSY2dE1} is then used to investigate the PSO splitting, but now in the SUSY representation of $\tilde{H}$ shown in Eq.~(\ref{Eq:4.3.SUSY2Htil}).
The corresponding operators include the kinetic term $-d^2/(2Mdr^2)$, the PCB $\kappa(\kappa-1)/(2Mr^2)$, and the central potential $\tilde{V}_\kappa(r)$.
The results for the $2\tilde{p}$ and $1\tilde{f}$ pseudospin doublets are listed in Table~\ref{Tab:4.3.SUSY2dE2}.
It is seen that for each pair of pseudospin doublets the energy contributions from the PSS-conserving terms, i.e., the kinetic and PCB, are very similar.
The PSO splitting $\Delta E_{\rm PSO}$ is mainly contributed by the difference in the central potentials $\Delta E_{\rm cen}$, which is due to the slight $\kappa$-dependence of $\tilde{V}_\kappa(r)$ as shown in Fig.~\ref{Fig:4.3.SUSY2V}.
In other words, the sophisticated cancellations among different terms in $H$ can be understood in a much clearer way by using a proper decomposition with the help of the SUSY quantum mechanics \cite{Liang2013_PRC87-014334}.

Finally, for the quantitative perturbation calculations, the Hamiltonian $\tilde{H}$ is split as
\begin{equation}\label{Eq:4.3.SUSY2HH0W}
  \tilde{H} = \tilde{H}^{\rm PSS}_0 + \tilde{W}^{\rm PSS}\,,
\end{equation}
where $\tilde{H}^{\rm PSS}_0$ and $\tilde{W}^{\rm PSS}$ are the corresponding PSS-conserving and PSS-breaking terms, respectively.
By assuming $\tilde{W}^{\rm PSS}$ proportional to $\kappa$, which is similar to the case of spin-orbit term in the conventional scheme, one has
\begin{subequations}
\begin{align}\label{Eq:4.3.SUSY2H0W}
    \tilde{H}^{\rm PSS}_0&=\frac{1}{2M}\ls-\frac{d^2}{dr^2}+\frac{\kappa(\kappa-1)}{r^2}\rs + \tilde{V}_{\rm PSS}(r)\,,\\
    \tilde{W}^{\rm PSS}&=\kappa \tilde{V}_{\rm PSO}(r)\,.
\end{align}
\end{subequations}
The PSS-conserving $\tilde{V}_{\rm PSS}(r)$ and PSS-breaking $\tilde{V}_{\rm PSO}(r)$ potentials are then determined as \cite{Liang2013_PRC87-014334}
\begin{equation}\label{Eq:4.3.SUSY2VPSS}
  \tilde{V}_{\rm PSS}(r) = \frac{\kappa_b\tilde{V}_{\kappa_a}(r) - \kappa_a\tilde{V}_{\kappa_b}(r)}{\kappa_a-\kappa_b}
\end{equation}
and
\begin{equation}\label{Eq:4.3.SUSY2VPSO}
  \tilde{V}_{\rm PSO}(r) =\frac{1}{M}\frac{q'_{\kappa_a}(r) - q'_{\kappa_b}(r)}{\kappa_a-\kappa_b}\,.
\end{equation}

\begin{figure}[tb]
\begin{center}
  \includegraphics[width=8cm]{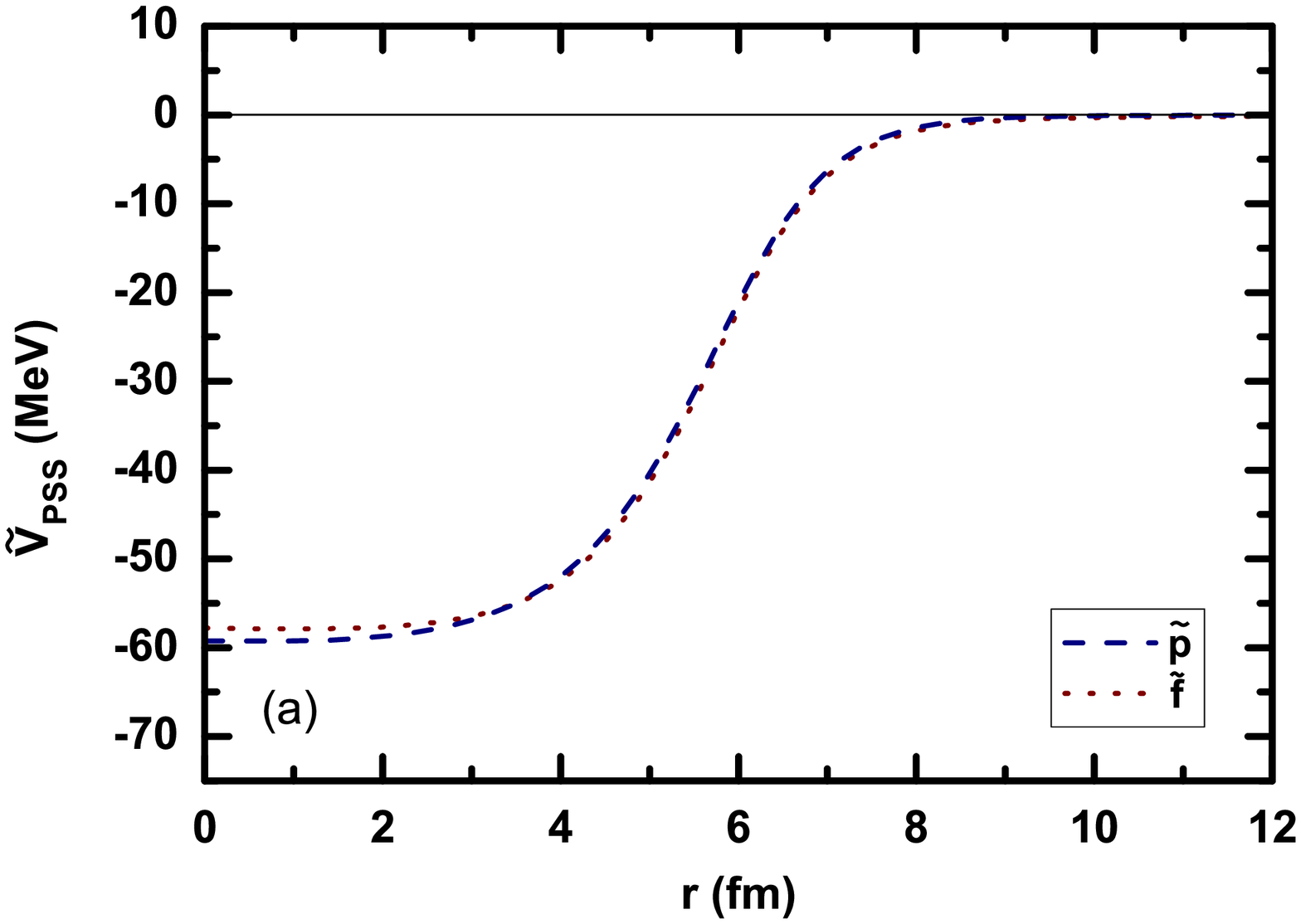}\\
  \includegraphics[width=8cm]{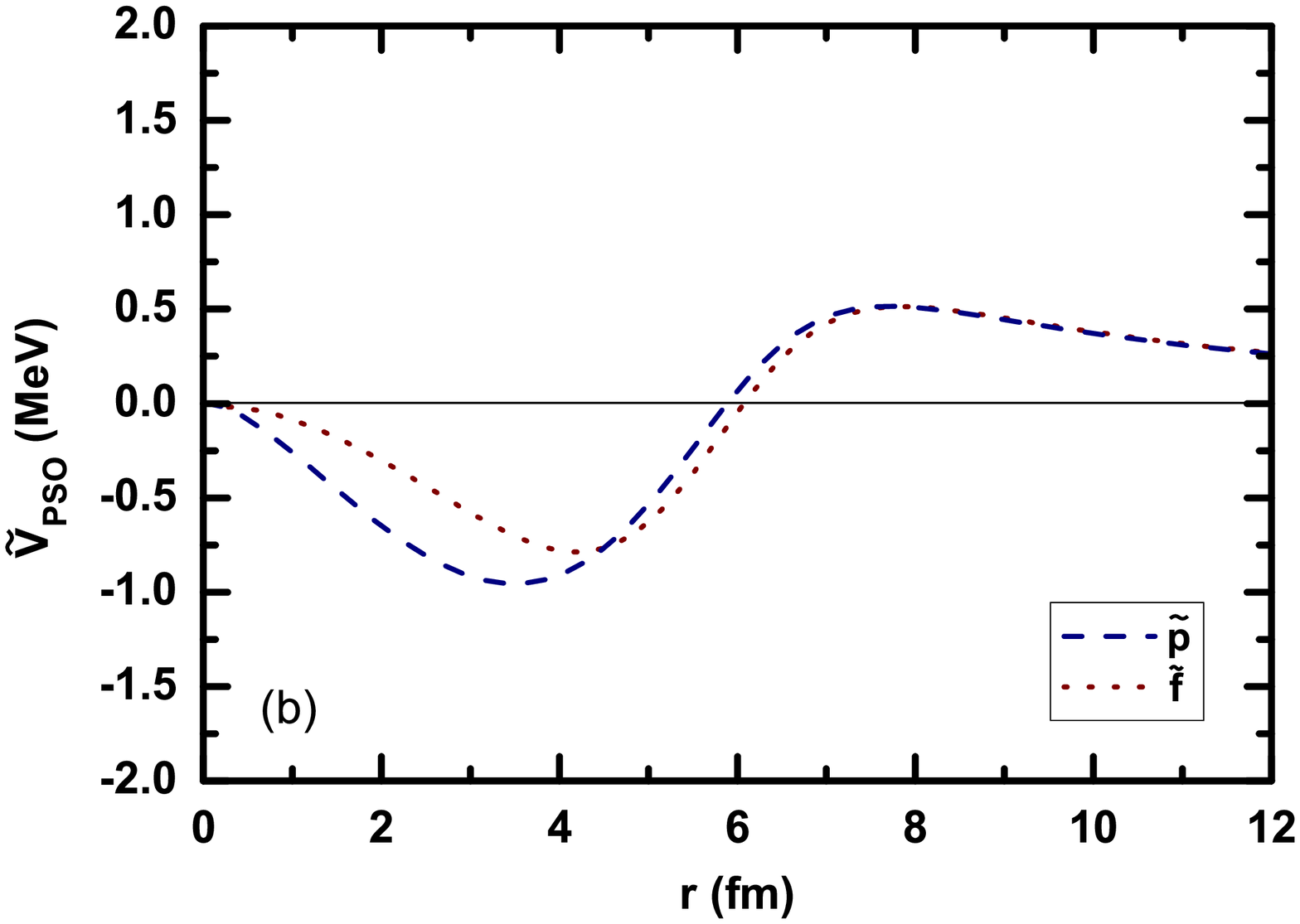}
\end{center}
\caption{(Color online) PSS-conserving potentials $\tilde{V}_{\rm PSS}(r)$ (the upper panel) and PSS-breaking potentials $\tilde{V}_{\rm PSO}(r)$ (the lower panel) for the $\tilde{p}$ and $\tilde{f}$ states.
Taken from Ref.~\cite{Liang2013_PRC87-014334}.}
\label{Fig:4.3.SUSY2H0W}
\end{figure}

In Fig.~\ref{Fig:4.3.SUSY2H0W}, the $\tilde{V}_{\rm PSS}(r)$ and $\tilde{V}_{\rm PSO}(r)$ potentials are shown by taking the $\tilde{p}$ and $\tilde{f}$ states as examples.
It can be seen that the PSS-conserving potentials $\tilde{V}_{\rm PSS}(r)$ remain an approximate Woods-Saxon shape, and they are $\kappa$-dependent to a small extent.
The PSS-breaking potentials $\tilde{V}_{\rm PSO}(r)$ show several special features:
(i) The PSS-breaking potentials are regular functions of $r$, in particular, they vanish at $r\rightarrow\infty$.
(ii) The amplitudes of $\tilde{V}_{\rm PSO}$ are around $1$~MeV that are relevant to the amplitudes of the reduced PSO splitting, e.g., $\Delta E_{\rm PSO}/(2\tilde l+1)\lesssim1$~MeV as shown in Fig.~\ref{Fig:4.3.SUSY2dE}.
(iii) More importantly, the PSO potentials $\tilde{V}_{\rm PSO}(r)$ change from negative to positive with a node at the surface region, which is totally different from the usual spin-orbit potentials with a surface-peaked shape.
Such a particular shape explains well the behavior that the PSO splitting decreases with the single-particle energy increases, see Ref.~\cite{Liang2013_PRC87-014334} for details.

\begin{figure}[tb]
\begin{center}
  \includegraphics[width=8cm]{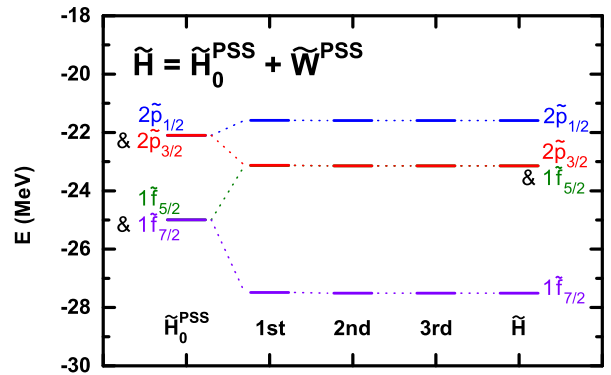}
\end{center}
\caption{(Color online) Single-particle energies obtained at the exact PSS limit $\tilde H^{\rm PSS}_0$, and their counterparts obtained by the first-, second-, and third-order perturbation calculations with $\tilde W^{\rm PSS}$, as well as those obtained with $\tilde H$.
Taken from Ref.~\cite{Liang2013_PRC87-014334}.}
\label{Fig:4.3.SUSY2PT}
\end{figure}

The perturbation calculations are performed based on the pseudospin-symmetric Hamiltonian $\tilde{H}^{\rm PSS}_0$ with the perturbation $\tilde{W}^{\rm PSS}$.
For the present decomposition, the largest perturbation correction $|\tilde{W}^{\rm PSS}_{mk}/(E_m-E_k)|$ is less than $0.03$ \cite{Liang2013_PRC87-014334}, which indicates that the criterion in Eq.~(\ref{Eq:4.1.PTcondition}) is satisfied to an excellent level.
It can be seen in Fig.~\ref{Fig:4.3.SUSY2PT} that the pseudospin doublets are exactly degenerate at the PSS limit $\tilde{H}^{\rm PSS}_0$, and the PSO splitting is excellently reproduced by the first-order perturbation calculations.

In such an explicit and quantitative way, the PSO splitting $\Delta E_{\rm PSO}$ can be directly understood by the PSS-breaking term $\tilde{W}^{\rm PSS}$ in the representation of the SUSY partner Hamiltonian $\tilde{H}$.
Furthermore, this symmetry-breaking term can be treated as a very small perturbation on the exact PSS limit $\tilde{H}^{\rm PSS}_0$.
This strongly confirms the pertubative nature of PSS \cite{Liang2013_PRC87-014334}.

Recently, it has been demonstrated in Ref.~\cite{Shen2013_PRC88-024311} that the perturbative nature of PSS maintains even when a substantial spin-orbit potential presents.

\section{Summary and Open Questions}\label{Sect:4}

In this Review, we mainly focus on the latest progress in the supersymmetric representation of pseudospin symmetry.
One of the key targets is to understand the origin of pseudospin symmetry and its symmetry-breaking mechanism in realistic nuclei in a quantitative and perturbative way.

It has been shown that, for the spherical case, the SU(2) spin symmetry and pseudospin symmetry of the Dirac Hamiltonian can be derived by using the supersymmetric quantum mechanics with the intertwining relation \cite{Leviatan2004_PRL92-202501}.
The spin-orbit splitting in realistic nuclei can be understood as a small perturbation around the SU(2) spin-symmetry limit.
Nevertheless, although the pseudospin-orbit splitting is in general smaller than the spin-orbit splitting in realistic nuclei, the pseudospin-orbit splitting behaves non-perturbatively if the Dirac Hamiltonian with the SU(2) pseudospin symmetry is regarded as its symmetry limit \cite{Liang2011_PRC83-041301R}.

Alternatively, the Dirac Hamiltonian holds another kind of symmetry---the U(3) symmetry---in which the single-particle energies of pseudospin doublets are also exactly degenerate \cite{Ginocchio2005_PRL95-252501}.
Within the scheme of perturbation theory, it has been proved that the pseudospin-orbit splitting in realistic nuclei can be understood as a result of small perturbation around such a U(3)-symmetry limit \cite{Liang2011_PRC83-041301R}.

Works are in progress for discovering the supersymmetric representation of this U(3) symmetry, although the complete answer is yet to be found.
One attempt is the supersymmetric quantum mechanics for the Schr\"odinger-like equation for the upper component of Dirac spinor \cite{Typel2008_NPA806-156}, and another is the supersymmetric quantum mechanics for the Dirac equation with the similarity renormalization group \cite{Guo2012_PRC85-021302R} evolution \cite{Liang2013_PRC87-014334, Shen2013_PRC88-024311}.
In particular, in the latter scheme, the origin of pseudospin symmetry and its symmetry-breaking mechanism, which are deeply hidden in the origin Hamiltonian, can be traced by using its supersymmetric partner Hamiltonian \cite{Liang2013_PRC87-014334, Shen2013_PRC88-024311}.

The pseudospin symmetry in deformed nuclei remains an important and open question.
As pointed out in the pioneering works by Bohr, Hamamoto, and Mottelson \cite{Bohr1982_PS26-267, Mottelson1990_NPA520-711c, Mottelson1991_NPA522-1c}, the origin of pseudospin symmetry is important to understand the nuclear (super)deformation and various nuclear rotation phenomena.

For the axially deformed case, the SU(2) spin and pseudospin symmetries of the Dirac Hamiltonian and their generators have been shown \cite{Ginocchio2004_PRC69-034303, Ginocchio2005_PR414-165}.
The supersymmetric representation of these SU(2) symmetries was discussed in Ref.~\cite{Leviatan2009_PRL103-042502}, together with additional symmetries when the scalar and vector potentials depend on different variables.
Nevertheless, the U(3) symmetry of the Dirac Hamiltonian was not included in the above studies.
At this symmetry limit the pseudospin-orbit splitting in realistic nuclei would be understood perturbatively and quantitatively.
For that, one of the possible ways is to investigate the Dirac equation with the similarity renormalization group evolution, which has been done in Ref.~\cite{Guo2014_PRL112-062502} for the deformed systems.
The supersymmetric representation that follows is probably nontrivial, because it involves multidimensional supersymmetric quantum mechanics \cite{Andrianov1984_TMP61-965, Ioffe2006_AP321-2552}.
Therefore, further progress along this direction is expected in the future.


\begin{ack}
I would like to express my gratitude to all the collaborators and colleagues who contributed to the investigations presented here, in particular to A. Arima, J.N. Ginocchio, J.Y. Guo, A. Leviatan, F.Q. Li, W.H. Long, J. Meng, P. Ring, S.H. Shen, S. Typel, N. Van Giai, S.Q. Zhang, Y. Zhang, P.W. Zhao, and S.G. Zhou.
I appreciate J. Meng and S.G. Zhou for the careful reading of the manuscript and the valuable suggestions.
This work was partly supported by the RIKEN iTHES Project.

\end{ack}


\bibliographystyle{iopart-num}
\bibliography{ref}

\providecommand{\newblock}{}
\begin{thebibliography}{100}
\expandafter\ifx\csname url\endcsname\relax
  \def\url#1{{\tt #1}}\fi
\expandafter\ifx\csname urlprefix\endcsname\relax\def\urlprefix{URL }\fi
\providecommand{\eprint}[2][]{\url{#2}}

\bibitem{Bohr1982_PS26-267}
Bohr A, Hamamoto I and Mottelson B~R 1982 {\em Phys. Scr.\/} {\bf 26} 267--272

\bibitem{Mottelson1990_NPA520-711c}
Mottelson B~R 1990 {\em Nucl. Phys. A\/} {\bf 520} 711c--722c

\bibitem{Mottelson1991_NPA522-1c}
Mottelson B~R 1991 {\em Nucl. Phys. A\/} {\bf 522} 1c--12c

\bibitem{Haxel1949_PR075-1766}
Haxel O, Jensen J~H~D and Suess H~E 1949 {\em Phys. Rev.\/} {\bf 75} 1766--1766

\bibitem{Mayer1949_PR075-1969}
Goeppert-Mayer M 1949 {\em Phys. Rev.\/} {\bf 75} 1969--1970

\bibitem{Nilsson1955_DMFM29-16}
Nilsson S~G 1955 {\em Dan. Mat. Fys. Medd.\/} {\bf 29} 16

\bibitem{Nilsson1969_NPA131-1}
Nilsson S~G, Tsang C~F, Sobiczewski A, Szymanski Z, Wycech S, Gustafson C, Lamm
  I~L, M\"oller P and Nilsson B 1969 {\em Nucl. Phys. A\/} {\bf 131} 1--66

\bibitem{Hecht1969_NPA137-129}
Hecht K~T and Adler A 1969 {\em Nucl. Phys. A\/} {\bf 137} 129--143

\bibitem{Arima1969_PLB30-517}
Arima A, Harvey M and Shimizu K 1969 {\em Phys. Lett. B\/} {\bf 30} 517--522

\bibitem{Liang2015_PR570-1}
Liang H, Meng J and Zhou S~G 2015 {\em Phys. Rep.\/} {\bf 570} 1--84

\bibitem{Ratna-Raju1973_NPA202-433}
Ratna~Raju R~D, Draayer J~P and Hecht K~T 1973 {\em Nucl. Phys. A\/} {\bf 202}
  433--466

\bibitem{Voigt1983_RMP55-949}
de~Voigt M~J~A, Dudek J and Szyma\'{n}ski Z 1983 {\em Rev. Mod. Phys.\/} {\bf
  55} 949--1046

\bibitem{Draayer1984_AP156-41}
Draayer J~P and Weeks K~J 1984 {\em Ann. Phys. (NY)\/} {\bf 156} 41--67

\bibitem{Blokhin1997_NPA612-163}
Blokhin A~L, Beuschel T, Draayer J~P and Bahri C 1997 {\em Nucl. Phys. A\/}
  {\bf 612} 163--203

\bibitem{Beuschel1997_NPA619-119}
Beuschel T, Blokhin A~L and Draayer J~P 1997 {\em Nucl. Phys. A\/} {\bf 619}
  119--128

\bibitem{Rosensteel1976AP96-1}
Rosensteel G and Rowe D~J 1976 {\em Ann. Phys. (NY)\/} {\bf 96} 1--42

\bibitem{Rowe1985RPP48-1419}
Rowe D~J 1985 {\em Rep. Prog. Phys.\/} {\bf 48} 1419--1480

\bibitem{Troltenier1994_NPA576-351}
Troltenier D, Draayer J~P, Hess P~O and Casta\~{n}os O 1994 {\em Nucl. Phys.
  A\/} {\bf 576} 351--386

\bibitem{Troltenier1995_NPA586-53}
Troltenier D, Bahri C and Draayer J~P 1995 {\em Nucl. Phys. A\/} {\bf 586}
  53--72

\bibitem{Iachello1981AP136-19}
Iachello F and Kuyucak S 1981 {\em Ann. Phys. (NY)\/} {\bf 136} 19--61

\bibitem{Dudek1987_PRL59-1405}
Dudek J, Nazarewicz W, Szymanski Z and Leander G~A 1987 {\em Phys. Rev.
  Lett.\/} {\bf 59} 1405--1408

\bibitem{Bahri1992_PRL68-2133}
Bahri C, Draayer J~P and Moszkowski S~A 1992 {\em Phys. Rev. Lett.\/} {\bf 68}
  2133--2136

\bibitem{Dudek1992_PPNP28-131}
Dudek J 1992 {\em Prog. Part. Nucl. Phys.\/} {\bf 28} 131--185

\bibitem{Molique2000_PRC61-044304}
Molique H, Dobaczewski J and Dudek J 2000 {\em Phys. Rev. C\/} {\bf 61} 044304

\bibitem{Dudek2005_APPB36-975}
Dudek J, Schunck N and Dubray N 2005 {\em Acta Phys. Pol. B\/} {\bf 36}
  975--1001

\bibitem{Byrski1990_PRL64-1650}
Byrski T {\em et~al.\/} 1990 {\em Phys. Rev. Lett.\/} {\bf 64} 1650--1653

\bibitem{Gelberg1990_JPG16-L143}
Gelberg A, von Brentano P and Casten R~F 1990 {\em J. Phys. G: Nucl. Part.
  Phys.\/} {\bf 16} L143--L148

\bibitem{Nazarewicz1990_PRL64-1654}
Nazarewicz W, Twin P~J, Fallon P and Garrett J~D 1990 {\em Phys. Rev. Lett.\/}
  {\bf 64} 1654--1657

\bibitem{Nazarewicz1990_NPA512-61}
Nazarewicz W, Riley M~A and Garrett J~D 1990 {\em Nucl. Phys. A\/} {\bf 512}
  61--96

\bibitem{Zeng1991_PRC44-R1745}
Zeng J~Y, Meng J, Wu C~S, Zhao E~G, Xing Z and Chen X~Q 1991 {\em Phys. Rev.
  C\/} {\bf 44} R1745--R1748

\bibitem{Stephens1990_PRL65-301}
Stephens F~S {\em et~al.\/} 1990 {\em Phys. Rev. Lett.\/} {\bf 65} 301--304

\bibitem{Xu2008_PRC78-064301}
Xu Q {\em et~al.\/} 2008 {\em Phys. Rev. C\/} {\bf 78} 064301

\bibitem{Hua2009_PRC80-034303}
Hua W {\em et~al.\/} 2009 {\em Phys. Rev. C\/} {\bf 80} 034303

\bibitem{Troltenier1994_NPA567-591}
Troltenier D, Nazarewicz W, Szymanski Z and Draayer J~P 1994 {\em Nucl. Phys.
  A\/} {\bf 567} 591--610

\bibitem{Ginocchio1999_PRC59-2487}
Ginocchio J~N 1999 {\em Phys. Rev. C\/} {\bf 59} 2487--2493

\bibitem{Neumann-Cosel2000_PRC62-014308}
von Neumann-Cosel P and Ginocchio J~N 2000 {\em Phys. Rev. C\/} {\bf 62} 014308

\bibitem{Jolos2012_PRC86-044320}
Jolos R~V, Shirikova N~Y and Sushkov A~V 2012 {\em Phys. Rev. C\/} {\bf 86}
  044320

\bibitem{Ginocchio1999_PRL82-4599}
Ginocchio J~N 1999 {\em Phys. Rev. Lett.\/} {\bf 82} 4599--4602

\bibitem{Leeb2000_PRC62-024602}
Leeb H and Wilmsen S 2000 {\em Phys. Rev. C\/} {\bf 62} 024602

\bibitem{Ginocchio2002_PRC65-054002}
Ginocchio J~N 2002 {\em Phys. Rev. C\/} {\bf 65} 054002

\bibitem{Leeb2004_PRC69-054608}
Leeb H and Sofianos S~A 2004 {\em Phys. Rev. C\/} {\bf 69} 054608

\bibitem{Long2010_PRC81-031302R}
Long W~H, Ring P, Meng J, Van~Giai N and Bertulani C~A 2010 {\em Phys. Rev.
  C\/} {\bf 81} 031302(R)

\bibitem{Jolos2007_PAN70-812}
Jolos R and Voronov V 2007 {\em Phys. At. Nucl.\/} {\bf 70} 812--817

\bibitem{Li2014_PLB732-169}
Li J~J, Long W~H, Margueron J and Van~Giai N 2014 {\em Phys. Lett. B\/} {\bf
  732} 169--173

\bibitem{Sorlin2008_PPNP61-602}
Sorlin O and Porquet M~G 2008 {\em Prog. Part. Nucl. Phys.\/} {\bf 61} 602--673

\bibitem{Wienholtz2013_Nature498-346}
Wienholtz F {\em et~al.\/} 2013 {\em Nature\/} {\bf 498} 346--349

\bibitem{Steppenbeck2013_Nature502-207}
Steppenbeck D {\em et~al.\/} 2013 {\em Nature\/} {\bf 502} 207--210

\bibitem{Gaudefroy2006_PRL97-092501}
Gaudefroy L {\em et~al.\/} 2006 {\em Phys. Rev. Lett.\/} {\bf 97} 092501

\bibitem{Bastin2007_PRL99-022503}
Bastin B {\em et~al.\/} 2007 {\em Phys. Rev. Lett.\/} {\bf 99} 022503

\bibitem{Tarpanov2008_PRC77-054316}
Tarpanov D, Liang H, Van~Giai N and Stoyanov C 2008 {\em Phys. Rev. C\/} {\bf
  77} 054316

\bibitem{Moreno-Torres2010_PRC81-064327}
Moreno-Torres M, Grasso M, Liang H, De~Donno V, Anguiano M and Van~Giai N 2010
  {\em Phys. Rev. C\/} {\bf 81} 064327

\bibitem{Nagai1981_PRL47-1259}
Nagai Y {\em et~al.\/} 1981 {\em Phys. Rev. Lett.\/} {\bf 47} 1259--1262

\bibitem{Long2007_PRC76-034314}
Long W, Sagawa H, Van~Giai N and Meng J 2007 {\em Phys. Rev. C\/} {\bf 76}
  034314

\bibitem{Long2009_PLB680-428}
Long W~H, Nakatsukasa T, Sagawa H, Meng J, Nakada H and Zhang Y 2009 {\em Phys.
  Lett. B\/} {\bf 680} 428--431

\bibitem{Jolos2001_PPN32-113}
Jolos R~V 2001 {\em Phys. Part. Nucl.\/} {\bf 32} 113--137

\bibitem{Castanos1992_PLB277-238}
Casta\~{n}os O, Moshinsky M and Quesne C 1992 {\em Phys. Lett. B\/} {\bf 277}
  238--242

\bibitem{Blokhin1995_PRL74-4149}
Blokhin A~L, Bahri C and Draayer J~P 1995 {\em Phys. Rev. Lett.\/} {\bf 74}
  4149--4152

\bibitem{Ring1996_PPNP37-193}
Ring P 1996 {\em Prog. Part. Nucl. Phys.\/} {\bf 37} 193--263

\bibitem{Vretenar2005_PR409-101}
Vretenar D, Afanasjev A~V, Lalazissis G~A and Ring P 2005 {\em Phys. Rep.\/}
  {\bf 409} 101--259

\bibitem{Meng2006_PPNP57-470}
Meng J, Toki H, Zhou S~G, Zhang S~Q, Long W~H and Geng L~S 2006 {\em Prog.
  Part. Nucl. Phys.\/} {\bf 57} 470--563

\bibitem{Niksic2011_PPNP66-519}
Nik\v{s}i\'{c} T, Vretenar D and Ring P 2011 {\em Prog. Part. Nucl. Phys.\/}
  {\bf 66} 519--548

\bibitem{Meng2011_PP31-199}
Meng J, Guo J~Y, Li J, Li Z~P, Liang H~Z, Long W~H, Niu Y~F, Niu Z~M, Yao J~M,
  Zhang Y, Zhao P~W and Zhou S~G 2011 {\em Prog. Phys.\/} {\bf 31} 199--336

\bibitem{Meng2013_FPC8-55}
Meng J, Peng J, Zhang S~Q and Zhao P~W 2013 {\em Front. Phys.\/} {\bf 8} 55--79

\bibitem{Meng2015_JPG42-093101}
Meng J and Zhou S~G 2015 {\em J. Phys. G: Nucl. Part. Phys.\/} {\bf 42} 093101

\bibitem{Meng2016}
Meng J (ed) 2016 {\em Relativistic Density Functional for Nuclear Structure\/}
  ({\em International Review of Nuclear Physics\/} vol~10) (World Scientific,
  Singapore)

\bibitem{Ginocchio1997_PRL78-436}
Ginocchio J~N 1997 {\em Phys. Rev. Lett.\/} {\bf 78} 436--439

\bibitem{Meng1998_PRC58-R628}
Meng J, Sugawara-Tanabe K, Yamaji S, Ring P and Arima A 1998 {\em Phys. Rev.
  C\/} {\bf 58} R628--R631

\bibitem{Meng1999_PRC59-154}
Meng J, Sugawara-Tanabe K, Yamaji S and Arima A 1999 {\em Phys. Rev. C\/} {\bf
  59} 154--163

\bibitem{Lalazissis1998_PRC58-R45}
Lalazissis G~A, Gambhir Y~K, Maharana J~P, Warke C~S and Ring P 1998 {\em Phys.
  Rev. C\/} {\bf 58} R45--R48

\bibitem{Sugawara-Tanabe1998_PRC58-R3065}
Sugawara-Tanabe K and Arima A 1998 {\em Phys. Rev. C\/} {\bf 58} R3065--R3068
  {Erratum:} \textit{ibid.} 60 (1999) 019901.

\bibitem{Sugawara-Tanabe2000_PRC62-054307}
Sugawara-Tanabe K, Yamaji S and Arima A 2000 {\em Phys. Rev. C\/} {\bf 62}
  054307

\bibitem{Sugawara-Tanabe2005_RMP55-277}
Sugawara-Tanabe K 2005 {\em Rep. Math. Phys.\/} {\bf 55} 277--286

\bibitem{Ginocchio1998_PRC57-1167}
Ginocchio J~N and Madland D~G 1998 {\em Phys. Rev. C\/} {\bf 57} 1167--1173

\bibitem{Ginocchio2002_PRC66-064312}
Ginocchio J~N 2002 {\em Phys. Rev. C\/} {\bf 66} 064312

\bibitem{Sugawara-Tanabe2002_PRC65-054313}
Sugawara-Tanabe K, Yamaji S and Arima A 2002 {\em Phys. Rev. C\/} {\bf 65}
  054313

\bibitem{Ginocchio2004_PRC69-034303}
Ginocchio J~N, Leviatan A, Meng J and Zhou S~G 2004 {\em Phys. Rev. C\/} {\bf
  69} 034303

\bibitem{Chen2003_CPL20-358}
Chen T~S, L\"{u} H~F, Meng J, Zhang S~Q and Zhou S~G 2003 {\em Chin. Phys.
  Lett.\/} {\bf 20} 358--361

\bibitem{Lisboa2003_PRC67-054305}
Lisboa R, Malheiro M and Alberto P 2003 {\em Phys. Rev. C\/} {\bf 67} 054305

\bibitem{Guo2003_CPL20-602}
Guo J~Y, Meng J and Xu F~X 2003 {\em Chin. Phys. Lett.\/} {\bf 20} 602--604

\bibitem{Berkdemir2006_NPA770-32}
Berkdemir C 2006 {\em Nucl. Phys. A\/} {\bf 770} 32--39 {Erratum:}
  \textit{ibid.} 821 (2009) 262.

\bibitem{Jia2007_EPJA34-41}
Jia C, Guo P, Diao Y, Yi L and Xie X 2007 {\em Eur. Phys. J. A\/} {\bf 34}
  41--48

\bibitem{Guo2005_PLA338-90}
Guo J~Y and Sheng Z~Q 2005 {\em Phys. Lett. A\/} {\bf 338} 90--96

\bibitem{Ginocchio2004_PRC69-034318}
Ginocchio J~N 2004 {\em Phys. Rev. C\/} {\bf 69} 034318

\bibitem{Asgarifar2013_PS87-025703}
Asgarifar S and Goudarzi H 2013 {\em Phys. Scr.\/} {\bf 87} 025703

\bibitem{Zhang2009_CEJP7-768}
Zhang M~C 2009 {\em Cent. Eur. J. Phys.\/} {\bf 7} 768--773

\bibitem{Bouyssy1987_PRC36-380}
Bouyssy A, Mathiot J~F, Van~Giai N and Marcos S 1987 {\em Phys. Rev. C\/} {\bf
  36} 380--401

\bibitem{Long2006_PLB639-242}
Long W~H, Sagawa H, Meng J and Van~Giai N 2006 {\em Phys. Lett. B\/} {\bf 639}
  242--247

\bibitem{Long2006_PLB640-150}
Long W~H, Van~Giai N and Meng J 2006 {\em Phys. Lett. B\/} {\bf 640} 150--154

\bibitem{Liang2008_PRL101-122502}
Liang H, Van~Giai N and Meng J 2008 {\em Phys. Rev. Lett.\/} {\bf 101} 122502

\bibitem{Long2010_PRC81-024308}
Long W~H, Ring P, Van~Giai N and Meng J 2010 {\em Phys. Rev. C\/} {\bf 81}
  024308

\bibitem{Liang2012_PRC86-021302R}
Liang H, Zhao P, Ring P, Roca-Maza X and Meng J 2012 {\em Phys. Rev. C\/} {\bf
  86} 021302(R)

\bibitem{Niu2013_PLB723-172}
Niu Z~M, Niu Y~F, Liang H~Z, Long W~H, Nik\v{s}i\'{c} T, Vretenar D and Meng J
  2013 {\em Phys. Lett. B\/} {\bf 723} 172--176

\bibitem{Lisboa2004_PRC69-024319}
Lisboa R, Malheiro M, de~Castro A~S, Alberto P and Fiolhais M 2004 {\em Phys.
  Rev. C\/} {\bf 69} 024319

\bibitem{Alberto2005_PRC71-034313}
Alberto P, Lisboa R, Malheiro M and de~Castro A~S 2005 {\em Phys. Rev. C\/}
  {\bf 71} 034313

\bibitem{deCastro2006_PRC73-054309}
de~Castro A~S, Alberto P, Lisboa R and Malheiro M 2006 {\em Phys. Rev. C\/}
  {\bf 73} 054309

\bibitem{Zhou2003_PRC68-034323}
Zhou S~G, Meng J and Ring P 2003 {\em Phys. Rev. C\/} {\bf 68} 034323

\bibitem{Zhou2003_PRL91-262501}
Zhou S~G, Meng J and Ring P 2003 {\em Phys. Rev. Lett.\/} {\bf 91} 262501

\bibitem{He2006_EPJA28-265}
He X~T, Zhou S~G, Meng J, Zhao E~G and Scheid W 2006 {\em Eur. Phys. J. A\/}
  {\bf 28} 265--269

\bibitem{Liang2010_EPJA44-119}
Liang H, Long W~H, Meng J and Van~Giai N 2010 {\em Eur. Phys. J. A\/} {\bf 44}
  119--124

\bibitem{Mishustin2005_PRC71-035201}
Mishustin I~N, Satarov L~M, B\"urvenich T~J, Stocker H and Greiner W 2005 {\em
  Phys. Rev. C\/} {\bf 71} 035201

\bibitem{Song2009_CPL26-122102}
Song C~Y, Yao J~M and Meng J 2009 {\em Chin. Phys. Lett.\/} {\bf 26} 122102

\bibitem{Song2010_ChinPhysC34-1425}
Song C~Y and Yao J~M 2010 {\em Chin. Phys. C\/} {\bf 34} 1425--1427

\bibitem{Song2011_CPL28-092101}
Song C~Y, Yao J~M and Meng J 2011 {\em Chin. Phys. Lett.\/} {\bf 28} 092101

\bibitem{Meng1996_PRL77-3963}
Meng J and Ring P 1996 {\em Phys. Rev. Lett.\/} {\bf 77} 3963--3966

\bibitem{Meng1998_PRL80-460}
Meng J and Ring P 1998 {\em Phys. Rev. Lett.\/} {\bf 80} 460--463

\bibitem{Meng1998_NPA635-3}
Meng J 1998 {\em Nucl. Phys. A\/} {\bf 635} 3--42

\bibitem{Zhou2010_PRC82-011301R}
Zhou S~G, Meng J, Ring P and Zhao E~G 2010 {\em Phys. Rev. C\/} {\bf 82}
  011301(R)

\bibitem{Chen2012_PRC85-067301}
Chen Y, Li L, Liang H and Meng J 2012 {\em Phys. Rev. C\/} {\bf 85} 067301

\bibitem{Kukulin1989}
Kukulin V~I, Krasnopol'sky V~M and Hor\'{a}cek J 1989 {\em Theory of
  Resonances: Principles and Applications\/} (Kluwer Academic, Dordrecht)

\bibitem{Yang2001_CPL18-196}
Yang S~C, Meng J and Zhou S~G 2001 {\em Chin. Phys. Lett.\/} {\bf 18} 196--198

\bibitem{Zhang2004_PRC70-034308}
Zhang S~S, Meng J, Zhou S~G and Hillhouse G~C 2004 {\em Phys. Rev. C\/} {\bf
  70} 034308

\bibitem{Zhang2008_PRC77-014312}
Zhang L, Zhou S~G, Meng J and Zhao E~G 2008 {\em Phys. Rev. C\/} {\bf 77}
  014312

\bibitem{Zhou2009_JPB42-245001}
Zhou S~G, Meng J and Zhao E~G 2009 {\em J. Phys. B: At. Mol. Opt. Phys.\/} {\bf
  42} 245001

\bibitem{Guo2010_PRC82-034318}
Guo J~Y, Fang X~Z, Jiao P, Wang J and Yao B~M 2010 {\em Phys. Rev. C\/} {\bf
  82} 034318

\bibitem{Liu2012_PRC86-054312}
Liu Q, Guo J~Y, Niu Z~M and Chen S~W 2012 {\em Phys. Rev. C\/} {\bf 86} 054312

\bibitem{Hagino2004_NPA735-55}
Hagino K and Van~Giai N 2004 {\em Nucl. Phys. A\/} {\bf 735} 55--76

\bibitem{Li2010_PRC81-034311}
Li Z~P, Meng J, Zhang Y, Zhou S~G and Savushkin L~N 2010 {\em Phys. Rev. C\/}
  {\bf 81} 034311

\bibitem{Guo2005_PRC72-054319}
Guo J~Y, Wang R~D and Fang X~Z 2005 {\em Phys. Rev. C\/} {\bf 72} 054319

\bibitem{Guo2006_PRC74-024320}
Guo J~Y and Fang X~Z 2006 {\em Phys. Rev. C\/} {\bf 74} 024320

\bibitem{Liu2013_PRA87-052122}
Liu Q, Niu Z~M and Guo J~Y 2013 {\em Phys. Rev. A\/} {\bf 87} 052122

\bibitem{Lu2012_PRL109-072501}
Lu B~N, Zhao E~G and Zhou S~G 2012 {\em Phys. Rev. Lett.\/} {\bf 109} 072501

\bibitem{Lu2013_PRC88-024323}
Lu B~N, Zhao E~G and Zhou S~G 2013 {\em Phys. Rev. C\/} {\bf 88} 024323

\bibitem{Liang2011_PRC83-041301R}
Liang H, Zhao P, Zhang Y, Meng J and Van~Giai N 2011 {\em Phys. Rev. C\/} {\bf
  83} 041301(R)

\bibitem{Li2011_ChinPhysC35-825}
Li F~Q, Zhao P~W and Liang H~Z 2011 {\em Chin. Phys. C\/} {\bf 35} 825--828

\bibitem{Cooper1995_PR251-267}
Cooper F, Khare A and Sukhatme U 1995 {\em Phys. Rep.\/} {\bf 251} 267--385

\bibitem{Cooper2001}
Cooper F, Khare A and Sukhatme U 2001 {\em Supersymmetry in Quantum
  Mechanics\/} (World Scientific, Singapore)

\bibitem{Leviatan2004_PRL92-202501}
Leviatan A 2004 {\em Phys. Rev. Lett.\/} {\bf 92} 202501

\bibitem{Typel2008_NPA806-156}
Typel S 2008 {\em Nucl. Phys. A\/} {\bf 806} 156--178

\bibitem{Leviatan2009_PRL103-042502}
Leviatan A 2009 {\em Phys. Rev. Lett.\/} {\bf 103} 042502

\bibitem{Guo2012_PRC85-021302R}
Guo J~Y 2012 {\em Phys. Rev. C\/} {\bf 85} 021302(R)

\bibitem{Li2013_PRC87-044311}
Li D~P, Chen S~W and Guo J~Y 2013 {\em Phys. Rev. C\/} {\bf 87} 044311

\bibitem{Guo2014_PRL112-062502}
Guo J~Y, Chen S~W, Niu Z~M, Li D~P and Liu Q 2014 {\em Phys. Rev. Lett.\/} {\bf
  112} 062502

\bibitem{Wegner1994_AP506-77}
Wegner F 1994 {\em Ann. Phys. (Berlin)\/} {\bf 506} 77--91

\bibitem{Bylev1998_PLB428-329}
Bylev A~B and Pirner H~J 1998 {\em Phys. Lett. B\/} {\bf 428} 329--333

\bibitem{Wegner2001_PR348-77}
Wegner F~J 2001 {\em Phys. Rep.\/} {\bf 348} 77--89

\bibitem{Liang2013_PRC87-014334}
Liang H, Shen S, Zhao P and Meng J 2013 {\em Phys. Rev. C\/} {\bf 87} 014334

\bibitem{Shen2013_PRC88-024311}
Shen S, Liang H, Zhao P, Zhang S and Meng J 2013 {\em Phys. Rev. C\/} {\bf 88}
  024311

\bibitem{Ginocchio2005_PR414-165}
Ginocchio J~N 2005 {\em Phys. Rep.\/} {\bf 414} 165--261

\bibitem{Guo2016}
Guo J~Y, Liang H~Z, Meng J and Zhou S~G 2016 Relativistic symmetries in nuclear
  single-particle spectra {\em Relativistic Density Functional for Nuclear
  Structure\/} ({\em International Review of Nuclear Physics\/} vol~10) ed Meng
  J (World Scientific, Singapore) pp 219--262

\bibitem{Varshalovich1988}
Varshalovich D~A, Moskalev A~N and Khersonskii V~K 1988 {\em Quantum theory of
  angular momentum\/} (World Scientific, Singapore)

\bibitem{Bell1975_NPB98-151}
Bell J~S and Ruegg H 1975 {\em Nucl. Phys. B\/} {\bf 98} 151--153

\bibitem{Zhang2010_IJMPE19-55}
Zhang Y, Liang H~Z and Meng J 2010 {\em Int. J. Mod. Phys. E\/} {\bf 19} 55--62

\bibitem{Zhang2009_ChinPC33S1-113}
Zhang Y, Liang H~Z and Meng J 2009 {\em Chin. Phys. C\/} {\bf 33(S1)} 113--115

\bibitem{Zhang2009_CPL26-092401}
Zhang Y, Liang H~Z and Meng J 2009 {\em Chin. Phys. Lett.\/} {\bf 26} 092401

\bibitem{Li2011_SciChinaPMA54-231}
Li F~Q, Zhang Y, Liang H~Z and Meng J 2011 {\em Sci. China-Phys. Mech.
  Astron.\/} {\bf 54} 231--235

\bibitem{Tanimura2015_PTEP2015-073D01}
Tanimura Y, Hagino K and Liang H~Z 2015 {\em Prog. Theor. Exp. Phys.\/} {\bf
  2015} 073D01

\bibitem{Leviatan2001_PLB518-214}
Leviatan A and Ginocchio J~N 2001 {\em Phys. Lett. B\/} {\bf 518} 214--220

\bibitem{Infeld1951_RMP23-21}
Infeld L and Hull T~E 1951 {\em Rev. Mod. Phys.\/} {\bf 23} 21--68

\bibitem{Ginocchio1998_PLB425-1}
Ginocchio J~N and Leviatan A 1998 {\em Phys. Lett. B\/} {\bf 425} 1--5

\bibitem{Nieto2003_AP305-151}
Nieto L~M, Pecheritsin A~A and Samsonov B~F 2003 {\em Ann. Phys. (NY)\/} {\bf
  305} 151--189

\bibitem{Marcos2001_PLB513-30}
Marcos S, L\'{o}pez-Quelle M, Niembro R, Savushkin L~N and Bernardos P 2001
  {\em Phys. Lett. B\/} {\bf 513} 30--36

\bibitem{Arima1999_RIKEN-AF-NP-276}
Arima A 1999 Dynamical symmetries and nuclear structure RIKEN-AF-NP-276

\bibitem{Alberto2001_PRL86-5015}
Alberto P, Fiolhais M, Malheiro M, Delfino A and Chiapparini M 2001 {\em Phys.
  Rev. Lett.\/} {\bf 86} 5015--5018

\bibitem{Alberto2002_PRC65-034307}
Alberto P, Fiolhais M, Malheiro M, Delfino A and Chiapparini M 2002 {\em Phys.
  Rev. C\/} {\bf 65} 034307

\bibitem{Long2004_PRC69-034319}
Long W, Meng J, Van~Giai N and Zhou S~G 2004 {\em Phys. Rev. C\/} {\bf 69}
  034319

\bibitem{Ginocchio2005_PRL95-252501}
Ginocchio J~N 2005 {\em Phys. Rev. Lett.\/} {\bf 95} 252501

\bibitem{Elliott1958PRSA245-128}
Elliott J~P 1958 {\em Proc. Roy. Soc. A\/} {\bf 245} 128--145

\bibitem{Bogner2010_PPNP65-94}
Bogner S~K, Furnstahl R~J and Schwenk A 2010 {\em Prog. Part. Nucl. Phys.\/}
  {\bf 65} 94--147

\bibitem{Hammer2013_RMP85-197}
Hammer H~W, Nogga A and Schwenk A 2013 {\em Rev. Mod. Phys.\/} {\bf 85}
  197--217

\bibitem{Koepf1991_ZPA339-81}
Koepf W and Ring P 1991 {\em Z. Phys. A\/} {\bf 339} 81--90

\bibitem{Lisboa2010_PRC81-064324}
Lisboa R, Malheiro M, Alberto P, Fiolhais M and de~Castro A~S 2010 {\em Phys.
  Rev. C\/} {\bf 81} 064324

\bibitem{Ginocchio2011_JPCS267-012037}
Ginocchio J~N 2011 {\em J. Phys: Conf. Ser.\/} {\bf 267} 012037

\bibitem{Andrianov1984_TMP61-965}
Andrianov A~A, Borisov N~V, Ioffe M~V and \'{E}ides M 1984 {\em Theor. Math.
  Phys.\/} {\bf 61} 965--972

\bibitem{Ioffe2006_AP321-2552}
Ioffe M~V, Guilarte J~M and Valinevich P~A 2006 {\em Ann. Phys. (NY)\/} {\bf
  321} 2552--2565

\end{thebibliography}



\end{document}